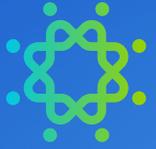

**Breakthrough Energy**
Sciences

# A 2030 United States Macro Grid

Unlocking Geographical Diversity to Accomplish Clean Energy Goals

January 2021

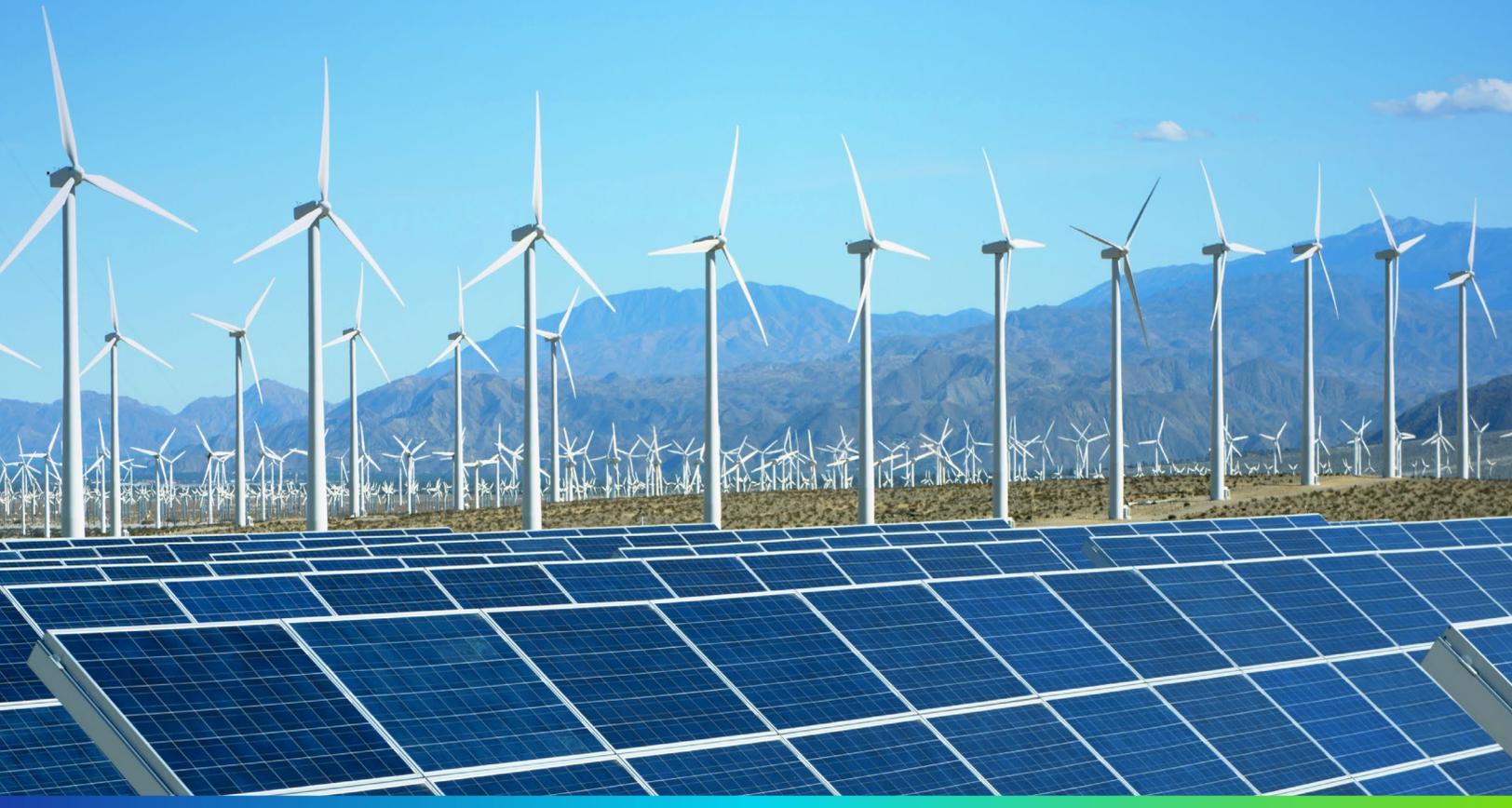

# Table of Contents

# Executive Summary

## Motivation and Contribution

Some U.S. states have set clean energy goals and targets in an effort to decarbonize their electricity sectors. There are many reasons for such goals and targets, including the increasingly apparent effects of climate change.[1] A handful of states (Washington, California, New York, and Virginia) are aiming for deep decarbonization by 2050 or earlier, a mere 30 years or less from today. The urgency of substantial carbon emissions reduction (50% or more by 2030) needed to avoid catastrophic climate impacts requires even more ambitious efforts than some of the original targets (e.g., a 30% renewable portfolio standard) set for between now and 2030.[2] With the cost of solar and wind energy falling faster than expected in recent years,[3] economics are also driving rapid expansion of clean energy investments.

With this in mind, this report examines combinations of interregional AC and High-Voltage DC (HVDC) transmission upgrades and additions to evaluate the benefits of large-scale transmission expansion.

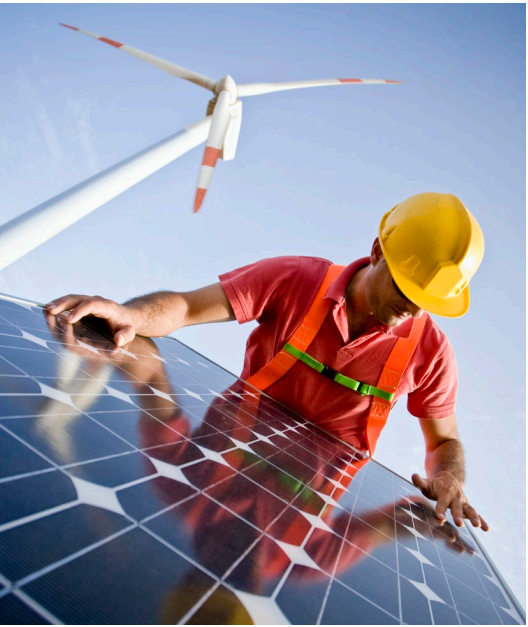

**More specifically, some of the major contributions explored in this Breakthrough Energy Sciences study are:**

1. Including all three interconnections (Western Interconnection, Eastern Interconnection, and Electric Reliability Council of Texas (ERCOT)) in a contiguous U.S. production cost model; and

2. Focusing on an ambitious goal of an electric grid powered by 70% clean energy* for the contiguous U.S. by the year 2030.

Considering the urgency of fighting climate change and the costs of solar and wind energy falling faster than expected, more progress must be made in the next ten years to ensure the U.S. is on a path to deep decarbonization by 2050.

Increased variable renewable energy on the grid presents challenges, such as maintaining system adequacy and operating the system economically. One way these challenges can be overcome is by improving collaboration across regional transmission organizations and the interconnections. Currently, the Western Interconnection, Eastern Interconnection, and ERCOT function effectively independently from one another. Stitching together the major regions of the grid through transmission upgrades and additions, thereby creating a Macro Grid, would allow the U.S. to further harness its abundant renewable resources and better balance electricity supply and demand across the country. In this study, the Breakthrough Energy Sciences team investigates four Macro Grid designs inspired by the National Renewable Energy Laboratory's (NREL's) Interconnections Seam Study[4] (referred to in this report as the 'Seams Study'). These Macro Grid designs illustrate how taking advantage of the benefits of geographical diversity helps address reliability and cost challenges.

> *The term Macro Grid is used by many groups in various ways. Dale Osborn's HVDC overlay design[5] is often referred to as THE Macro Grid. For this report, a Macro Grid is defined as any large-scale AC or DC transmission expansion design that stitches together the major regions of the grid.*

Data and models will play an important role across academia, industry, and governments in energy system planning and operation, policy implementation, and rule making. These data and models will be critical to achieving the U.S. energy system's deep decarbonization goals. Transparency and accessibility of this kind of research is crucial, and is a prime motivation for the approach taken herein. This report entails its own open-source data, models, and studies for maximum accessibility and transparency. With the open-source data and models developed by the Breakthrough Energy Sciences team, similar studies can be conducted by a much broader research community.

---

\* In this study, renewable energy resources include solar, wind, and geothermal. Clean energy resources include hydro, nuclear, solar, wind, and geothermal. While there are many other resources that are also renewable or clean energy sources, those resources are not considered in the model at this time. With their large-scale integration potential, solar and wind energy resources are the only generation types for which capacity is expanded in this study.

# Study Design

There are numerous ways to achieve the ambitious goal of an electric grid powered by 70% clean energy for the contiguous U.S. by 2030. This study investigates how the ambitious goals are met by adding a combination of new renewable generation capacity to existing generation capacity and expanding the transmission infrastructure (which includes lines rated at 69 kV and higher in this study) via one of four Macro Grid designs as shown in Figure 1. Other technologies, including energy storage technologies, are not considered in this study but will be investigated in future studies.

The efficacy of each design is determined using Breakthrough Energy Sciences' open-source production cost model. Using this full-year hourly model, details pertaining to energy generation, power transfer, and investment costs are able to be studied for each Macro Grid design.

**Design 1** refers to a scenario with no upgrades to the HVDC infrastructure. Only existing AC transmission systems are upgraded to accommodate high renewable energy penetration.

**Design 2a** refers to a scenario in which the existing back-to-back (B2B) HVDC converter stations, which are used to bridge two interconnections, are upgraded to allow substantially more power to transfer across an interconnection 'seam'.

**Design 2b** refers to a scenario in which the existing back-to-back HVDC converter stations are upgraded and three new long-distance HVDC lines connecting the Eastern and Western Interconnections are added.

**Design 3** refers to a scenario in which the existing back-to-back converter stations are not upgraded, but a new 16-line HVDC network spanning all three interconnections is added.

FIG. 01

## The Four Macro Grid Designs Considered in This Study

Transmission system upgrades are represented as comparisons against infrastructure in 2020.

**DESIGN 1**

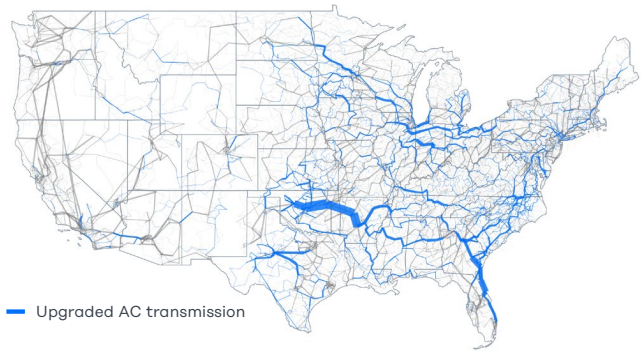

— Upgraded AC transmission

**DESIGN 2A**

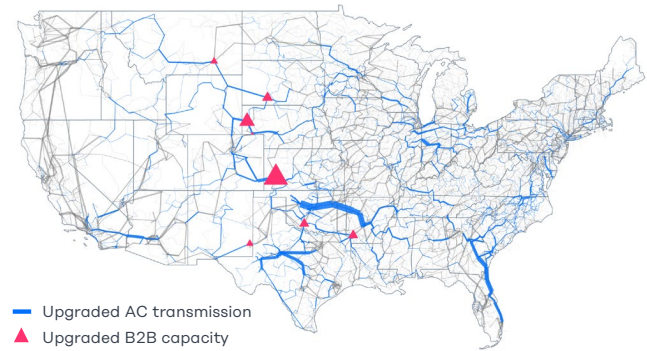

— Upgraded AC transmission
▲ Upgraded B2B capacity

**DESIGN 2B**

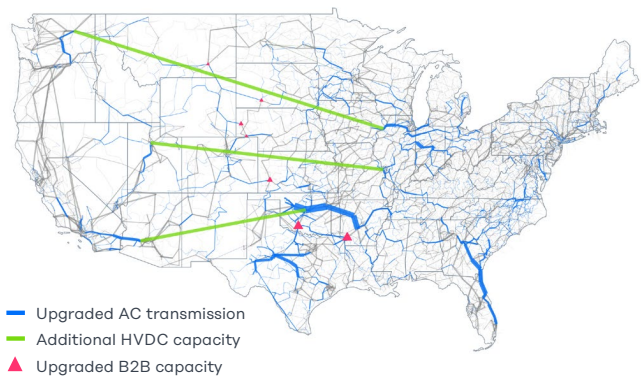

— Upgraded AC transmission
— Additional HVDC capacity
▲ Upgraded B2B capacity

**DESIGN 3**

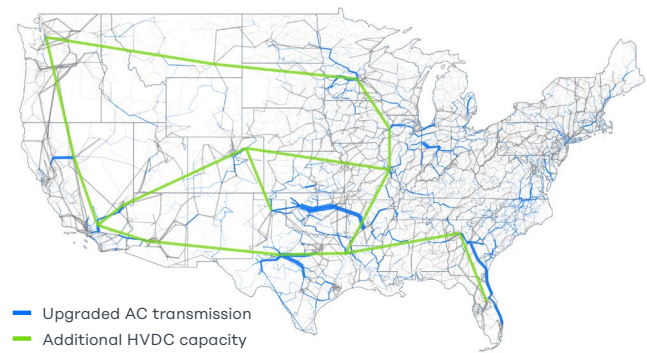

— Upgraded AC transmission
— Additional HVDC capacity

## Key Results

Macro Grids enable carbon-free electricity from renewable energy resources to reach geographically distant demand centers. When a Macro Grid includes new HVDC connections that span interconnection 'seams,' the interconnections can share renewable resources that in aggregate will be less variable and more reliable. New HVDC infrastructure is also beneficial in that clean energy goals can be achieved with fewer upgrades to AC transmission.

Achieving ambitious 70% clean energy goals results in substantial reductions in fuel costs, greenhouse gas (GHG) emissions, and other air pollutants. As shown in Table 1, the investments to attain these ambitious goals are not small; when compared with the investment costs of the 'current goals' scenario, each Macro Grid design is over four times as costly. Despite these costs, the benefits that would be realized by a proposed transformation of the U.S. electric grid are also substantial. The Macro Grid designs proposed to meet the 'ambitious goals' scenarios each yield large reductions in emissions when compared against the 'current goals' scenario; emissions reductions of over 42%, 37%, and 29% are observed for carbon dioxide, nitrogen oxide, and sulfur dioxide, respectively, for each design. The mass integration of renewable energy also produces reductions in fuel costs of over 46% across each of the four Macro Grid designs.

TABLE 01

### Investment Costs of Ambitious Goals vs. Current Goals

Achieving ambitious clean energy goals substantially reduces fuel costs, GHG emissions, and other air pollutants but requires considerable investments for each design ($B USD).

| DESIGN | WIND CAPACITY | SOLAR CAPACITY | AC LINES | AC TRANSFORMERS | HVDC TRANSMISSION | HVDC B2B STATIONS | TOTAL |
|---|---|---|---|---|---|---|---|
| Current Goals | $185 | $164 | $9 | $0.63 | $0 | $0 | $359 |
| Design 1 | $745 | $574 | $213 | $7.59 | $0 | $0 | $1,539 |
| Design 2a | $745 | $574 | $196 | $7.14 | $0 | $9.04 | $1,530 |
| Design 2b | $745 | $574 | $179 | $6.76 | $24.6 | $4.06 | $1,533 |
| Design 3 | $745 | $574 | $152 | $5.87 | $65.5 | $0 | $1,542 |

For the U.S. and each interconnection, Figure 2 shows the energy generation by resource for the scenarios explored in this report. The substantial increase in solar and wind generation is clear when comparing the current goals to the ambitious goals. Similarly, coal and natural gas generation are also substantially reduced for all four Macro Grid designs. Although these patterns are consistent across the U.S., the four designs studied in this report show slight variations in their generation mixes within each interconnection.

FIG. 02

## Energy Generation by Resource

Solar and wind generation increases substantially to meet the ambitious clean energy goals, reducing the need for coal and natural gas.

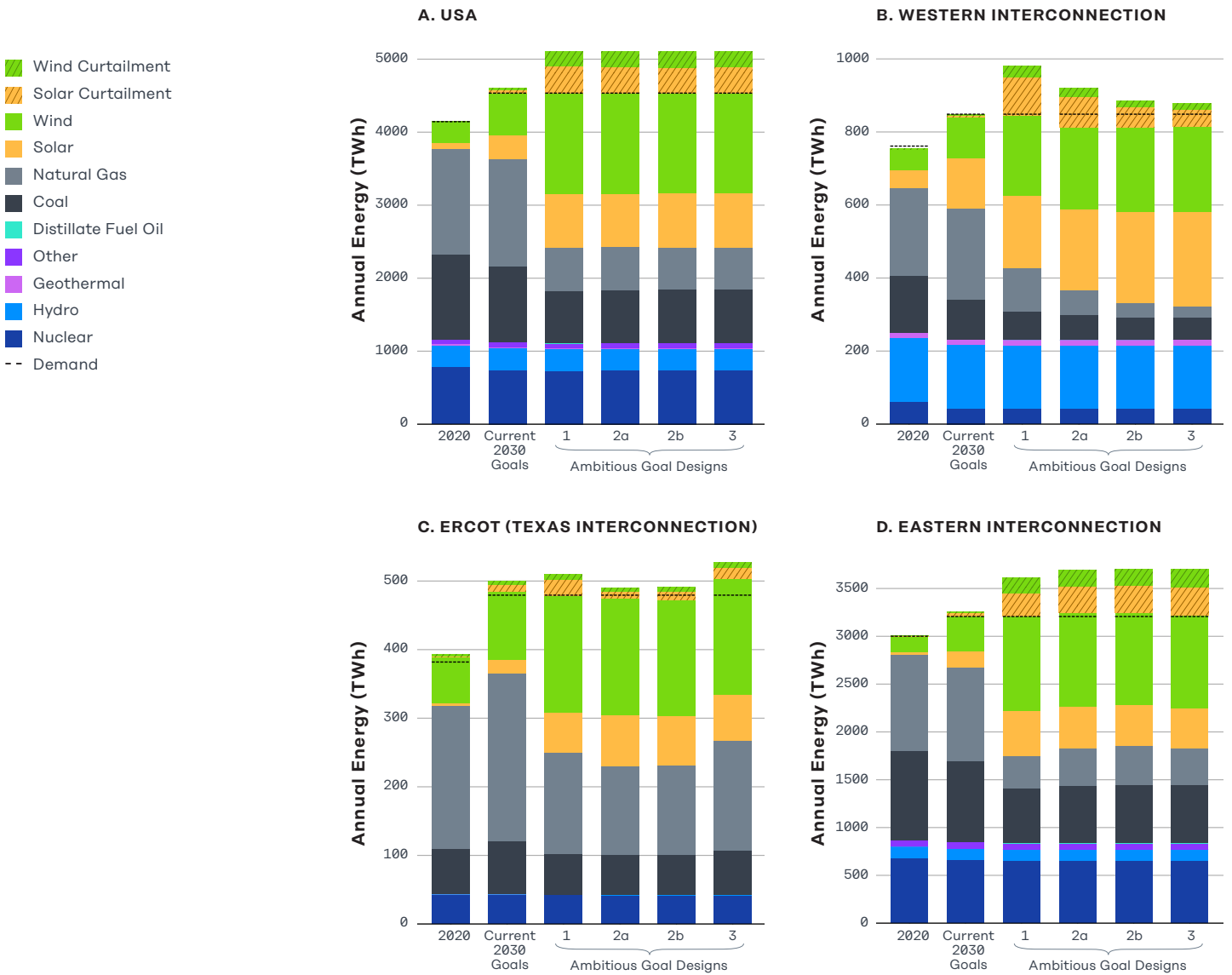

For each of the designs with HVDC transmission upgrades, the net energy transfer across the East-West seam is an export from the Eastern Interconnection to the Western Interconnection. As Figure 3 shows, the Eastern Interconnection is typically sending power West during the early morning and late evening hours when there is usually a wind energy surplus, and the Western Interconnection is typically sending power East during the daytime hours when there is an abundance of solar energy. The HVDC infrastructure in each design, including properly upgraded surrounding subsystems, is effectively utilized, with the overall capacity factors being greater than 65%. The instantaneous power flow across the East-West seam is highly correlated with the difference in instantaneous renewable penetration (i.e., renewable generation in the interconnection as a fraction of demand): higher renewable penetration in each interconnection typically results in higher power export from that interconnection.

FIG. 03

## Power Transfer East to West Across the Interconnection Seam

The Eastern Interconnection is typically sending power West during the early morning and late evening hours when there is usually a wind energy surplus, and the Western Interconnection is typically sending power East during the daytime hours when there is an abundance of solar energy. Positive power flow indicates exports from East to West and negative power flows indicate exports from West to East.

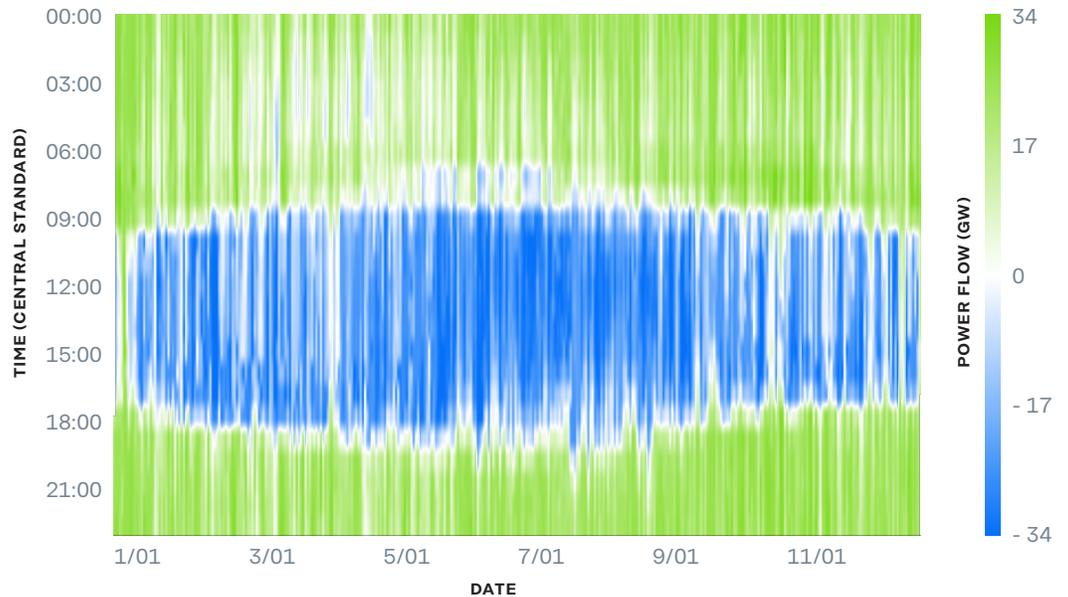

Although there are many similarities between the operation of the network across the four Macro Grid designs, there are also some key differences:

**01** Total AC transmission upgrades decrease as more HVDC capacity is added. Design 1 requires a 36% increase over current AC transmission capacity, Design 2a requires a 30% increase, Design 2b requires a 26% increase, and Design 3 requires a 23% increase.

**02** In Design 2a, the Eastern Interconnection exports, net, more than twice as much energy to the West as compared to Design 2b and 50% more as compared to Design 3. The cross-seam lines are utilized to similar degrees (65% absolute capacity factor in Design 2a, 70% in Design 2b, 68% in Design 3), but the broader reach of Design 3 allows states in the Southeast to import power from elsewhere in the country (e.g., Georgia and Florida are strong power importers, but are not well connected via the HVDC infrastructure in either Design 2a or Design 2b).

**03** In Design 1, where ERCOT is connected to the East only via two small back-to-back stations as they exist today, the energy transfer, net, is roughly equivalent in each direction. In Designs 2a and 2b, the existing back-to-back stations are expanded by a factor of ten in total; in these cases, ERCOT is a net importer of power from the East. However, in Design 3, with a broader HVDC network, ERCOT is a net exporter to both the Western and Eastern Interconnections.

## Policy Implications

Strong transmission policies are required to achieve the Macro Grid needed to meet the ambitious goal of an electric grid powered by 70% clean energy by 2030. The Federal Energy Regulatory Commission should take a greater role in coordinating regional, interregional, and interconnection-level transmission planning, in cooperation with the Department of Energy's existing Power Marketing Administrations. New policies can ensure that the costs of new transmission lines are distributed among all beneficiaries, and that jurisdictions hosting new transmission lines without a direct benefit are still compensated for their contribution. The federal government could also support the financing of these new lines via a combination of tax credits and loan programs. In concert, a suite of new federal policies could improve the efficiency of transmission planning and markets and enable robust participation by both the public and private sectors.

# Names and Affiliations of Authors and Contributors

**The following people designed and conducted scenario studies, and wrote this report:**
Yixing Xu,[a] Daniel Olsen,[a] Bainan Xia,[a] Dan Livengood,[a] Victoria Hunt,[a] Yifan Li,[a] Lane Smith[b]

**The following people built the software, website, and visualizations used in this study:**
Jon Hagg,[a] Victoria Hunt,[a] Anna Hurlimann,[a] Kamilah Jenkins,[a] Kaspar Mueller,[a] Daniel Muldrew,[a] Daniel Olsen,[a] Merrielle Ondreicka,[a] Benjamin Rouillé-d'Orfeuil,[a] Bainan Xia,[a] Nina Vincent,[b] Lane Smith[b]

**The following people contributed to the policy content:**
Robin Millican,[a] Rob Gramlich[c]

---

a. Breakthrough Energy Sciences, b. University of Washington, c. Grid Strategies LLC

# Acknowledgements

*Thanks to Dhileep Sivam for his guidance and support in conducting this research and producing this report.*

**The following people from Breakthrough Energy provided crucial strategic guidance and support in making this report possible:**

Jonah Goldman, Robin Millican, Ken Caldeira

# Technical Reviewers

The following people provided a technical review of this report and the scenario results. They were not involved in the study design and evaluation — for that, the authors bear sole responsibility. Similarly, the content and conclusions of this report, including any errors and omissions, are the sole responsibility of the authors. The technical reviewers' affiliations in no way imply that those organizations support or endorse this work in any way.

**Jay Caspary** - *Grid Strategies LLC (formerly a Director at the Southwest Power Pool)*

**Eli Massey** - *Midwest Independent System Operator*

**Aaron Bloom -** *Chair, System Planning Working Group, Energy Systems Integration Group*

## About Breakthrough Energy Sciences

Breakthrough Energy Sciences develops rigorous analytic tools to determine the best evidence-based approaches to meet zero-emission targets in a reliable and economic way. Breakthrough Energy Sciences' open-source data, models, and conducted studies support energy system planning and operation, policy-making, and investments.

Open-source data, models, the software framework, interactive data visualizations, and supporting materials can be found at:

https://science.breakthroughenergy.org/

Questions or comments about this report are welcome via email at:

sciences@breakthroughenergy.org

## Reference

*Please cite as:*

Yixing Xu, Daniel Olsen, Bainan Xia, Dan Livengood, Victoria Hunt, Yifan Li, and Lane Smith. 2021. "A 2030 United States Macro Grid: Unlocking Geographical Diversity to Accomplish Clean Energy Goals." Seattle, WA: Breakthrough Energy Sciences.

https://science.breakthroughenergy.org/publications/MacroGridReport.pdf

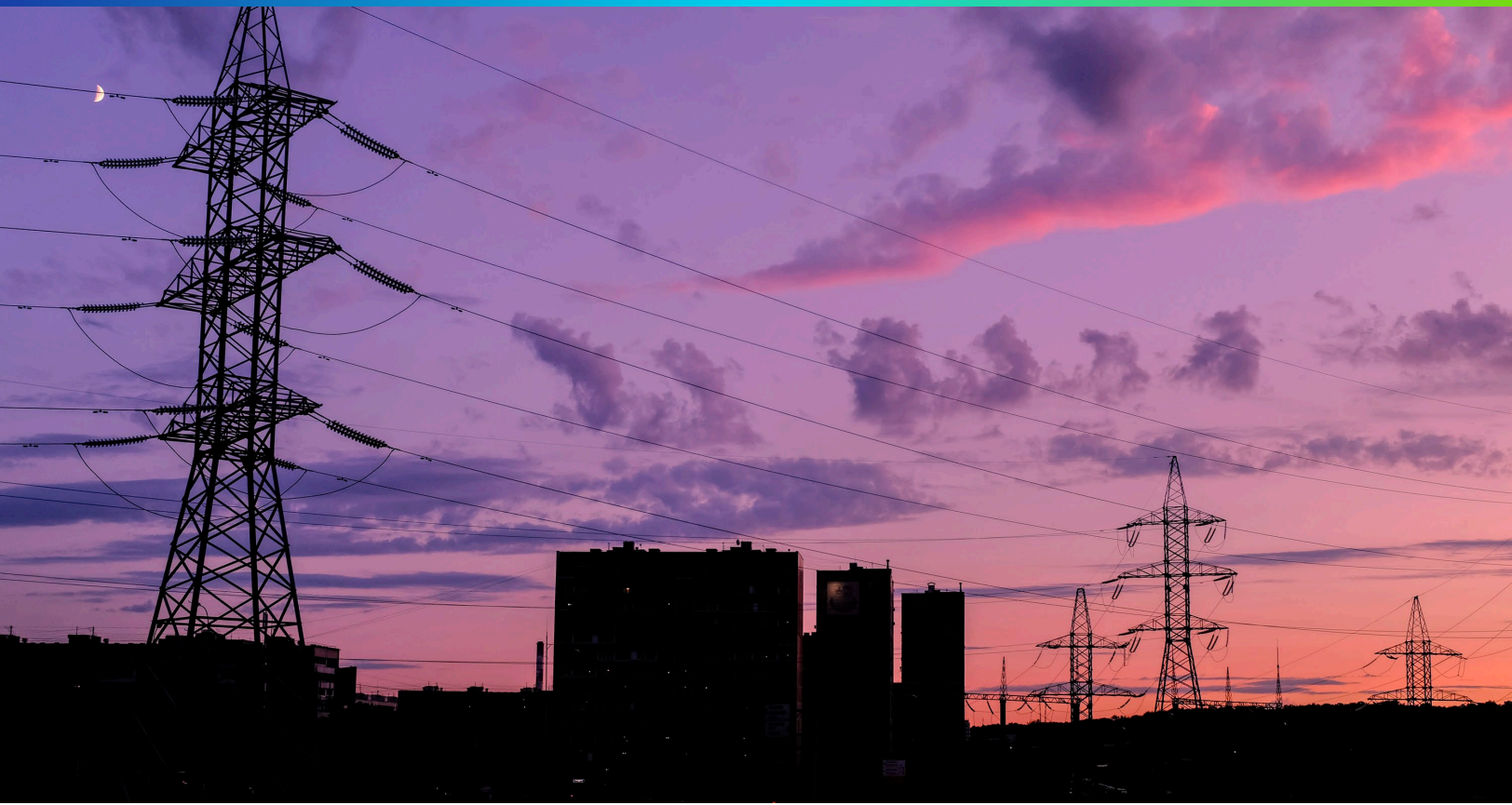

# 1.0 Introduction — The Motivation for a U.S. Macro Grid

Some U.S. states have set clean energy goals and targets in an effort to decarbonize their electricity sectors. There are many reasons for such goals and targets, including the increasingly apparent effects of climate change.[1] A handful of states (Washington, California, New York, and Virginia) are aiming for deep decarbonization by 2050 or earlier, a mere 30 years or less from today. The urgency of substantial carbon emissions reduction (50% or more by 2030) needed to avoid catastrophic climate impacts requires even more ambitious efforts than some of the original targets (e.g., a 30% renewable portfolio standard) set for between now and 2030.[2]

Economics are also driving rapid expansion of clean energy investments on the grid even without ambitious goals. In recent years, with the cost of solar and wind falling faster than expected,[3] some states without ambitious renewable energy goals have nonetheless experienced rapid renewable capacity expansion. Texas, for instance, has already exceeded its 2025 goal and has long decided to let the market decide what is best for its grid. This has led to 20% of its energy coming from wind and a rapid growth of solar installations coming online in the coming years.[6]

Breakthrough Energy Sciences

A 2030 United States Macro Grid | 12

Increased clean energy on the grid does come with a few challenges. Ensuring reliability and satisfying peaks in demand are not trivial tasks, but are not insurmountable either. One way these challenges can be overcome is by improving collaboration across regional transmission organizations (RTOs) and the interconnections. Currently the Western Interconnection, Eastern Interconnection, and Electric Reliability Council of Texas (ERCOT) effectively function independently from one another. Building a Macro Grid (i.e., improving the connectivity between the major regions of the grid through transmission upgrades and additions) would unlock the benefits of geographical diversity towards meeting these capacity and reliability challenges.

*The term Macro Grid is used by many groups in various ways. Dale Osborn's HVDC overlay design[5] is often referred to as THE Macro Grid. For this report, a Macro Grid is defined as any large-scale AC or DC transmission expansion design that stitches together the major regions of the grid.*

Unfortunately, investment in transmission has been lagging behind generation expansion and load growth for many years. In an attempt to encourage more investment in transmission, the Federal Energy Regulatory Commission (FERC) established Order Nos. 890 and 1000. Order No. 890, issued in 2007, outlined general requirements for local as well as regional transmission planning practices and procedures. Order No. 1000, issued in 2011, laid out specific requirements for:

1. Regional transmission planning;
2. Consideration of transmission needs driven by public policy requirements;
3. Non-incumbent transmission development;
4. Interregional transmission coordination; and
5. Cost allocation procedures for transmission facilities that have been selected in a regional transmission plan.

In 2016, FERC issued an initial report on transmission investment performance metrics to inform whether additional FERC action would be necessary beyond Order 890 and Order 1000 to facilitate more efficient or cost-effective transmission development to sufficiently satisfy the transmission needs.[7] The report, along with current and former commissioners, acknowledges that these Orders might not be working as intended and that it may be time to return to this discussion. The continued rapid pace of renewable generation deployment combined with the aging transmission infrastructure has made the need for transmission development more urgent if renewable growth is to be maintained at its projected pace over the next few decades.

## 1.1 - Macro Grid Interest

The interest in expanding and upgrading the transmission system to support higher levels of clean energy has been growing for years, if not decades. In a presentation about the National Renewable Energy Laboratory's (NREL's) Interconnections Seam Study[8,9] (referred to in this report as the 'Seams Study'), Aaron Bloom noted interest in similar concepts and projects in the following publications over the last century:

- 1923: Chicago Tribune's "Tying the Seasons to Industry"
- 1952: Bureau of Reclamation's "Super Transmission System"
- 1979: Bonneville Power Administration's "Interconnection of the Eastern and Western Interconnections"
- 1994: Western Area Power Administration's "East/West AC Intertie Feasibility Study"
- 2002: Department of Energy's "National Transmission Study"

Building upon this interest in a Macro Grid, NREL completed the Eastern Renewable Generation Integration Study,[10] the latest in a series of studies exploring how to increase clean energy penetration on the grid; additional studies in this series include the Eastern Wind Integration and Transmission Study and the Western Wind and Solar Integration Studies.[11,12,13] Ongoing work at NREL expands on that foundation via the North American Renewable Integration Study[14] (results pending) and the Seams Study[4,9] (preliminary results released in October 2020 as a journal article preprint).

Outside of NREL, many other groups are joining the discussion. One such effort is the Macro Grid Initiative, a joint initiative of the American Council on Renewable Energy and Americans for a Clean Energy Grid, which launched in 2020 with philanthropic support from Breakthrough Energy. The Macro Grid Initiative seeks to expand and upgrade the nation's transmission network to deliver job growth, economic development, a cleaner environment, and lower costs for consumers. As stated on their website:[15]

> *"The 15 states between the Rockies and the Mississippi River account for 88 percent of the nation's wind technical potential and 56 percent of solar technical potential. However, this region is home to only 30 percent of expected 2050 electricity demand. Through a transmission Macro Grid, centers of high renewable resources can be connected with centers of high electric demand, which can enhance grid resiliency and dramatically reduce carbon emissions."*
>
> *— Macro Grid Initiative*

**The Macro Grid Initiative has three primary objectives:**

- Expand and upgrade interregional transmission lines;
- Increase transmission development at the 'seams' between regions; and
- Build a nationwide high-voltage direct current (HVDC) network, optimized for connecting generation from the nation's best solar and wind resources to load centers.

Aligning with these objectives, this Breakthrough Energy Sciences report evaluates the benefits of large-scale transmission expansion on a simulated U.S. electric grid. More specifically, some of the major contributions explored in this study are:

1. Including all three interconnections (Western Interconnection, Eastern Interconnection, and ERCOT) in a contiguous U.S. production cost model;* and
2. Focusing on an ambitious goal of an electric grid powered by 70% clean energy** for the contiguous U.S. by the year 2030.

Considering the urgency of fighting climate change and the cost of solar and wind falling faster than expected, more progress needs to be made in the next ten years to ensure the U.S. is on a path to deep decarbonization by 2050.

## 1.2 - Value of Open-Source Modeling

Data and models will play an important role across academia, industry, and governments in energy system planning and operation, policy implementation, and rule making. These data and models will be critical to achieving the U.S. energy system's deep decarbonization goals. Transparency and accessibility of this research are crucial and a prime motivator for this and many other open-source energy modeling efforts.[16,17]

There have been efforts by government agencies — such as the U.S. Department of Energy — to commission and publish important research on the energy system for use in the public domain. However, these efforts are not always open source and all information is not always made fully available to the public. The high cost of proprietary data, models, and consulting fees makes it very difficult for non-government entities to replicate or conduct similar studies. Free open-source data and models provide any person the opportunity to conduct similar studies at a much lower cost. Proprietary models and associated research and consulting fees can be cost prohibitive to many researchers, experts, and policy makers in the clean energy space. Compounding this problem, the dearth of accurate open-source energy data and models has failed to provide stakeholders with a reasonable alternative to proprietary models.

---

\* A Production Cost Model handles the process of allocating the required demand between the available generation units such that the cost of operation is minimized, considering technical constraints such as transmission capacities.

\*\* In this study, renewable energy resources include solar, wind, and geothermal. Clean energy resources include hydro, nuclear, solar, wind, and geothermal. While there are many other resources that are also renewable or clean energy sources, those resources are not considered in the model at this time. With their large-scale integration potential, solar and wind energy resources are the only generation types for which capacity is expanded in this study.

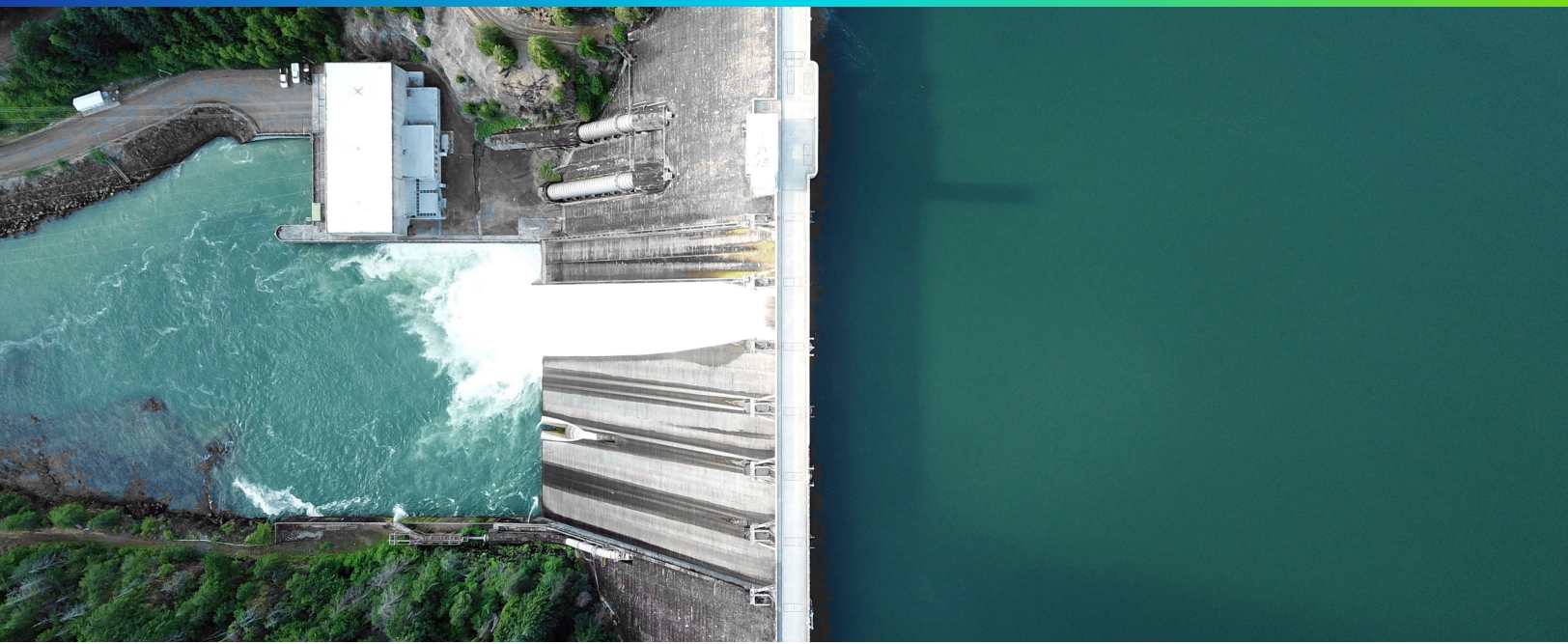

> Compared to the blackbox of proprietary data and models, open-source energy data and models provide transparency and reproducibility, which are particularly important in the energy policy-making process.

Open-source data and models have the ability to provide an alternative to their proprietary counterparts. Using the limited publicly available data, including proposed HVDC designs from NREL's Seams Study,[4,5,18,19] the Breakthrough Energy Sciences team was able to conduct a simulation of high-resolution operations for the described ambitious designs using Breakthrough Energy's open-source data and model. As is further discussed in Section 4, the Breakthrough Energy Sciences team found similar high-level conclusions to those reported in the conference presentations[8] and journal articles[4,18,19] related to NREL's Seams Study. Additionally, Section 4 presents the detailed results from the single-year hourly production cost simulation for the three interconnections.

Compared to the blackbox of proprietary data and models, open-source energy data and models provide transparency and reproducibility, which are particularly important in the energy policy-making process. These qualities allow groups representing different interests or perspectives to scrutinize and validate the data, assumptions, methodologies, and results to reach common ground or identify differences. Such openness may help provide shared information that can act as a basis for new policies. Thus, the Breakthrough Energy Sciences team has made available all input and output data and the model for this study.[20]

Making data and models open source reduces repetitive efforts among different groups and allows researchers worldwide to contribute and collaborate more effectively. Currently, collaborations building upon Breakthrough Energy's open-source data and model are in place with researchers from Columbia University, University of Washington, Texas A&M University, University of California Irvine, and University of Houston. These collaborations will add more modeling features to study future energy system operations and planning, including transportation and building electrification, demand flexibility, and system reliability.

## 1.3 - Previous Work

To conduct an investigation into what the U.S. power grid may look like and how it may operate in the year 2030, it is first necessary to have a good model of the current grid infrastructure and profiles of demand and renewable resource availability. Several previous efforts have produced models with some of these characteristics, but to the best of the authors' knowledge there are no publicly available data sets providing granular temporal and spatial data of the full U.S. power grid. Some previous efforts include:

- 'Synthetic' models of the U.S. power grid, which provide high spatial resolution but not high temporal resolution of time-varying parameters (e.g., NREL's ReEDS model[21,22] with 134 model balancing authorities, and Texas A&M University's Electric Grid Test Case Repository,[23] which provided the basis for this study's transmission network).

- Models with high temporal resolution, but not high geographic coverage (e.g., the Reliability Test System of the Grid Modernization Lab Consortium).[24]

- Proprietary models (e.g., PLEXOS from Energy Exemplar and WIS:dom from Vibrant Clean Energy), which are typically bundled with production cost model or capacity expansion model software,* whose prices can make them inaccessible to many potential users.

As will be discussed in Section 2, this study leverages the 82,000-node model of the U.S. power grid developed and published by the Breakthrough Energy Sciences team. Details of the development of this 82,000-node data set are available in Sections II and III of Breakthrough Energy Sciences' 2020 IEEE Power & Energy Society General Meeting paper.[25] This data set is further developed to simulate scenarios of future grids, featuring grid modifications intended to evaluate various policy choices, as will be discussed in Section 3.

---

\* A Capacity Expansion Model optimizes not only the operation of existing grid infrastructure, but also the installation of new infrastructure or the retirement of existing infrastructure, given assumptions about future electricity demand, fuel prices, technology costs and performance, policies and regulations, etc.

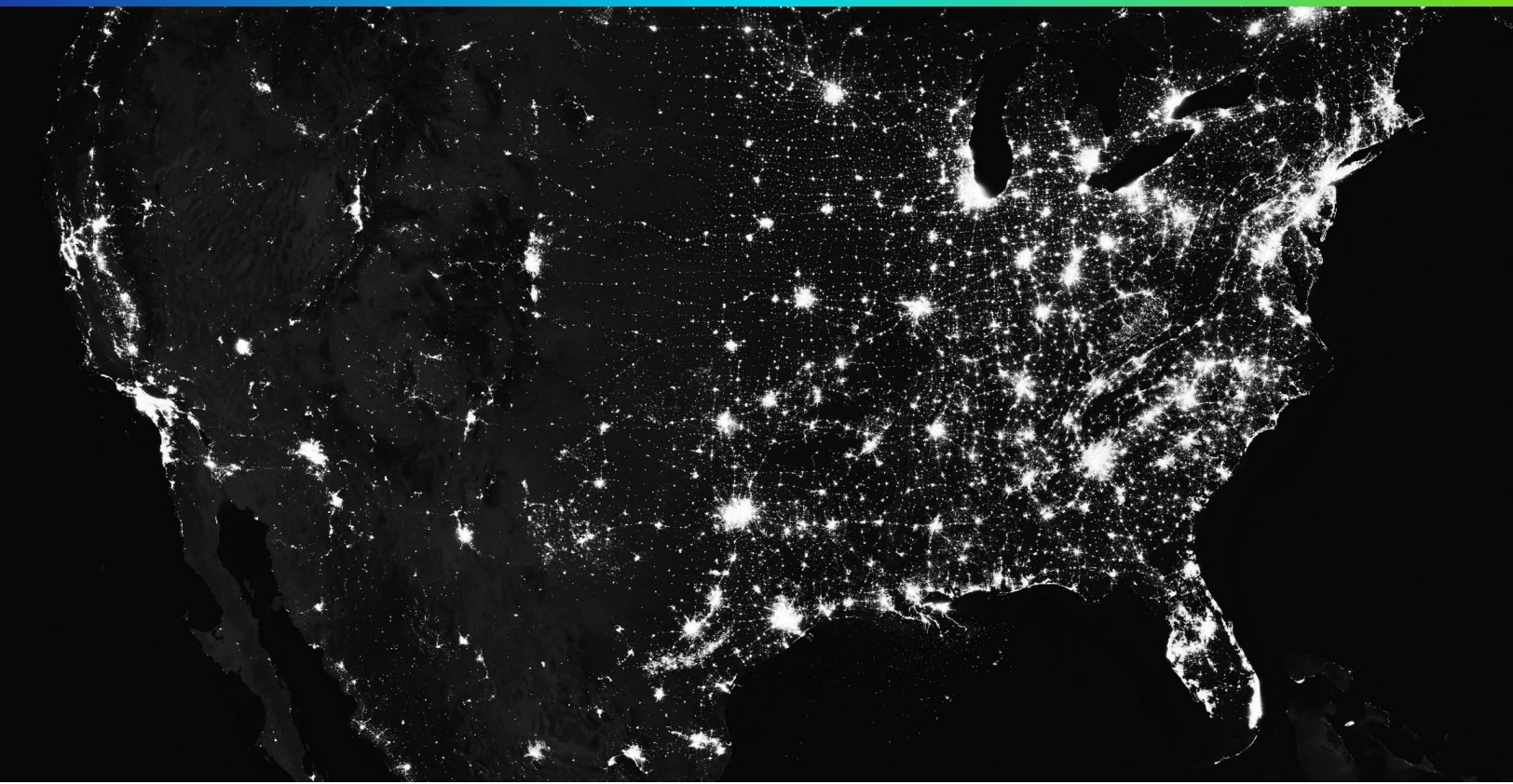

# 2.0 Model Overview and Performance

The model of the U.S. power system that was used as the basis of this study is high-resolution in terms of both space and time, with 82,000 nodes, over 104,000 branches, and hourly resolution profiles for both electric demand and generation potential from wind, solar, and hydro resources. The aggregate transmission capacity is 307 TW-miles, connecting 14,000 generators to 38,000 demand buses. This model was assembled using only publicly available information, allowing it to be released publicly without confidentiality concerns.

The network topology represents the lines on the U.S. transmission system at 69 kV and higher. It is "designed to be statistically and functionally similar to actual electric grids while containing no confidential critical energy infrastructure information,"[23] and is updated to reflect the generation capacities and power flow patterns of the year 2016 (chosen as the validation year). A detailed description of the source data that were used to develop this model and the corresponding synchronized weather and demand profiles has been published,[25] and the data set itself is available for download.[26]

The network model and hourly profiles are input into a production cost model and run as a series of 24-hour multi-period Optimal Power Flow problems, using the Direct Current power flow approximation (this sort of problem is often referred to as a 'DCOPF'). The initial conditions of each problem are constrained by the final conditions of the preceding problem, enabling temporal constraints such as generator ramp rate limits to 'bridge' the individual optimization problems.

Although this approach does not model several important aspects of power system operation (e.g., security-constrained unit commitment, ancillary services, weather uncertainty), omitting these details for a high-level analysis of Macro Grid designs is assumed to be an acceptable tradeoff for the reduced computational complexity. A further discussion of these aspects can be found in Section 6.

The model and profiles were evaluated by comparing the results from subsequent simulations to the historical results from the year 2016.[25] The primary comparison metric was the amount of energy generated by generators of each type, in each state. Cost curve coefficients for each class of generator were scaled based primarily on reported fuel expenditures in 2016 and secondarily tuned to target historical energy production values.[27] Simultaneously, various system parameters were tuned to better reflect congestion levels observed in historical real-time and day-ahead markets. Finally, once the generation patterns within each interconnection were representative of the historical data, the cost curves for each interconnection were scaled to produce prices at transmission hubs that are consistent with the historical records.

Once the model was tuned to be representative of the year 2016, it was then updated to reflect the year 2020. In accordance with EIA Form 860, generators are retired and removed from the system based on their recorded retirements from 2017 through 2019. Retirements across the U.S. from 2017 through 2019 primarily consisted of conventional steam coal power plants (35 GW), natural gas steam turbines (12 GW), and a few nuclear power plants (2 GW). Similarly, generators that were added from 2017 through 2019 or planned for the early 2020s are also added to the system. Additions across the U.S. from 2017 through 2019 primarily consisted of solar photovoltaics (15 GW), onshore wind turbines (22 GW), and natural gas combined-cycle power plants (33 GW).[28] Demand profile shapes were kept the same as in 2016 but scaled to account for load growth from 2016 to 2020, which ranged from -0.4% to 1.7% per year across the U.S.[29,30,31,32,33,34,35] Some additional transmission upgrades were required to ensure network feasibility and prevent severe congestion from appearing when demand is scaled up to 2020 levels. With these updates in place, the 2020 electric grid representation is established.

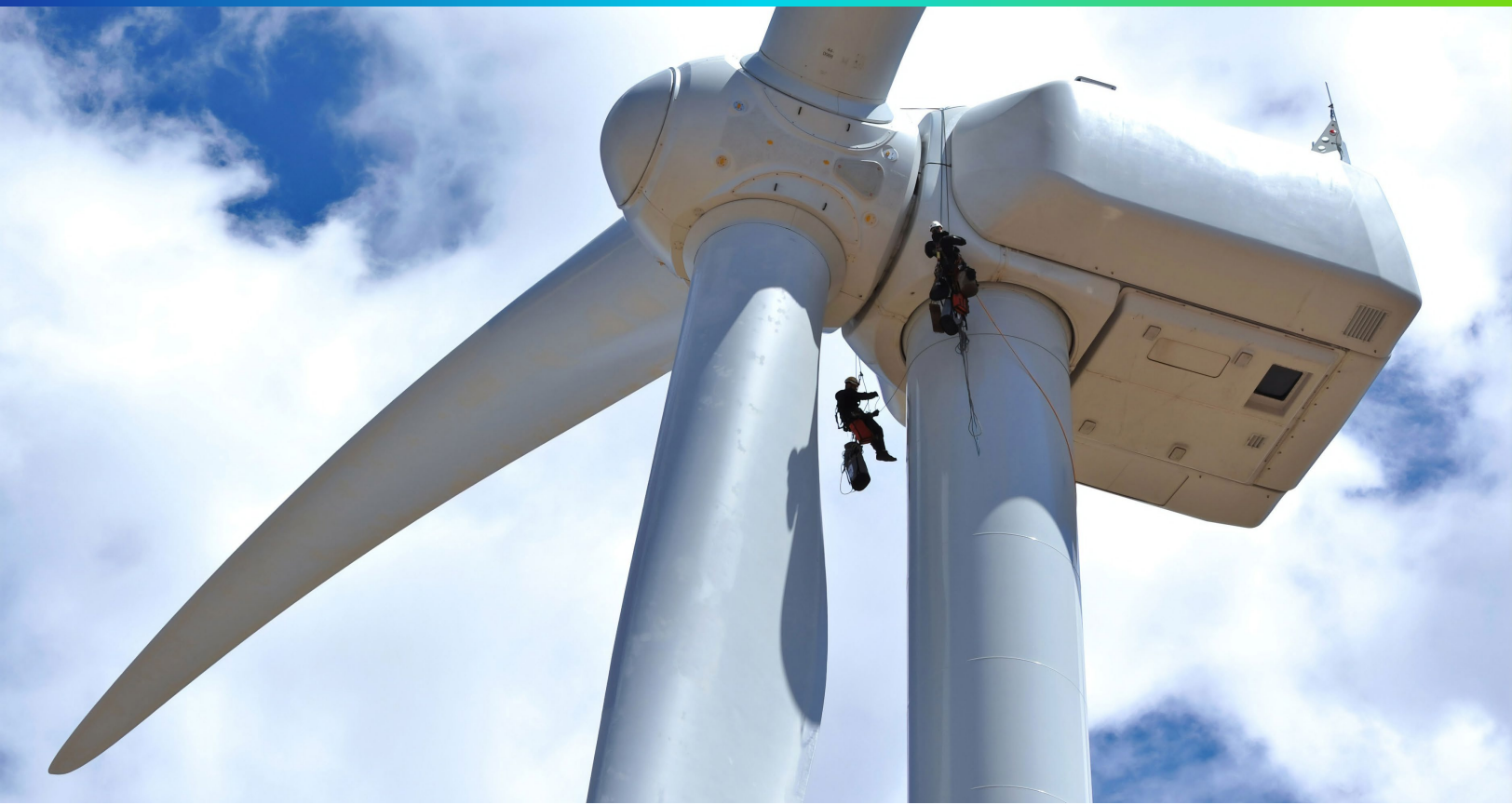

# 3.0   2030 U.S. Scenario Studies

As discussed in Section 2, this project focuses on the development of an open-source production cost model with publicly available data. In the fall of 2018, California's legislature passed a bill to set a goal of an electric grid powered by 60% renewable energy by 2030.[36] During the following years, additional states expanded upon or introduced their own renewable and clean energy goals, further solidifying the need to plan for an electric grid powered by substantial penetrations of clean energy.[2] These renewable and clean energy goals formed the basis for a series of case studies that were ultimately used to showcase this production cost tool's functionality. As large electricity infrastructure projects often take several years to design, permit, and construct, planning infrastructure to meet 2030 goals is an urgent priority.

## 3.1 - 2030 U.S. Scenario Setup

To establish the 2030 scenarios, the 2020 electric grid representation described in Section 2 is first updated based on any announced generator retirements from 2020 to 2030.[28] Wide-scale expected changes in coal and natural gas capacities from 2020 to 2030 were based on the NREL ReEDS Mid-Case projections[37] and achieved by scaling down or up the capacities of the entire generator fleet on a per-state basis. Cost curve coefficients for all generators were kept the same from 2016 to 2030, although this is certainly an area for future sensitivity analysis.

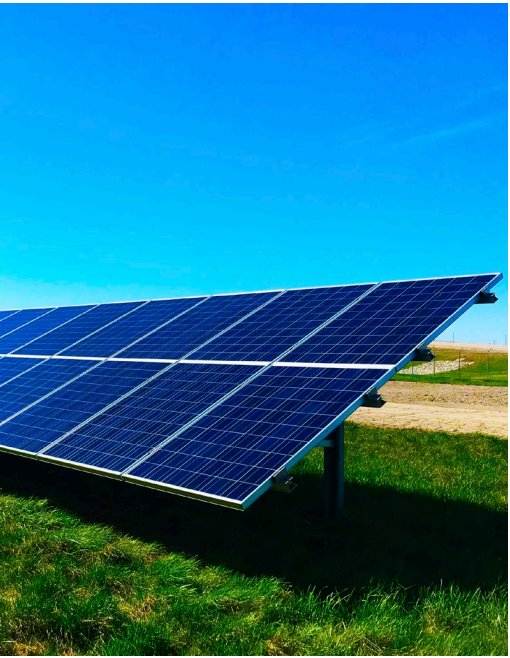

Last but not least, demand curves were scaled up by zone based on the following sources:

– Western Interconnection: California load growth is projected using the California Energy Demand 2018-2030 Revised Forecast,[29] while all other states feature 1% growth per year;

– Eastern Interconnection: Load growth is determined using projections from ISO-NE, MISO, NERC, NYISO, PJM, and several other source;[31,32,33,34,35] and

– ERCOT: Load growth is determined using projections from ERCOT.[30]

Each of these load growth projections result in an aggregate load growth of approximately 10% from 2020 to 2030.

For each state with a renewable energy goal, a total clean energy generation target is determined based on each state's 2030 goal (e.g., 60% in California) and their projected 2030 demand (e.g., 339 TWh in California).[29] If there was no curtailment of clean energy sources, it would be possible to estimate the capacity of new clean energy resources required to reach the total clean energy generation target. Instead, the increased penetration of renewable sources will result in increased curtailment unless there are also substantial increases in transmission capacity and energy storage.* The total clean energy targets are treated collectively, assuming that there is some mechanism for sharing clean electricity obligations so that states are not forced to meet their target solely from generation inside their borders.

To reach renewable energy goals, along with adding more renewable energy capacity, the production cost model is run iteratively and the transmission system is upgraded before each run based on the following:

1  Identifying the branches that experienced frequent, substantial congestion in the previous run; and

2  Estimating a benefit per MW-mile of increased transmission capacity.

This process is repeated until sufficient renewable energy is no longer curtailed so that all states with goals meet the total clean energy target. Although this method is a heuristic approach to transmission expansion, it maintains tractability when used on such a detailed network model and roughly reflects current industry practice for market efficiency planning studies to identify and evaluate transmission projects. This method is able to give an estimate of total investment cost that is closer to what is likely to happen in reality under current business planning processes, as opposed to just a purely academic exercise. Economic transmission planning in today's independent system operators (ISOs), RTOs, and utilities are usually performed in such a way that individual congested flowgates are studied one-by-one (via identification and congestion relief analysis) and solution candidates are evaluated piece-by-piece using an 8,760-hour production cost simulation.[38,39]

---

* For this study, the focus is exclusively on the transmission system and renewable energy capacity (e.g., solar and wind generation). The impact of energy storage, hydrogen production and storage, and other technologies will be investigated in future studies.

It is worth mentioning that although the solution's optimality could improve with an advanced capacity expansion model based on strict mathematical programming, the general trends and major observations within this report would still hold true. Similarly, adding energy storage and demand flexibility, which are omitted from this study to reduce computational effort, could help integrate variable renewable sources without changing many of the report's general conclusions. These and other opportunities for future work are further discussed in Section 6.

Given current assumptions about the future, substantial amounts of transmission investment cannot be avoided for the system to be able to guarantee renewable generation deliverability, system adequacy, and reliability. New system operating patterns, interregional exchanges, and major investments on severely congested paths will be needed regardless of what choices are made with respect to generation. Further, the performance of AC or HVDC solutions will likely yield even greater benefits when simulated using a capacity expansion model.

## 3.2 - Current Goals

For the scenarios targeting the current goals of 2030, this study models the required expansion of new renewable generation capacity and AC transmission to meet the states' existing renewable energy goals, as shown in Figure 4 and listed in Table 2. For the scenarios presented in this report, states with goals are allowed to collaborate within each interconnection with other states that have goals. In order to 'meet the goal', enough clean energy must be produced within each interconnection's participating states to satisfy the sum of the interconnection's clean energy goals. New solar and wind capacity is added in each participating state proportional to its 2020 capacities.

FIG. 04

### Goals up to 2030

To decrease carbon emissions, save money, improve air quality, and reduce dependence on imported fuels, many U.S. states have established clean energy goals of varying ambition and scale.

- No goals
- 0-15%
- 15-20%
- 20-30%
- 30-50%
- 50-70%
- 70-100%

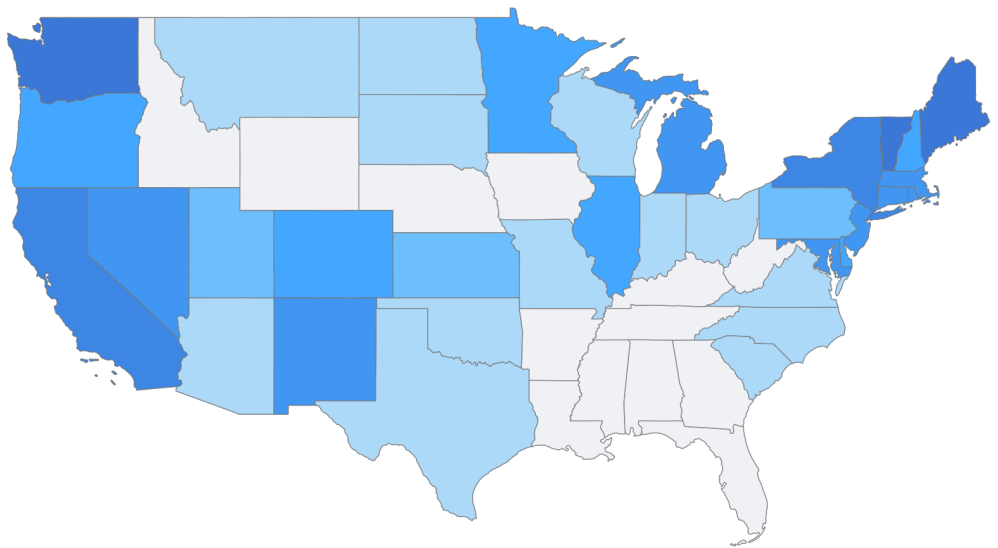

As mentioned above, some states do not have 2030 clean electricity goals. In the Western Interconnection, the two states without goals, Idaho and Wyoming, are nonetheless assumed to collaborate with the rest of the Western Interconnection, and therefore their capacity is also scaled up proportionally. This modeling decision is due to their substantial renewable buildouts despite the lack of a mandate, as well as their active participation in interconnection-wide renewable energy integration efforts.[40] On the other hand, in the Eastern Interconnection, there are a substantial number of states without 2030 goals and with limited participation in interstate renewable coordination efforts. For this study, those states are assumed to not collaborate with the states with goals. The growth of renewables in these states is projected based on the Mid-Case projections in NREL's Standard Scenarios,[37] reflecting deployment based on economics rather than clean electricity mandates. Last but not least, for ERCOT, solar and wind capacity projections are based on reports from ERCOT itself.[6,41]

To determine whether the renewable energy goals have been met, the model sums up the total renewable energy that is generated and delivered to consumers (i.e., energy that is not curtailed) from all states included in the collaborative efforts within each interconnection. Once this sum reaches the combined total of each state's renewable energy goals, the heuristic process described in Section 3.1 ends and the new renewable generation and transmission capacity is deemed to be the amount of expansion necessary to meet the 2030 goals for that particular scenario. By aggregating the different states' current goals, the contiguous U.S. is considered to have a current renewable energy goal (i.e., solar and wind energy*) of 20% by 2030 and a current clean energy goal (i.e., wind, solar, hydro, and nuclear energy) of 43% by 2030.

\* This study does not consider geothermal, biomass, or other renewable resources at this time as the economics of these resources are not yet competitive with other energy sources.

TABLE 02

## Clean Energy Goals by State

U.S. states with clean energy goal targets between the years of 2020 and 2030 (current as of October 2020).[2]

| STATE | GOAL BY YEAR |
|---|---|
| AZ | 15% by 2025 |
| CA | 60% by 2030 |
| CO | 30% by 2020 |
| CT | 44% by 2030 |
| DE | 25% by 2025 |
| IL | 25% by 2025 |
| IN | 10% by 2025 |
| KS | 20% by 2020 |
| ME | 80% by 2030 |
| MD | 50% by 2030 |
| MA | 35% by 2030 |
| MI | 35% by 2025 |
| MN | 25% by 2025 |
| MO | 15% by 2020 |
| MT | 15% by 2020 |
| NV | 50% by 2030 |
| NH | 25% by 2025 |
| NJ | 50% by 2030 |
| NM | 40% by 2025 |
| NY | 70% by 2030 |
| NC | 12.5% by 2020 |
| ND | 10% by 2020 |
| OH | 8.5% by 2025 |
| OK | 15% by 2020 |
| OR | 25% by 2025 |
| PA | 18% by 2020 |
| RI | 31% by 2030 |
| SC | 2% by 2020 |
| SD | 10% by 2020 |
| TX | 10 GW by 2025 |
| UT | 20% by 2025 |
| VT | 75% by 2030 |
| VA | 15% by 2025 |
| WA | 80% by 2030 |
| WI | 10% by 2020 |

## 3.3 - Ambitious Goals

Although some states have mandated ambitious clean energy goals by 2030 (most notably California and New York), many states have modest goals and some do not have any goals, which will complicate efforts to reduce overall electricity sector emissions consistent with reaching net-zero emissions by 2050.[2] As a result, meeting the current clean energy goals would only produce a modest reduction in overall greenhouse gas (GHG) emissions from the electric power sector. In part because demand for electricity is expected to grow by nearly 10% over the next decade, this study shows that the current goals across the U.S. would only achieve a 6% carbon dioxide ($CO_2$) emissions reduction in 2030 relative to 2020. To investigate more ambitious deep decarbonization efforts (e.g., a reduction in $CO_2$ emissions of 50% or more from 2020 to 2030), which are more aligned with what is called for by the United Nations' Intergovernmental Panel on Climate Change (UN IPCC) in their 2018 Special Report,[1] this study envisions several scenarios with 'ambitious goals' for 2030.

For these 'ambitious goals' scenarios, states in the Western Interconnection with existing goals match California's goal (60% solar and wind\*), states in the Eastern Interconnection with existing goals match New York's goal (70% wind, solar, and hydro\*\*), and ERCOT reaches 45% renewable energy via a tripling of their projections of solar and wind growth by 2030.[6,41] States without goals continue to follow the capacity growth projections from NREL's Standard Scenarios Mid-Case projections.[37] By aggregating the different states' 'ambitious goals,' the contiguous U.S. is deemed to have an ambitious renewable energy goal (i.e., solar and wind energy) of 47% by 2030 and an ambitious clean energy goal (i.e., wind, solar, hydro, and nuclear energy) of 70% by 2030.

Similar to the 'current goals' scenarios, states with ambitious goals are modeled such that they can collaborate with other states that have goals. However, the 'ambitious goals' scenarios differ by allowing states to collaborate across different interconnections. This is made possible through a series of proposed Macro Grid designs, details of which are discussed in Section 3.4. The Macro Grid designs explored in this study traverse RTOs and the interconnection seams, allowing geographical diversity to be better leveraged via these resulting cross-country power flows that connect areas of complementary, high renewable energy potential to demand centers, as will be discussed in Section 4.

> Meeting the current clean energy goals would only produce a modest reduction in overall greenhouse gas emissions from the electric power sector.

\*  Although California includes other renewable sources when meeting the 60% goal, this study only focuses on adding solar and wind as these are currently the most cost competitive.

\*\* Although New York includes other renewable sources when meeting the 70% goal, this study only focuses on adding solar and wind (and counts existing hydro) as these are currently the most cost competitive.

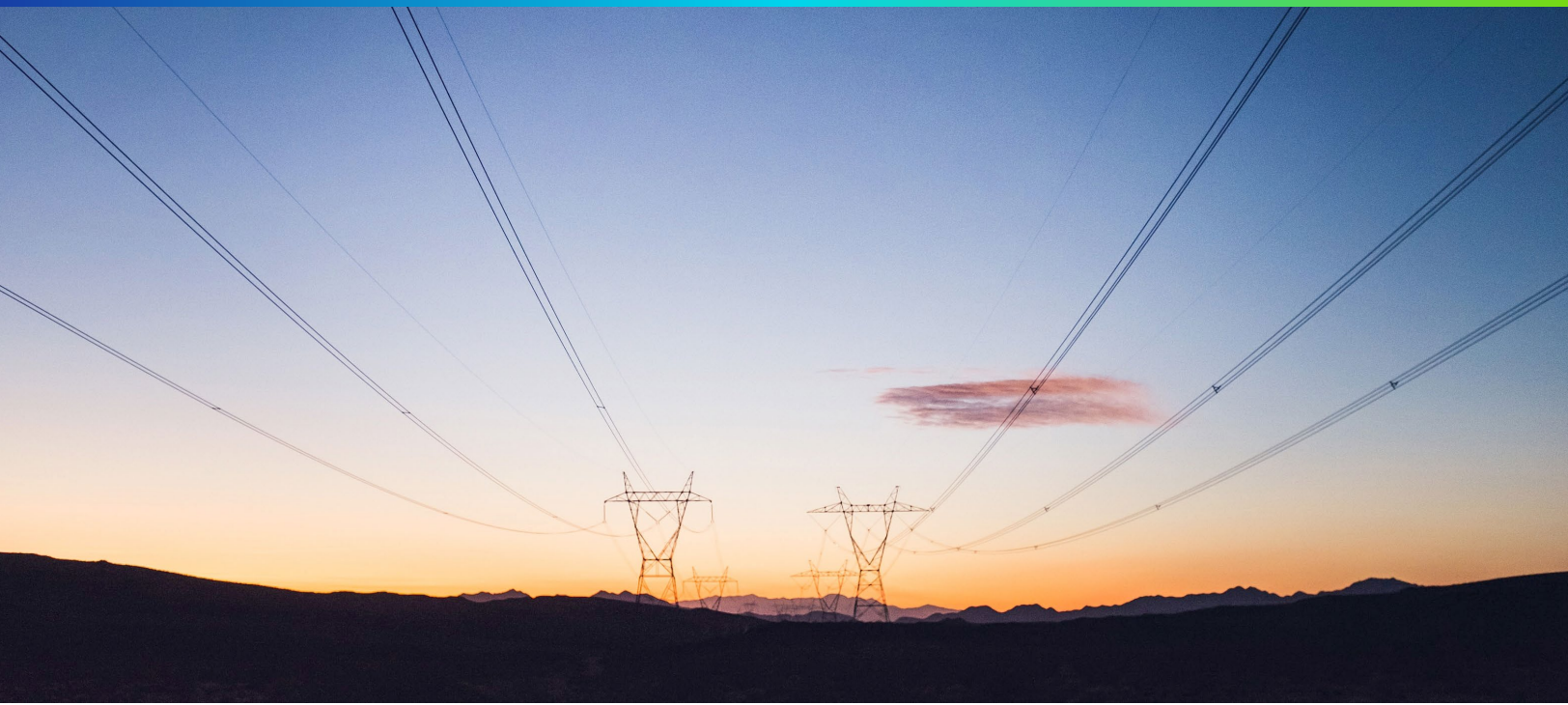

> Improving the connectivity of the major regions of the grid through transmission upgrades and additions, thereby creating a Macro Grid, would allow the U.S. to further harness its abundant renewable resources and better balance electric demand across the country.

## 3.4 - Achieving Ambitious Goals with Macro Grid Designs

There are numerous ways to achieve the ambitious clean energy goals described in Section 3.3. This study was inspired by the approaches of upgrading and adding HVDC and AC transmission lines seen in NREL's Seams Study[4,42] and research from Iowa State University.[18,19] The ambitious goals are met with a combination of new renewable generation capacity, AC transmission upgrades, and HVDC upgrades and additions. Other technologies, such as energy storage, are not considered in this study but will be investigated in future studies. In the U.S., the locations of strong renewable resources and high electric demand are often not within the same grid regions. Improving the connectivity of the major regions of the grid through transmission upgrades and additions, thereby creating a Macro Grid, would allow the U.S. to further harness its abundant renewable resources and better balance electric demand across the country.

The most ambitious Macro Grid design considered in this report is a country-spanning HVDC design. The original design was created by Dale Osborn during his time with MISO.[5] Armando Luis Figueroa-Acevedo then tested that design and two other HVDC concepts during his doctoral research at Iowa State University; results from that work have been published in his Ph.D. dissertation and a 2020 IEEE Transactions on Power Systems article.[18,19] All three of Figueroa-Acevedo's HVDC design concepts are included in this report and further discussed in Section 4. These same concepts are also an integral part of NREL's Seams Study.[4,42] In addition to the three HVDC designs, this study includes one Macro Grid design with no HVDC upgrades, but with considerable AC transmission upgrades that substantially improve the connections between regions within interconnections. Each Macro Grid design is discussed in further detail in the following sections.

**DESIGN 1**

This design features no upgrades to the existing HVDC infrastructure. However, there are substantial upgrades to AC transmission within interconnections (more than in any of the designs with increased HVDC expansion), connecting areas of high renewable capacity to areas of high demand. One note across all designs: the path from Oklahoma to Memphis is one of the same corridors that Clean Line Energy was targeting for developing a large transmission line, as discussed in Russell Gold's *Superpower: One Man's Quest to Transform American Energy*.[43] As is further discussed in Section 4.8, certain transmission corridors, including the Oklahoma-Memphis path, will require capacity upgrades regardless of the selected Macro Grid design. To adequately deliver clean energy to major U.S. demand centers, substantial transmission system investments are needed to unlock the geographical diversity of renewable resources.

FIG. 05

## Design 1

Transmission upgrades for Design 1 yield 36% more transmission capacity (measured in MW-miles) for the simulated 2030 grid compared to the current 2020 grid.

— Upgraded AC transmission

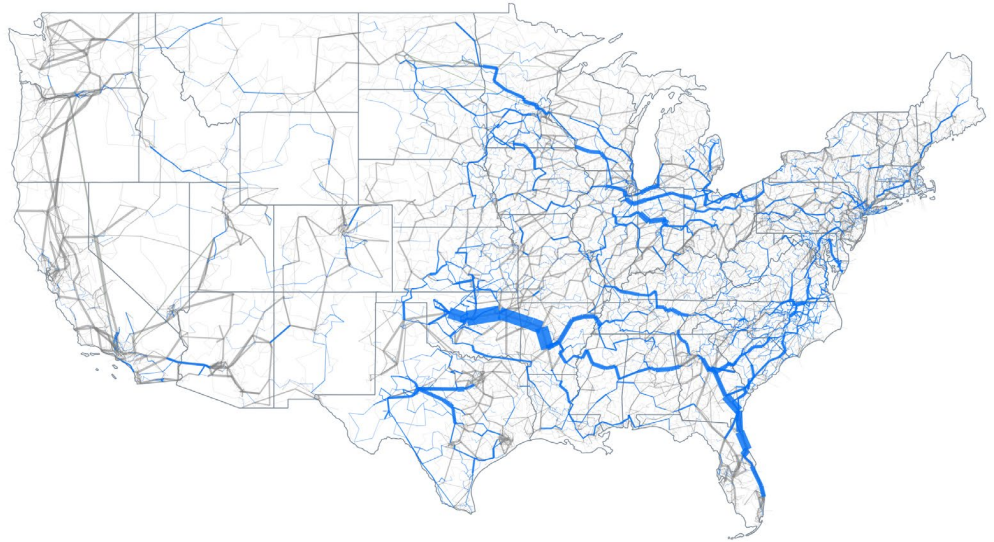

**DESIGN 2A**

Upgrades to the existing HVDC back-to-back (B2B) converter stations between the Eastern and Western Interconnections are taken from the preliminary findings from the NREL Seams Study,[4,42] specifically the Design 2a 'high variable generation' (VG) scenario (upgrades to the renewable generation capacities are more ambitious in the 'high VG' scenario than in NREL's 'base case' scenario). The NREL Seams Study did not include ERCOT, but two back-to-back converter stations do exist connecting ERCOT to the Eastern Interconnection; these stations are upgraded by the average capacity upgrade of the East-West converter stations (3,671 MW each). These upgrades are summarized in Table 3.

FIG. 06

## Design 2a

Transmission upgrades for Design 2a yield 30% more AC transmission capacity (measured in MW-miles) for the simulated 2030 grid compared to the current 2020 grid, plus 33 GW of back-to-back converter upgrades.

— Upgraded AC transmission
▲ Upgraded B2B capacity

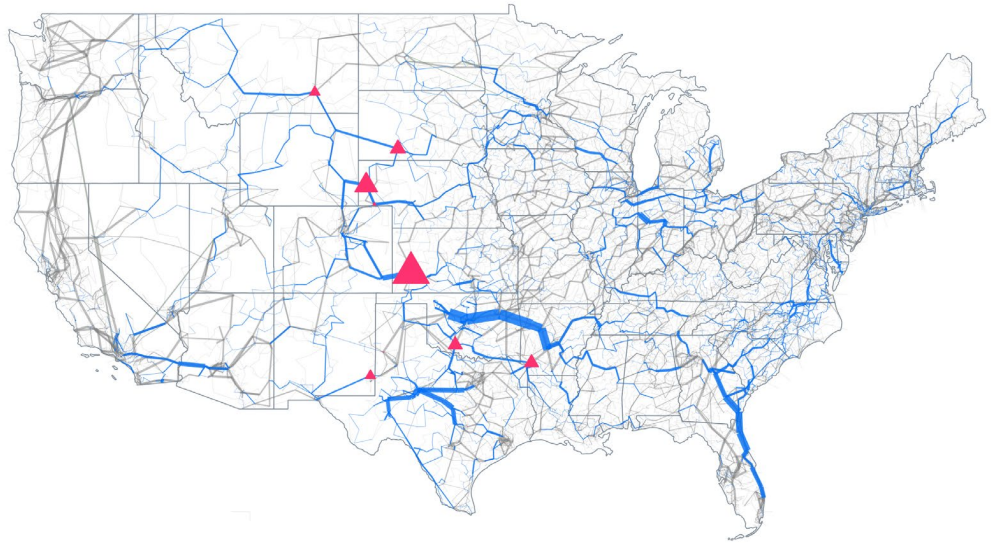

TABLE 03

## Design 2a HVDC Upgrades

Design 2a upgrades the existing HVDC back-to-back converter stations that cross the East-West seam and the East-ERCOT seam.

| BACK-TO-BACK STATION | SEAM | PREVIOUS CAPACITY (MW) | NEW CAPACITY (MW) |
|---|---|---|---|
| Blackwater | East/West | 200 | 399 |
| Eddy | East/West | 200 | 2,895 |
| Lamar | East/West | 210 | 9,541 |
| Miles City | East/West | 200 | 2,957 |
| Oklaunion | East/ERCOT | 200 | 3,871 |
| Rapid City | East/West | 200 | 4,166 |
| Sidney | East/West | 200 | 1,108 |
| Stegal | East/West | 100 | 5,943 |
| Welsh | East/ERCOT | 600 | 4,271 |
| **Total** | | **2,110** | **35,151** |

### DESIGN 2B

Similarly, the East-West HVDC back-to-back converter stations are upgraded by the capacities presented in the preliminary findings of the NREL Seams Study's Design 2b 'high VG' scenario. Three new HVDC lines are added at a capacity of 9,500 MW each (compared to the 9,481-MW HVDC lines in the Seams Study) and added at high-voltage locations within this study's synthetic network that roughly correspond to the locations shown in the Seams Study.[4,42] Since no new HVDC lines are added between the Eastern Interconnection and ERCOT, the back-to-back converter stations are upgraded by the same 3,671 MW as in Design 2a. These upgrades are summarized in Table 4.

FIG. 07

## Design 2b

Transmission upgrades for Design 2b yield 26% more AC transmission capacity (measured in MW-miles) for the simulated 2030 grid compared to the current 2020 grid, plus 15 GW of back-to-back converter upgrades and three new 9.5-GW HVDC lines.

— Additional HVDC capacity
— Upgraded AC transmission
▲ Upgraded B2B capacity

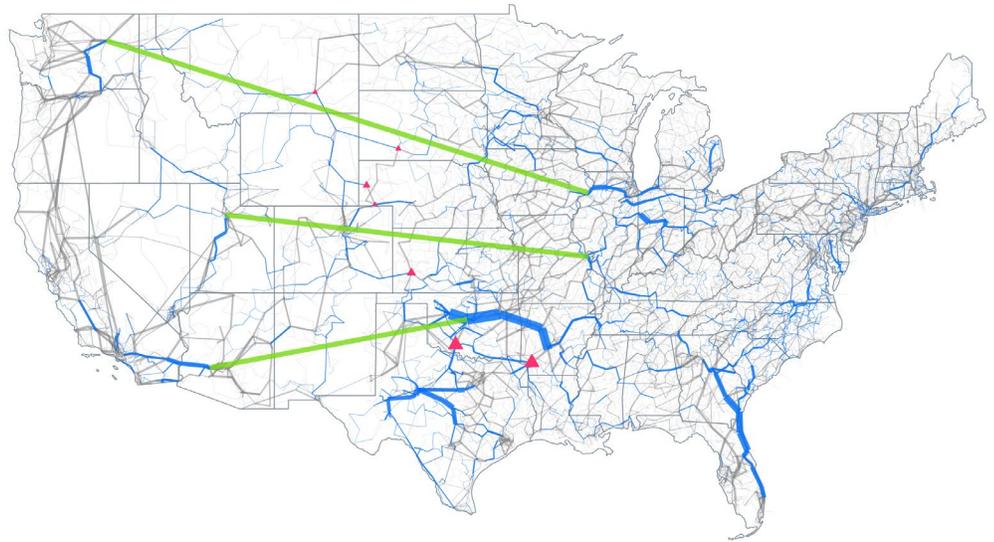

TABLE 04

## Design 2b HVDC Upgrades

Design 2b upgrades the existing HVDC back-to-back converter stations and adds three new HVDC lines across the East-West seam.

| HVDC ELEMENT | SEAM | PREVIOUS CAPACITY (MW) | NEW CAPACITY (MW) |
|---|---|---|---|
| Blackwater B2B | East/West | 200 | 234 |
| Eddy B2B | East/West | 200 | 338 |
| Lamar B2B | East/West | 210 | 2,285 |
| Miles City B2B | East/West | 200 | 1,319 |
| Oklaunion B2B | East/ERCOT | 200 | 3,871 |
| Rapid City B2B | East/West | 200 | 1,589 |
| Sidney B2B | East/West | 200 | 1,255 |
| Stegal B2B | East/West | 100 | 1,782 |
| Welsh B2B | East/ERCOT | 600 | 4,271 |
| New HVDC Washington-Iowa | East/West | - | 9,500 |
| New HVDC Utah-Missouri | East/West | - | 9,500 |
| New HVDC Arizona-Oklahoma | East/West | - | 9,500 |
| **Total** |  | **2,110** | **45,444** |

**DESIGN 3**

In this design, sixteen new HVDC lines are added at high-voltage locations within this study's synthetic network that roughly correspond to the locations shown in the NREL Seams Study.[4,42] These new line capacities are 8,000 MW, as compared to the 8,389-MW capacities in the Seams Study. Importantly, this study does include an HVDC terminal location within ERCOT — near Sweetwater, TX — that is connected to both the Eastern and Western Interconnections via new HVDC lines. These upgrades are summarized in Table 5.

FIG. 08

## Design 3

Transmission upgrades for Design 3 yield 23% more AC transmission capacity (measured in MW-miles) for the simulated 2030 grid compared to the current 2020 grid, plus sixteen new 8-GW HVDC lines.

— Additional HVDC capacity
— Upgraded AC transmission

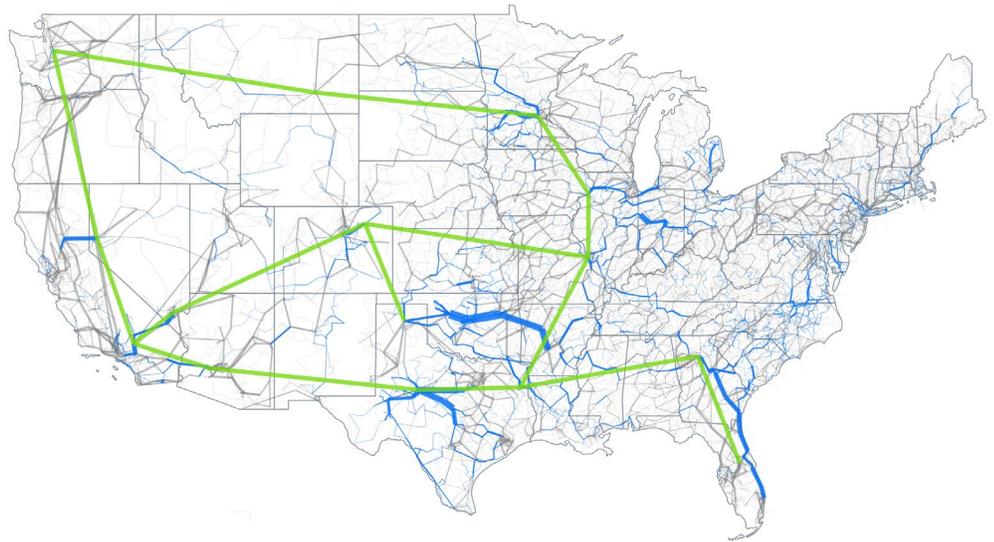

TABLE 05

## Design 3 HVDC Upgrades

Design 3 adds sixteen new HVDC lines, each with a capacity of 8,000 MW, connecting the following locations and crossing all three interconnection seams.

| FROM LOCATION | FROM INTERCONNECT | TO LOCATION | TO INTERCONNECT |
|---|---|---|---|
| Orlando, FL | Eastern | Atlanta, GA | Eastern |
| Atlanta, GA | Eastern | Panola, TX | Eastern |
| Panola, TX | Eastern | St. Louis, MO | Eastern |
| Panola, TX | Eastern | Sweetwater, TX | ERCOT |
| St. Louis, MO | Eastern | Brush, CO | Western |
| St. Louis, MO | Eastern | Davenport, IA | Eastern |
| Davenport, IA | Eastern | Minneapolis, MN | Eastern |
| Minneapolis, MN | Eastern | Colstrip, MT | Western |
| Colstrip, MT | Western | Seattle, WA | Western |
| Seattle, WA | Western | Reno, NV | Western |
| Reno, NV | Western | Victorville, CA | Western |
| Victorville, CA | Western | Las Vegas, NV | Western |
| Las Vegas, NV | Western | Brush, CO | Western |
| Brush, CO | Western | Amarillo, TX | Eastern |
| Victorville, CA | Western | Palo Verde, AZ | Western |
| Palo Verde, AZ | Western | Sweetwater, TX | ERCOT |

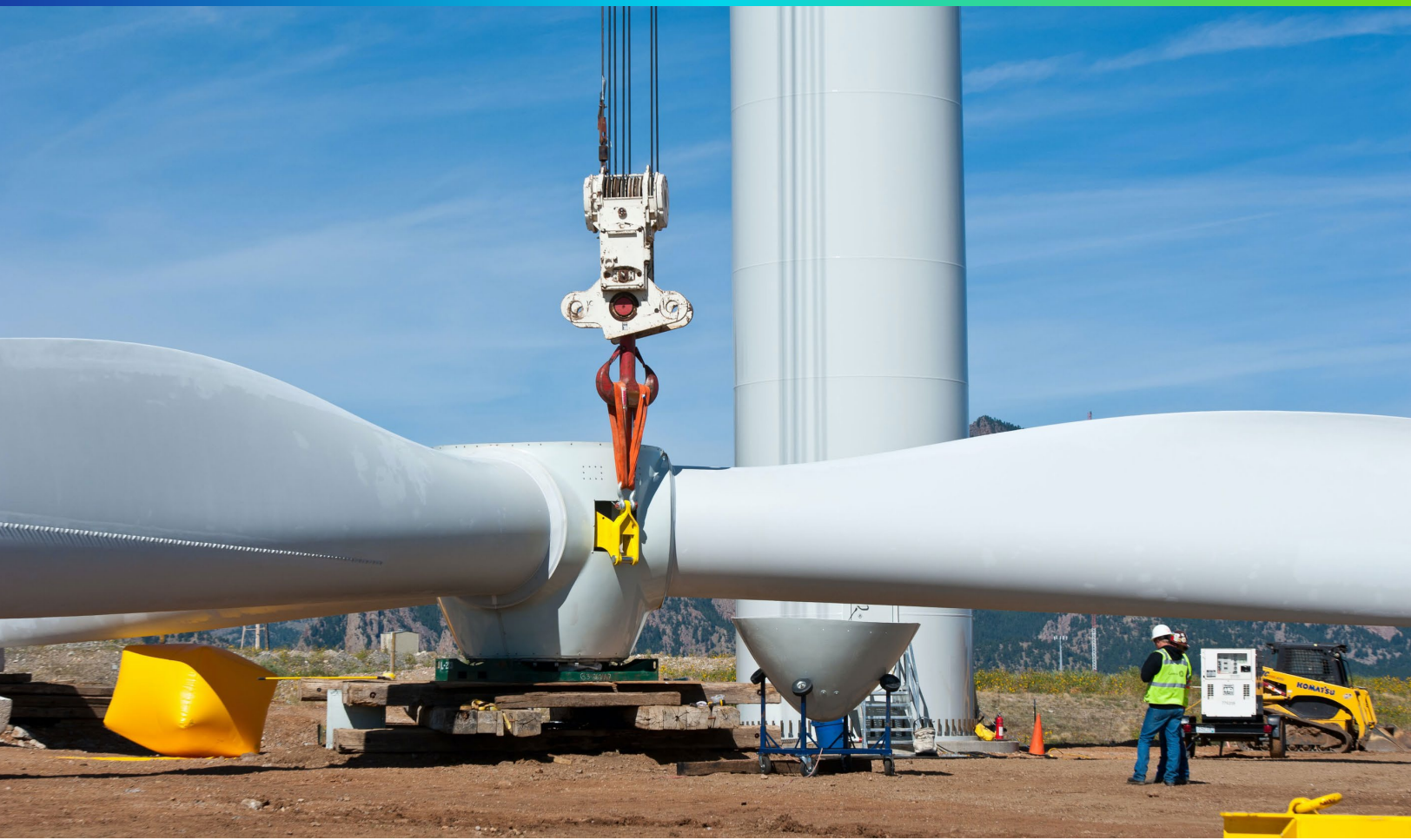

## 3.5 - Investment Cost Calculations

Investment costs are calculated separately for new generation capacity, new AC transmission capacity, and new HVDC capacity. In general, cost estimates and regional multipliers follow the methodology from the NREL ReEDS documentation:[21,22] $/MW for generators, transformers, and AC/DC converter stations, and $/(MW-mile) for transmission lines, with regional cost multipliers. Generation capacity costs are calculated using NREL's 2020 ATB.[3] In particular, this study uses base cost estimates for the year 2025 (a midpoint between 2020 and 2030) in the 'Moderate' cost scenario. Costs for new HVDC lines follow the per-MW-mile calculations from the NREL ReEDS report,[21,22] with terminal costs set at 0.135 $M/MW based on estimates from MISO.[44]

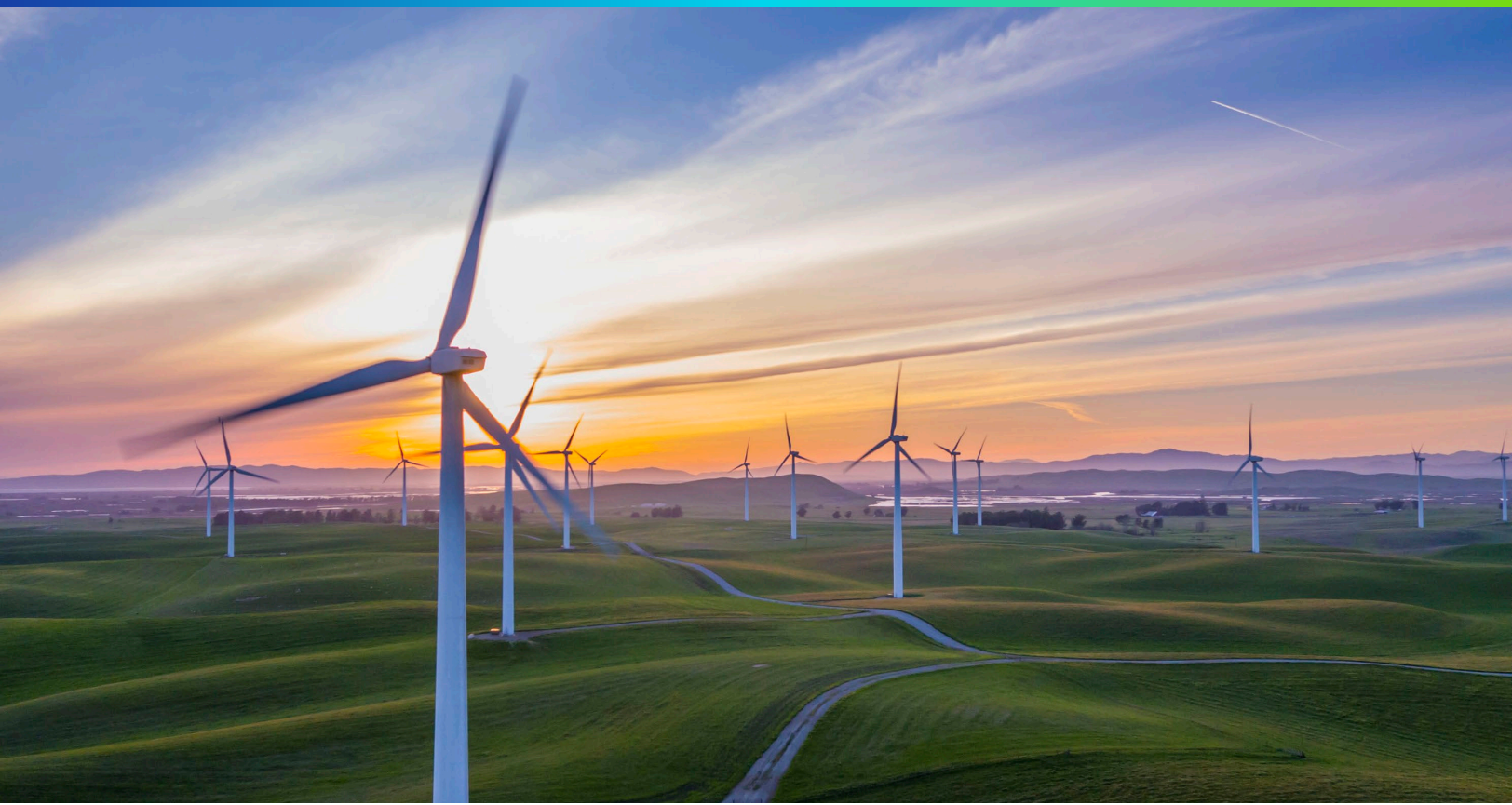

# 4.0 Results

Each of the four transmission expansion designs* presented in Section 3.4 has the ability to achieve the ambitious clean energy goals described in Section 3.3; this can be seen in Table 6, where the percentages of renewable energy, clean energy, and fossil fuel generation are listed for each scenario. To better analyze the impacts of achieving the ambitious goals, Table 6 also includes total fuel costs and the amount of $CO_2$, nitrogen oxide ($NO_x$), and sulfur dioxide ($SO_2$) emissions.

For a deeper exploration of the results of these scenarios, Section 4.1 first presents the potential emissions reductions and changes to generation penetration that are realized under the proposed designs. Section 4.2 then describes the infrastructure investments needed to meet the ambitious goals, followed by a discussion about the scenario design with no HVDC upgrades in Section 4.3. Some common results across all of the designs with HVDC transmission upgrades are reviewed in Section 4.4, with further exploration into the different designs' unique results discussed in Sections 4.5, 4.6, and 4.7. Section 4.8 discusses some of the common AC transmission corridors that receive upgrades across each of the four Macro Grid designs. Finally, Section 4.9 examines a few impacts of the generation mix at the state level.

\* As a reminder, the transmission network model used in this study is a 'synthetic' network, designed not to exactly represent existing infrastructure, but to represent a statistically similar network with electricity demand and generation capacity at their approximate locations. Therefore, although the specific transmission corridors may not match reality, the overall patterns are intended to illustrate the effects of high-level grid infrastructure expansion decisions.

TABLE 06
## Generation Mix and Emissions Across Designs

Each of the four transmission expansion designs has the ability to achieve the ambitious clean energy goals.

|  | 2020 | CURRENT 2030 GOALS | AMBITIOUS 2030 GOALS | | | |
| --- | --- | --- | --- | --- | --- | --- |
|  |  |  | DESIGN 1 | DESIGN 2A | DESIGN 2B | DESIGN 3 |
| Renewable Energy | 9.3% | 20.2% | 46.6% | 46.5% | 46.6% | 47.0% |
| Clean Energy | 35.4% | 43.1% | 69.6% | 69.6% | 69.6% | 69.7% |
| Fossil Fuel | 63.0% | 55.4% | 29.0% | 29.0% | 28.9% | 28.9% |
| Fuel Cost ($B) | $106.55 | $102.91 | $54.74 | $55.11 | $55.06 | $54.43 |
| $CO_2$ (MMmt) | 1,841.7 | 1,729.5 | 997.5 | 1,003.9 | 1,004.1 | 1,004.6 |
| $NO_X$ (MMmt) | 1.023 | 0.946 | 0.578 | 0.584 | 0.585 | 0.586 |
| $SO_2$ (MMmt) | 1.135 | 1.015 | 0.698 | 0.709 | 0.712 | 0.714 |

## 4.1 - Generation and Emissions

For the U.S. and each interconnection, Figure 9 shows the energy generation by resource for the scenarios tested in this report. The considerable increase in solar and wind generation is apparent when comparing the 'current goals' scenario to the 'ambitious goals' scenarios. Resulting from the influx of clean energy and the upgraded infrastructure, there is a substantial reduction in coal and natural gas generation across all four Macro Grid designs. Although these patterns are consistent across the U.S., the four designs studied in this report show slight variations in their generation mixes within each interconnection. These variations are discussed in more detail in Sections 4.5, 4.6, and 4.7.

The high spatial and temporal resolution time series results in Figure 10 reveal operational patterns in the 'current goals' and 'ambitious goals' scenarios, again showing that the dependency on fossil fuels substantially reduces in the 'ambitious goals' scenarios. However, a large amount of natural gas generation is still needed during the summer when both the wind generation drops and the highest demand of the year arrives. On the other hand, the fall and spring show very small amounts of natural gas and coal generation. Despite the reduced fossil fuel generation, curtailment of solar and wind generation is at its highest in those same months. The sizeable renewable energy penetration in the fall and spring leads to some of the largest variations in power flow, with the Macro Grid helping distribute the abundant solar and wind energy across the country to meet demand. A few examples of these variations from the simulated day of November 2nd are discussed in Sections 4.4 and 4.7. Additional details of this and other operational patterns are available from Breakthrough Energy Sciences' open-source data and visualization tools.[20]

FIG. 09
## Energy Generation by Resource
Solar and wind generation increases substantially to meet the ambitious clean energy goals, cutting into the levels of coal and natural gas generation.

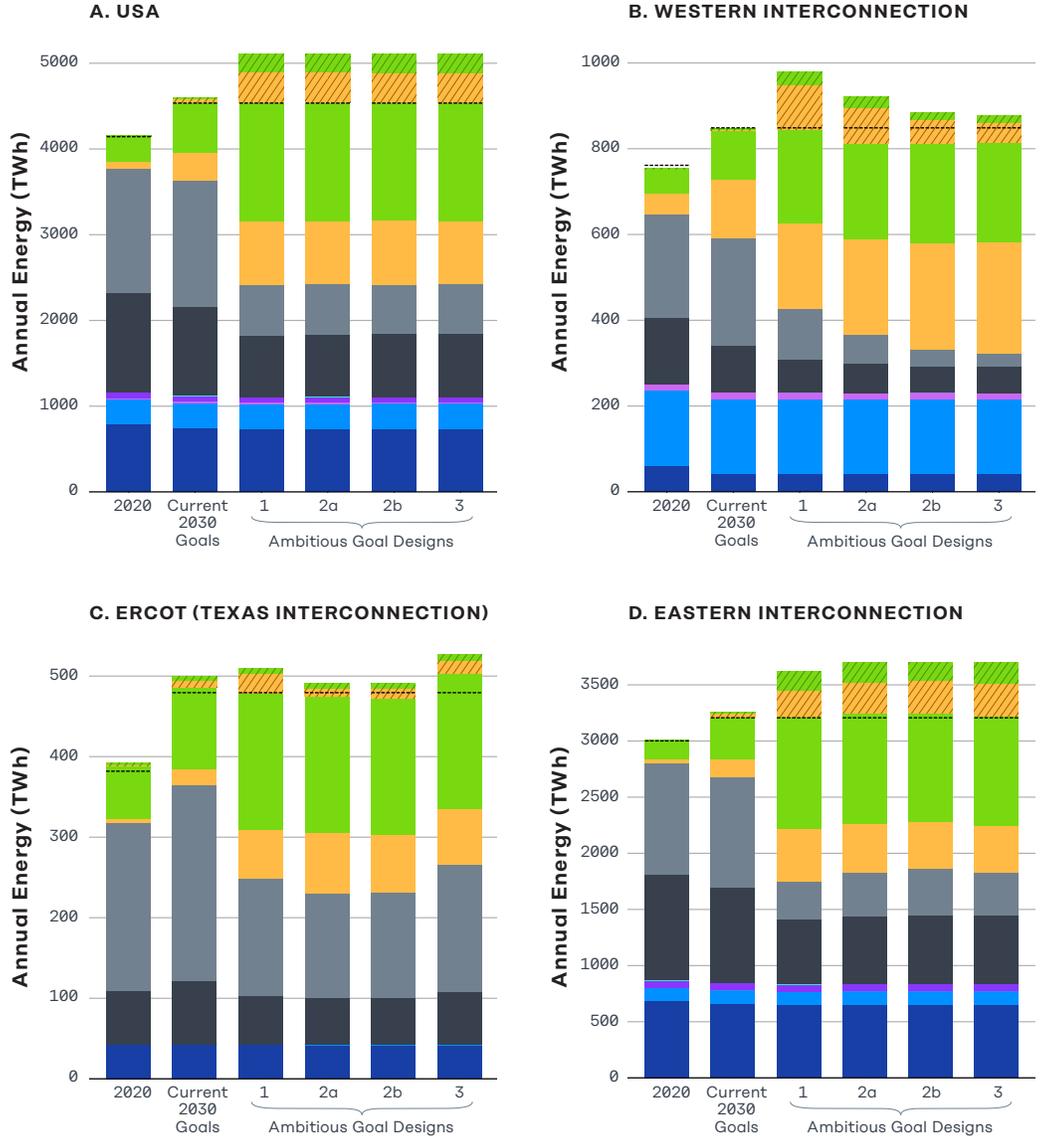

FIG. 10

## Generation Mix - Current vs. Ambitious Goals for 2030

High spatial and temporal resolution time series results reveal operational patterns in the 'current goals' and 'ambitious goals' scenarios. The 'ambitious goals' scenarios have remarkably less natural gas and coal and substantially more solar and wind (some of which is curtailed).

- Wind Curtailment
- Solar Curtailment
- Wind
- Solar
- Natural Gas
- Coal
- Distillate Fuel Oil
- Other
- Geothermal
- Hydro
- Nuclear
- Demand

### 'Current Goals' Scenario in 2030

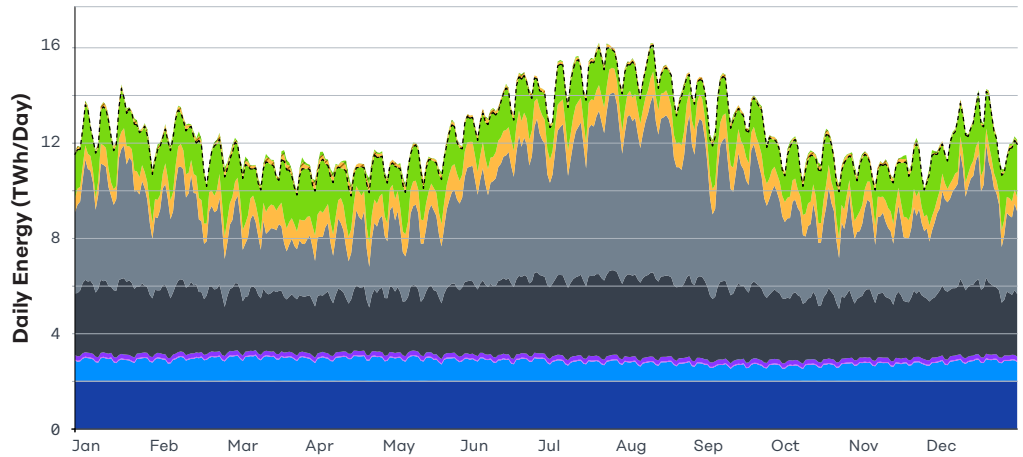

### 'Ambitious Goals' Scenarios in 2030

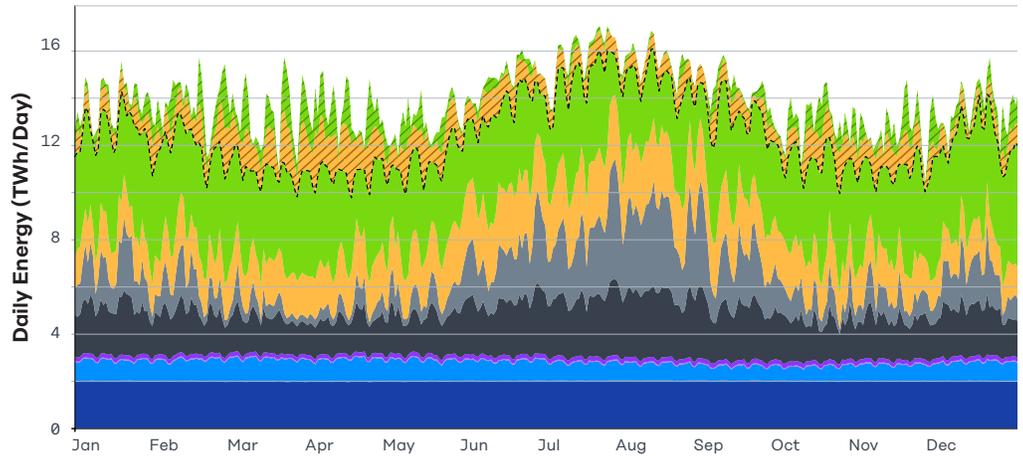

Common across all of the designs is a sizeable increase in the curtailment of solar and wind generation, averaging 21% for each scenario design. Even with the Macro Grid supporting energy exchanges over longer distances, the timing of solar and wind relative to the demand does not always align. With the cross-seam HVDC lines in Designs 2b and 3, there is a noticeable decrease in the curtailment of ERCOT's and the Western Interconnection's renewable resources; this of course means the curtailment of solar and wind resources in the Eastern Interconnection increases.

This curtailment reveals both diurnal and seasonal patterns, as shown in Figure 11. On the seasonal timescale, the curtailment is greatest during spring, when heating and cooling loads are generally lower and when significant hydropower is available, and to a lesser extent in autumn. On a daily timescale, curtailment is greatest between the hours of 8am and 3pm CST, indicative of excess solar power ahead of the typical afternoon demand peaks. There is also significant week-to-week variability, with even some weeks in the spring having very little curtailment, due to either low renewable generation or high demand; conversely, there are also some weeks with significant curtailment throughout the day and night. Therefore, there appear to be opportunities for energy storage and demand flexibility to help mitigate curtailment on daily, weekly, and seasonal timescales.

Although curtailment can reach as high as 500 GW in the worst hour (a particularly sunny and windy hour in April), most of the time the curtailment is much more modest, as shown in Figure 12. The median absolute curtailment is only 31 GW, and the median curtailment as a share of available solar and wind energy is only 12%. Based on the patterns of sun and wind availability and their imperfect alignment with electric demand, some amount of curtailment is inevitable in a power system with very high renewable penetration, even with a perfect transmission network. Since hours with significant curtailment will feature very low real-time prices (at least near the solar and wind sources), by the time the grid features this degree of renewable energy, there may be sufficient energy storage and demand flexibility to make good use of this available energy and prevent it from being wasted (as discussed further in Section 6).

Table 6 shows that the four designs each enable $CO_2$ emissions reductions of over 45% when compared to 2020 emission levels. A geographical depiction of the current state of $CO_2$ emissions (totaling 1,841 million metric tons (MMmt) of $CO_2$ as seen in Table 6) is included in Figure 13A. The changes in $CO_2$ emissions for the 'current goals' scenario, shown in Figure 13B, cuts only 112 MMmt, or 6% of 2020 emissions. However, the 'ambitious goals' scenarios yield $CO_2$ emissions reductions of around 840 MMmt, which is a reduction of 42% compared to the 'current goals' scenario or 45% of 2020 emissions, as shown in Figure 13C. Emissions of $NO_x$ and $SO_2$ are reduced by 37-39% and 29-31%, respectively, compared to the 'current goals' scenario, with slight differences based on the selected Macro Grid design. These differences are attributed to different transmission network upgrades that connect the new renewable capacity to different demand centers, thereby displacing fossil fuel generators.

FIG. 11

## Distribution of Renewable Curtailment

Renewable curtailment is generally greater during daytime hours and in spring, and lower at night and during the summer.

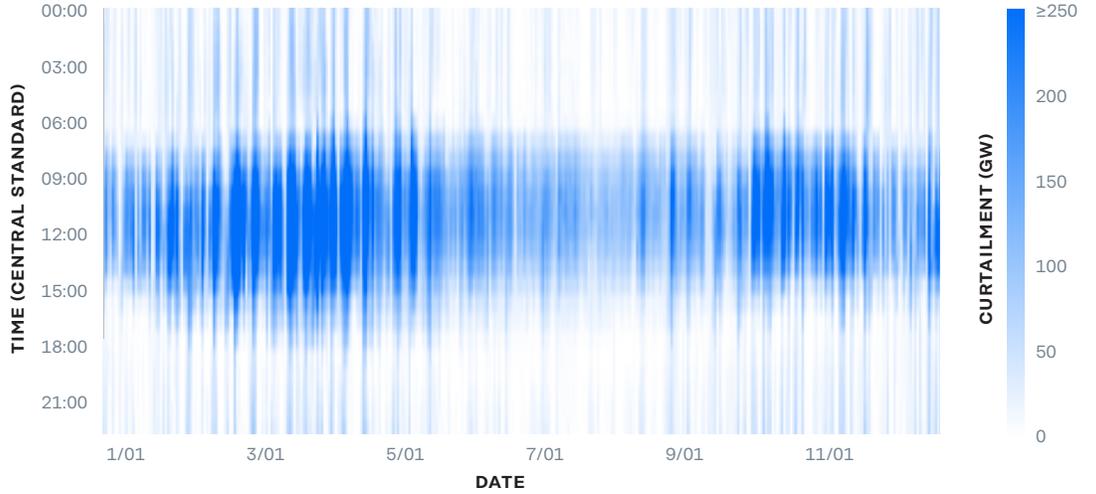

FIG. 12

## Frequency of Renewable Curtailment

Renewable curtailment is typically low, with the worst curtailment only occurring for a few hours each year.

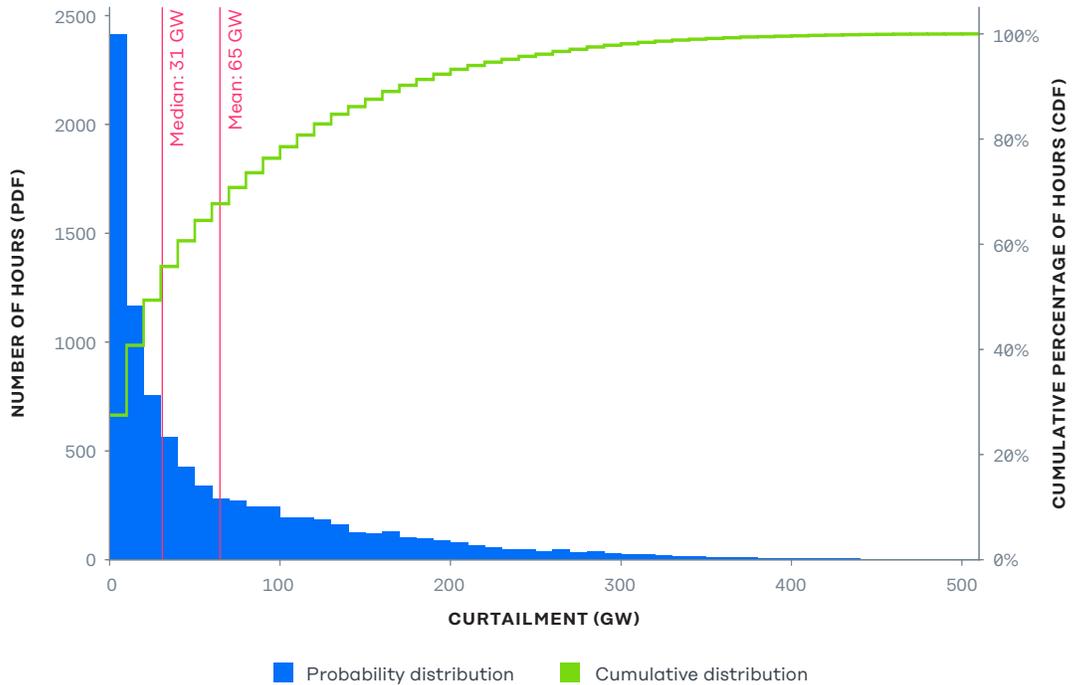

FIG. 13

# Geographical Depiction of $CO_2$ Emissions Changes from 2020 to 2030

$CO_2$ emission changes from the current state of the grid in 2020 to both the 'current goals' and 'ambitious goals' scenarios in 2030.

**(A) CURRENT GRID (2020)**

Locations of the 1,841 MMmt of $CO_2$ emissions from the current state of the grid in 2020.

- ● 10 MMmt $CO_2$ emissions, coal
- ● 10 MMmt $CO_2$ emissions, natural gas

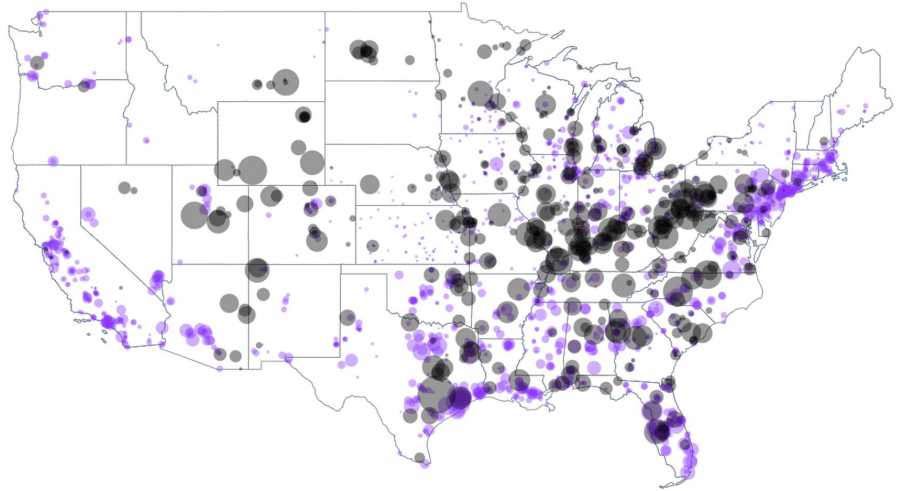

**(B) CURRENT 2030 GOALS**

6% reduction to 1,730 MMmt of $CO_2$ emissions when states meet their current goals in 2030.

- ● 10 MMmt less $CO_2$ emissions
- ● 10 MMmt more $CO_2$ emissions

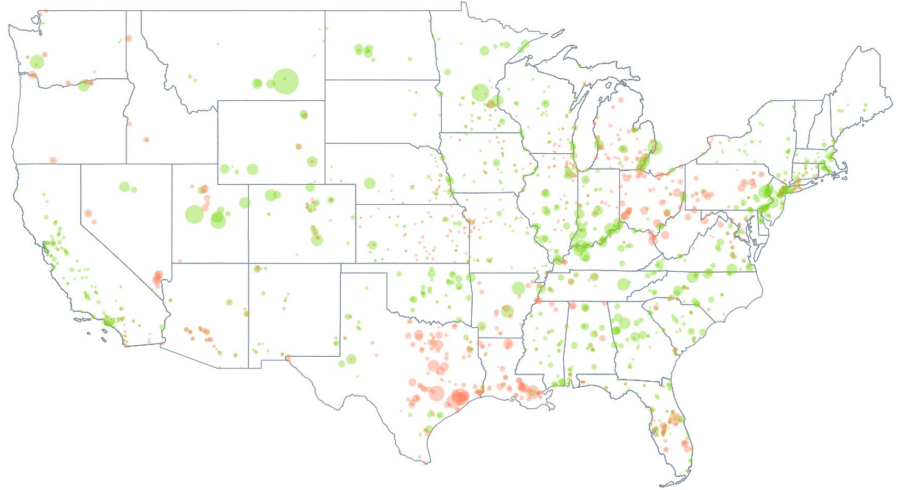

**(C) AMBITIOUS 2030 GOALS**

45% reduction to 1,004 MMmt of $CO_2$ emissions when states meet the ambitious goals in 2030.

- ● 10 MMmt less $CO_2$ emissions
- ● 10 MMmt more $CO_2$ emissions

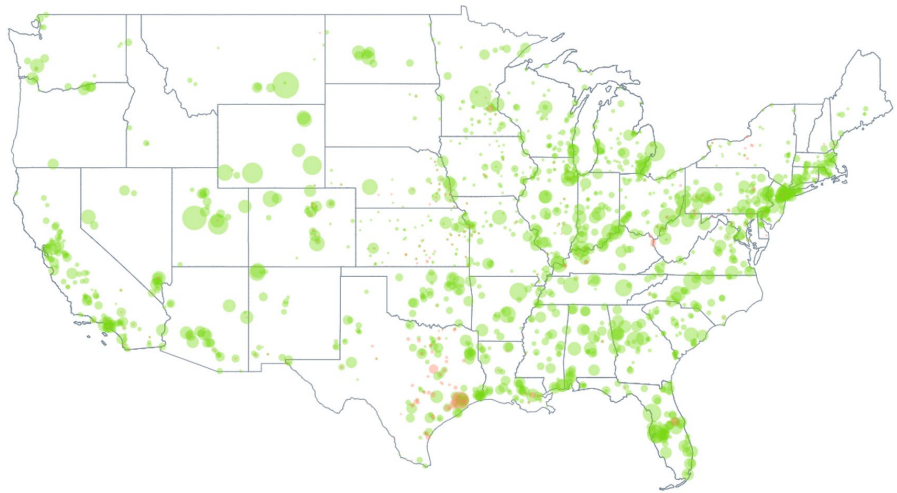

## 4.2 - Infrastructure Investments for Ambitious Goals

Reaching the ambitious goals will require substantial investments in infrastructure,* as shown in Tables 7 and 8. Although the investments in the NREL Seams Study were decided differently — using a reduced-network capacity expansion model rather than the high-resolution network (representing an estimated 307 TW-miles of transmission lines down to 69 kV) and production-cost-model-informed network tuning used by Breakthrough Energy Sciences — there are still some similarities between the results from this study and those in the Seams Study's 'high VG' scenario (the most similar to this study's 'ambitious goals' scenario). In both cases, the generation cost investments are substantially greater than the transmission upgrade investments, with generation representing 92-93% of total investment in the Seams study,[4,42] compared to 85-87% in this study.

TABLE 07

### Additions to Renewable Generation and Transmission Capacity

Achieving ambitious clean energy goals requires substantial additions in both renewable generation capacity and transmission capacity, whether all AC as in Design 1 or a combination of AC and HVDC as in Designs 2a, 2b, and 3.

| DESIGN | WIND CAPACITY (GW) | SOLAR CAPACITY (GW) | AC UPGRADES (TW-MILES)** | AC UPGRADES (PERCENT) | HVDC UPGRADES (TW-MILES) | HVDC B2B STATIONS (GW) |
|---|---|---|---|---|---|---|
| Current Goals | 204 | 172 | 2.48 | 0.81% | 0 | 0 |
| Design 1 | 541 | 529 | 112 | 36.4% | 0 | 0 |
| Design 2a | 541 | 529 | 92.1 | 30.0% | 0 | 33.0 |
| Design 2b | 541 | 529 | 81.0 | 26.4% | 33.0 | 14.8 |
| Design 3 | 541 | 529 | 69.4 | 22.6% | 60.2 | 0 |

The increased transmission expansion costs in this study can be partially explained by the inclusion of a much more detailed network model, which requires increased transmission capacity to adequately transmit power from the new renewable generation added across the country, as depicted in Figure 14. In addition, the Seams Study objective is a 35-year overall cost minimization, while the objective in this study is designing transmission to achieve ambitious renewable penetration goals.

Of note, the renewable capacity additions in this report of 1,070 GW over a decade are ambitious and intended to be provocative. While this may not be achievable by 2030, it is certainly an aspirational goal and aligns with the renewable energy capacity recommended by the 2035 Report by UC Berkeley's Goldman School of Public Policy and GridLab.[45] Whichever year this renewable energy capacity buildout is completed, a Macro Grid will be essential in delivering this energy to the major demand centers.

* There are many other costs and benefits to take into consideration when deciding which infrastructure design choice is ultimately preferred, but those are outside the scope of this report.

** Calculated relative to the model of the current 2020 grid, estimated at 307 TW-miles of aggregate transmission capacity.

TABLE 08

## Investment Costs of Ambitious Goals vs. Current Goals

Achieving ambitious clean energy goals results in substantial reductions in fuel costs, GHG emissions, and other air pollutants, but requires considerable investment costs for each design ($B USD).

| DESIGN | WIND CAPACITY | SOLAR CAPACITY | AC LINES | AC TRANSFORMERS | HVDC TRANSMISSION | HVDC B2B STATIONS | TOTAL |
|---|---|---|---|---|---|---|---|
| Current Goals | $185 | $164 | $9 | $0.63 | $0 | $0 | $359 |
| Design 1 | $745 | $574 | $213 | $7.59 | $0 | $0 | $1,539 |
| Design 2a | $745 | $574 | $196 | $7.14 | $0 | $9.04 | $1,530 |
| Design 2b | $745 | $574 | $179 | $6.76 | $24.6 | $4.06 | $1,533 |
| Design 3 | $745 | $574 | $152 | $5.87 | $65.5 | $0 | $1,542 |

FIG. 14

## New Solar and Wind Capacity

A geographical depiction of the 1,070 GW of solar and wind capacity added, along with a Macro Grid, to meet the ambitious goals in 2030.

- 1 GW new solar capacity
- 1 GW new wind capacity

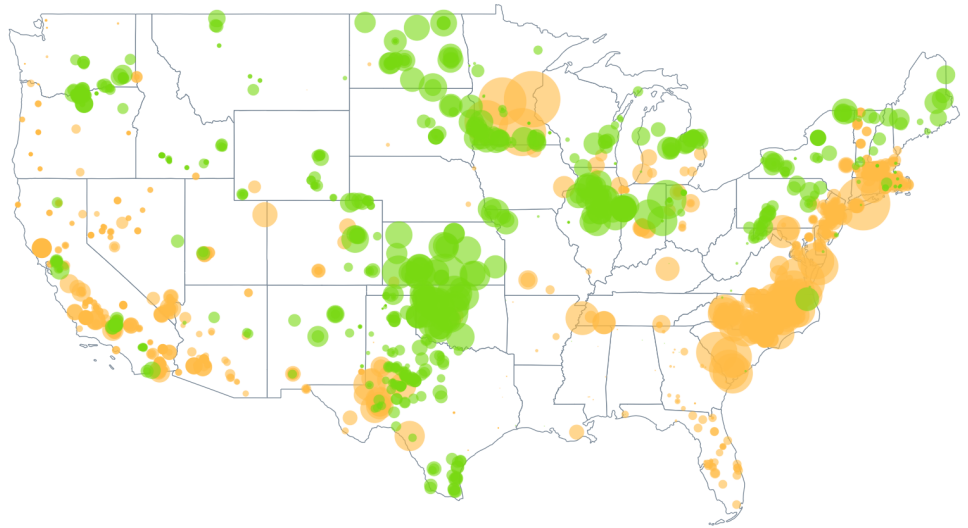

By combining the information presented in Tables 6 and 8, simple payback periods can be calculated for each of the Macro Grid designs, as shown in Table 9. These values are calculated relative to the current goals, as the ratio of increased investment costs to reductions in operational costs. All four designs have nearly equivalent paybacks, since they have the same renewable capacity, and transmission networks are upgraded until they hit the same overall renewable penetration. Thus, from an energy delivery perspective, the distinction between an HVDC or an AC Macro Grid design is less important than the conclusion that a Macro Grid is indeed necessary to unlock the geographic diversity of U.S. renewable energy resources to meet these ambitious clean energy goals.

Looking only at fuel costs, the simple payback periods are in the range of 24 to 25 years, a range that typically precludes investment from non-governmental entities. However, there are substantial co-benefits to increasing the penetration of renewable energy. Table 9 shows an example of considering these co-benefits by including several representative prices on emissions of $CO_2$. This example reveals that including the co-benefit of GHG emissions reductions can substantially reduce the payback period. Other non-$CO_2$ pollution reduction co-benefits may be even more considerable, by avoiding premature deaths caused by increases in local concentrations of $PM_{2.5}$ and ozone.[46,47]

TABLE 09

### Simple Payback Periods

Looking only at fuel costs, the simple payback period of the Macro Grid investments are in the range of 24 to 25 years, but factoring in the co-benefit of GHG emissions reductions substantially reduces that payback period length.

| DESIGN | $0/TON | $25/TON | $50/TON | $75/TON | $100/TON |
| --- | --- | --- | --- | --- | --- |
| 1 | 24.5 yr | 17.8 yr | 13.9 yr | 11.4 yr | 9.7 yr |
| 2a | 24.5 yr | 17.8 yr | 13.9 yr | 11.5 yr | 9.7 yr |
| 2b | 24.5 yr | 17.8 yr | 14.0 yr | 11.5 yr | 9.7 yr |
| 3 | 24.4 yr | 17.8 yr | 14.0 yr | 11.5 yr | 9.8 yr |

## 4.3 - Design 1: No HVDC Upgrades

For this scenario design, only AC transmission is built, as shown in Figure 15. This upgrade yields 36% more transmission capacity (measured in MW-miles) for the simulated 2030 grid compared to the current 2020 grid. Transmission upgrades are more concentrated in the Eastern Interconnection, where the electric grid is already denser, and where the states are typically further away from reaching the ambitious clean energy goals than the states in the West.

As a reminder, in the Western Interconnection, almost every state matches California's ambitious goal, the exceptions being Wyoming and Idaho since they do not have existing 2030 goals. In the Eastern Interconnection, only states with existing 2030 goals are assumed to upgrade their goals to match those of New York; this modeling assumption creates a pronounced geographical imbalance between areas with high shares of renewable energy (generally the Northeast and the Plains) and areas without such shares (generally the Southeast).

Without HVDC lines crossing the interconnection seams, power transfer mainly occurs locally and within each interconnection. As can be seen in Figure 16, large amounts of wind energy from Oklahoma and nearby states flow eastward to serve demand centers from Memphis to New Orleans. The path from Oklahoma to Memphis is one of the same corridors that Clean Line Energy was targeting for developing a large transmission line, as discussed in Russell Gold's *Superpower: One Man's Quest to Transform American Energy.*[43] Solar energy from North Carolina flows southward to support demand in Florida. Along the West coast, local demand is primarily met via solar energy from California and Arizona, along with wind energy in Washington and Oregon. In ERCOT, wind energy from western Texas flows eastward to serve the demand centers in eastern Texas.

FIG. 15

## Design 1 - Transmission Upgrades

Transmission upgrades for Design 1 yield 36% more transmission capacity (measured in MW-miles) for the simulated 2030 grid compared to the current 2020 grid.

— Upgraded AC transmission

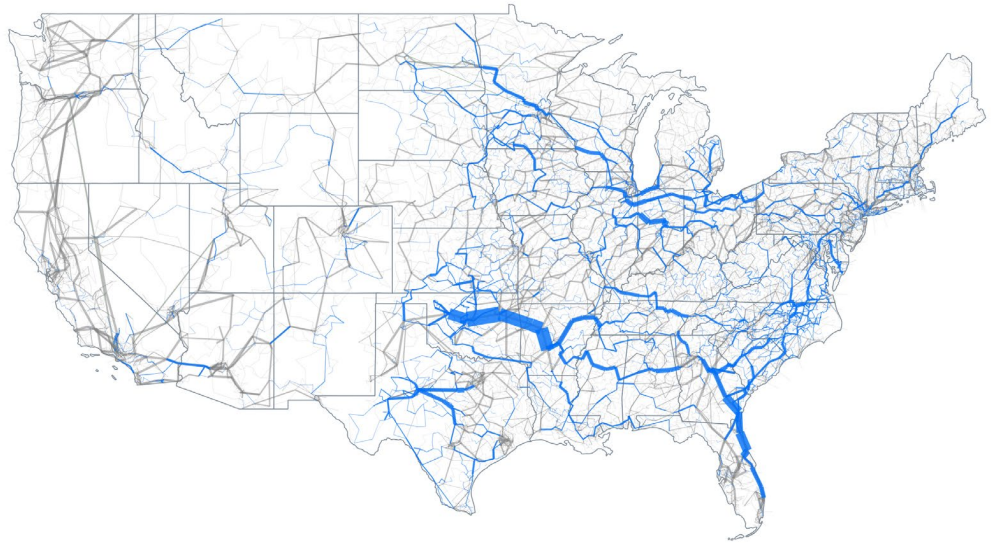

FIG. 16

## Design 1 - Net Energy Flow

The net energy flow in Design 1 shows large amounts of wind energy from the Plains and West Texas flowing eastward within their respective interconnections.

— HVDC power flow
— AC power flow

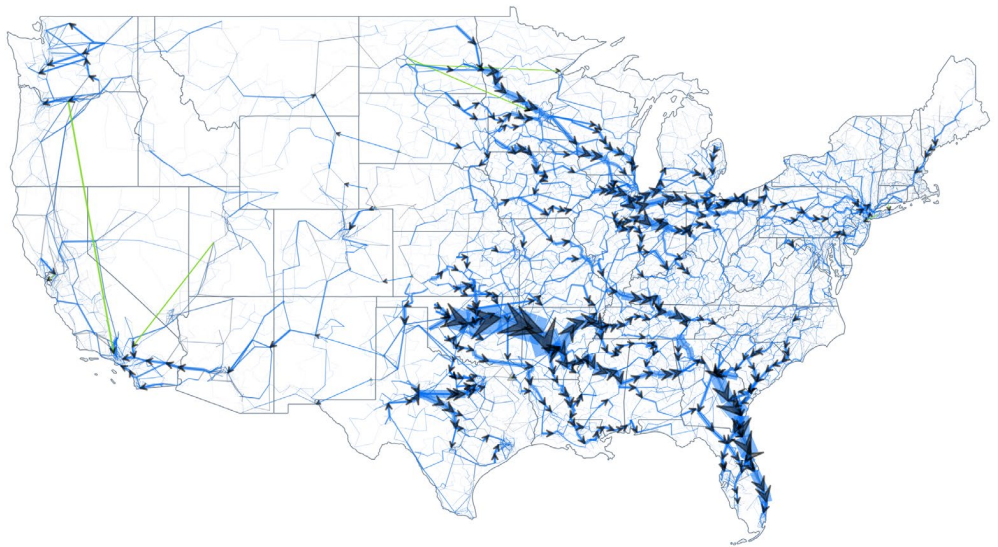

## 4.4 - Common HVDC Operation Patterns

With no upgrades to the HVDC infrastructure in Design 1, only 10.7 TWh of electricity are transferred across the East-West seam and 6.9 TWh are transferred across the East-ERCOT seam. For each of the upgraded HVDC scenarios (Designs 2a, 2b, and 3), energy transfer is dramatically increased: approximately 160-240 TWh are exchanged across the East-West seam, and approximately 60-70 TWh flow into and out of ERCOT during the simulated year. The overall capacity factor for the HVDC infrastructure is greater than 65% across all HVDC upgrade scenarios. A summary of the annual energy transfers for all designs is given in Table 10.

TABLE 10

### Annual Energy Transfers for Each Macro Grid Design

For each of the upgraded HVDC scenarios (Designs 2a, 2b, and 3), energy transfer is dramatically increased: approximately 160-240 TWh are exchanged across the East-West seam and approximately 60-70 TWh flow into and out of ERCOT during the simulated year.

| METRIC | MACRO GRID DESIGNS | | | |
|---|---|---|---|---|
| | 1 | 2A | 2B | 3 |
| East-to-West (TWh) | 6.9 | 95.6 | 130.0 | 112.0 |
| West-to-East (TWh) | 3.8 | 59.5 | 112.3 | 87.6 |
| East-to-West to West-to-East ratio | 1.83:1 | 1.61:1 | 1.16:1 | 1.28:1 |
| East-to-West Capacity Factor | 93% | 65% | 70% | 68% |
| East-to-ERCOT (TWh) | 4.0 | 31.8 | 32.0 | 7.6 |
| ERCOT-to-East (TWh) | 2.9 | 27.9 | 25.3 | 24.2 |
| East-to-ERCOT to ERCOT-to-East ratio | 1.38:1 | 1.14:1 | 1.26:1 | 1:3.2 |
| East-to-ERCOT Capacity Factor | 98% | 83% | 80% | 45% |
| West-to-ERCOT (TWh) | - | - | - | 15.7 |
| ERCOT-to-West (TWh) | - | - | - | 25.3 |
| West-to-ERCOT to ERCOT-to-West ratio | - | - | - | 1:1.61 |
| West-to-ERCOT Capacity Factor | - | - | - | 58% |
| ERCOT 'pass through' Capacity Factor | - | - | - | 27% |
| ERCOT-to-Elsewhere Capacity Factor | 98% | 83% | 80% | 79% |

Across all three upgraded HVDC scenarios, the pattern of power transfer across the East-West* Interconnection seam is similar, as shown in Figure 17. The net power exchange flows from East to West, but there is a strong diurnal pattern of the West exporting power during the daytime hours and the East exporting power during most other hours. There is also a strong seasonal pattern, with power transfer nearly balanced during the spring and summer, but with the East as a strong exporter during the fall and winter.

These East-to-West power flows are strongly correlated with the difference in instantaneous renewable generation among the interconnections; when the Eastern Interconnection is generating more renewable energy as a share of its own demand than the Western Interconnection, it is very likely that the HVDC network will be transferring power from East to West, and vice versa. This effect is shown in Figure 18.

\* The power transfers into and out of ERCOT differ depending on the HVDC design and are discussed in Sections 4.5, 4.6, and 4.7.

FIG. 17
# Patterns of Power Transfer Across the East-West Interconnection Seam

Across the three upgraded HVDC designs, the Eastern Interconnection is typically sending power West during the early morning and late evening hours when there is usually a wind energy surplus, and the Western Interconnection is typically sending power East during the daytime hours when there is an abundance of solar energy. Positive power flow indicates exports from East to West and negative power flows indicate exports from West to East.

**DESIGN 2A**

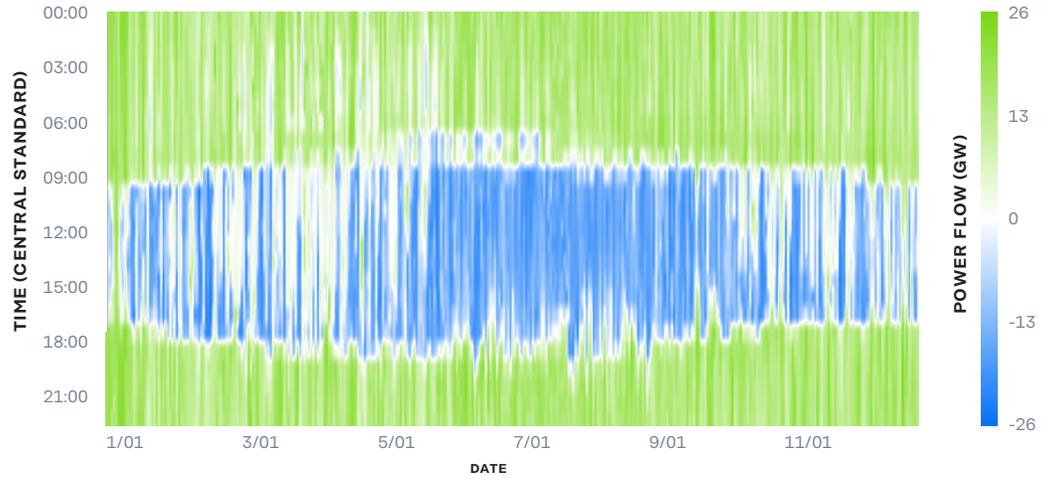

**DESIGN 2B**

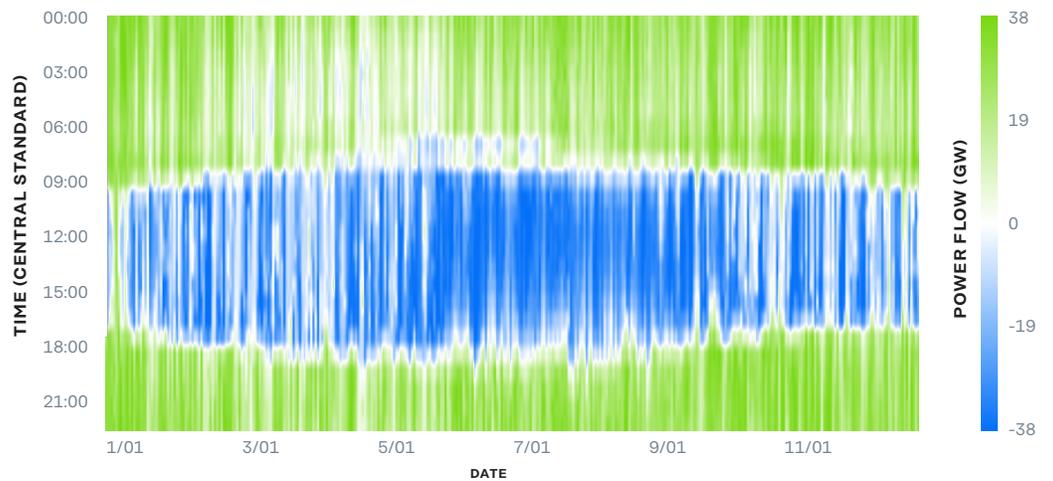

**DESIGN 3**

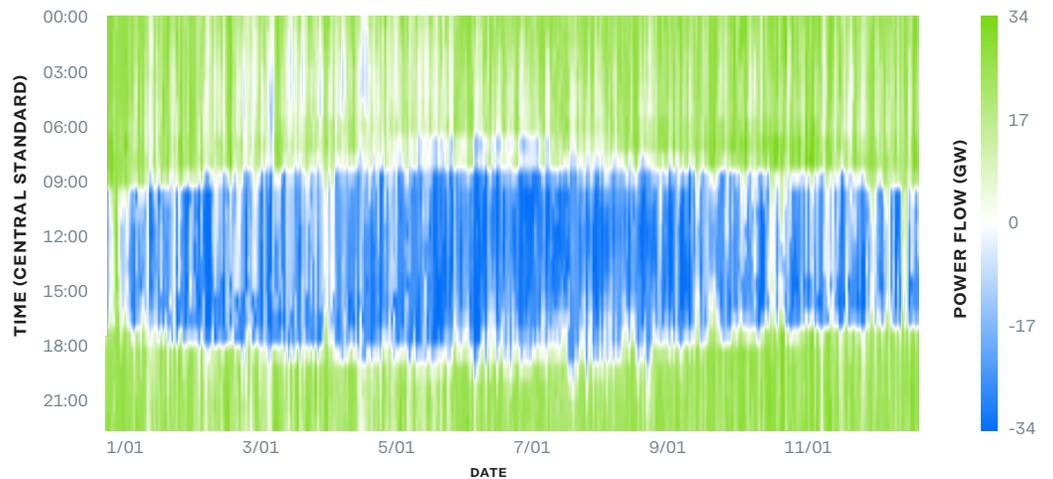

FIG. 18

# Correlation Between Renewable Generation Differences and Power Flow Across East-West Interconnection Seam

Across the three upgraded HVDC designs, the East-to-West power flows are strongly correlated with the difference in instantaneous renewable generation among the interconnections. The purple circles indicate the hours of November 2nd, a particularly variable day. Data are identical between the left and right columns; the only difference is a shift in perspective.

● = November 2nd

**DESIGN 2A**

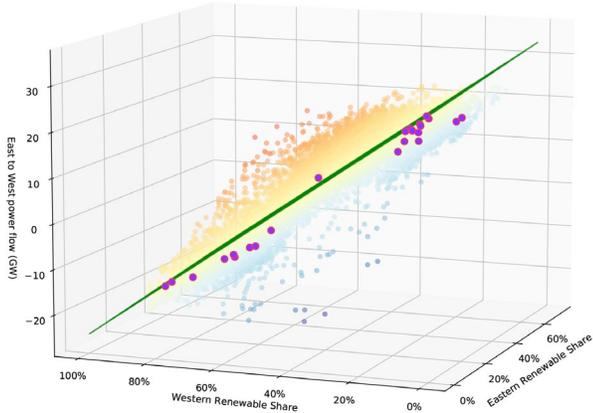
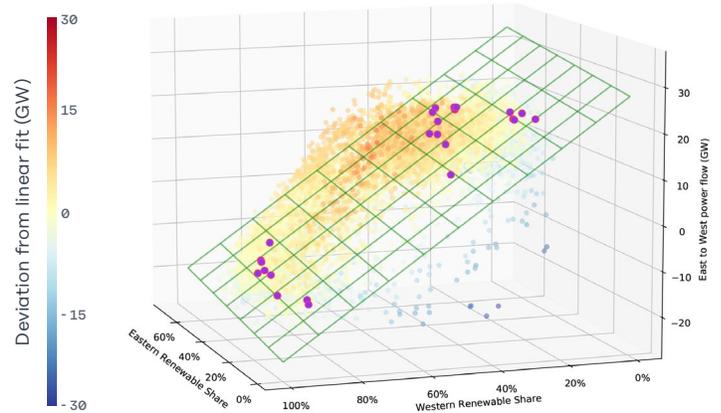

**DESIGN 2B**

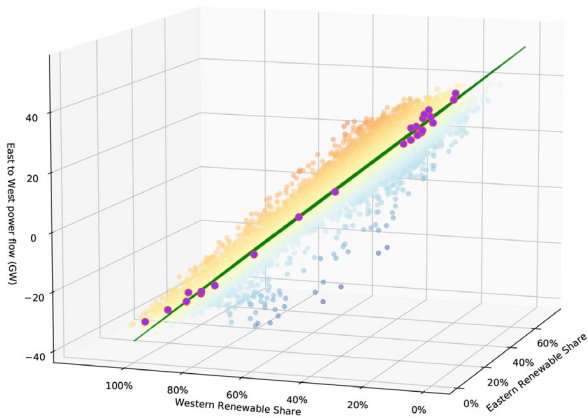
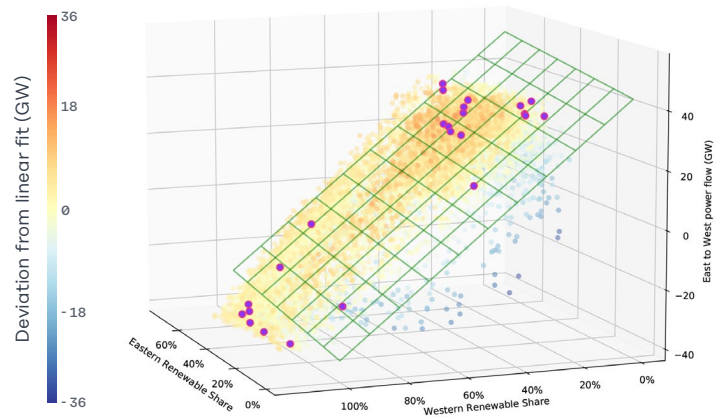

**DESIGN 3**

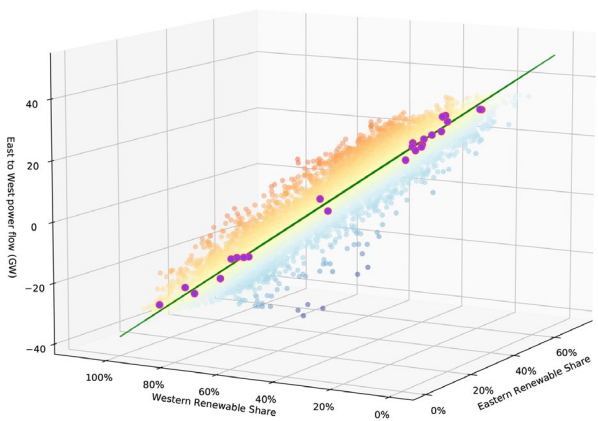
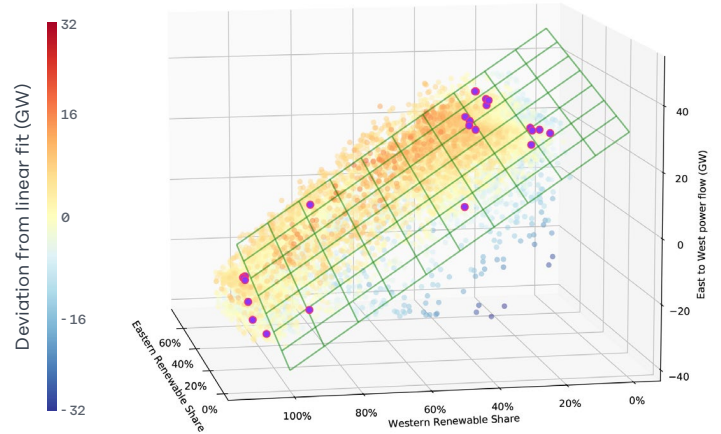

As seen in each of the sub-figures in Figure 18, the planes represent the best fit of the underlying data; as expected, there is an increase in the East-to-West power flow when the Western renewable share decreases and the Eastern renewable share increases. The associated linear coefficients are determined by an ordinary-least-squares fit to a multiple linear regression model, with a constant term added to orient the regression line to the origin. The raw coefficients of these model fits are shown in Table 11.

TABLE 11

## Correlation Values Between Renewable Generation Differences and Power Flow Across East-West Interconnection Seam

These are the correlation values between the instantaneous renewable generation by inter-connection as a share of its own demand and the cross-seam power flows. As expected, there is an increase in the East-to-West power flow when the Western renewable share decreases and the Eastern renewable share increases.

| DESIGN | CONSTANT TERM (GW) | EASTERN RENEWABLE SHARE COEFFICIENT (GW) | WESTERN RENEWABLE SHARE COEFFICIENT (GW) | $r^2$ |
|---|---|---|---|---|
| 2a | 27.6 | 8.49 | -52.4 | 0.862 |
| 2b | 43.2 | 13.1 | -79.8 | 0.937 |
| 3 | 29.4 | 24.1 | -65.6 | 0.917 |

To demonstrate one particular instance of large power flow swings within a single day (highlighted via the purple circles in Figure 18), consider Figures 19 and 20, which show the power flow patterns at 8:00am and at 4:00pm Central Standard Time (CST) on November 2nd in the model.

At 8:00am CST on November 2nd, the wind energy from the Plains is abundant, whereas the solar energy from the West is not yet available. With Design 1 lacking any HVDC additions, Figure 19A shows how the wind energy from the Plains traverses primarily to the demand centers ranging from Detroit to New Orleans. For Design 2a, where upgrades are made to the back-to-back HVDC tie lines (resulting in a total capacity of 35 GW) and the AC networks (particularly in the Western Interconnection), Figure 19B shows how wind energy from the Plains is able to support demand all the way from Detroit to the Rocky Mountains. Similarly, the moderate back-to-back HVDC tie line upgrades and three new 9,500-MW long-distance HVDC lines prescribed by Design 2b are fully utilized in sending wind energy from the Plains to the West, as is shown in Figure 19C. Finally, Figure 19D shows how the HVDC network in Design 3 also distributes the abundant wind energy from the Plains to the different Western demand centers.

By contrast, at 4:00pm CST on November 2nd, the sun has set in the East (notice that the solar energy from North Carolina to New England is no longer available) but is still up in the West, leading to large amounts of solar energy being available in the Western Interconnection. Without any HVDC additions in Design 1, Figure 20A shows how this solar energy is spread around the West, but is unable to reach the demand centers in the Eastern Interconnection. This leads to a substantial amount (i.e., 20-30 GW) of that solar energy being curtailed. On the other hand, as is shown in Figure 20B, the upgraded back-to-back HVDC tie lines modeled in Design 2a create a path through which abundant solar energy from the Western Interconnection can pass to support late afternoon demand in the Eastern Interconnection. Similarly, Figures 20C and 20D show how the three long-distance HVDC lines of Design 2b and the HVDC network of Design 3, respectively, enable surplus solar energy to flow from the West to Eastern demand centers.*

* In both cases for Design 3, the HVDC lines that connect in ERCOT pass the full power capacity (8 GW) from East to West in the morning and then West to East in the evening. This is not always the case and will be discussed further in Section 4.7.

FIG. 19

# Making Use of Wind Energy from the Plains in the Morning

At 8:00am CST on November 2nd, the wind energy from the Plains is abundant. With pathways available across the East-West seam in Designs 2a, 2b, and 3, this wind energy can support demand centers from coast to coast, depending on the design.*

**(A) DESIGN 1**

With a limited cross-seam exchange in Design 1, wind energy from the Plains traverses primarily to the demand centers ranging from Detroit to New Orleans.

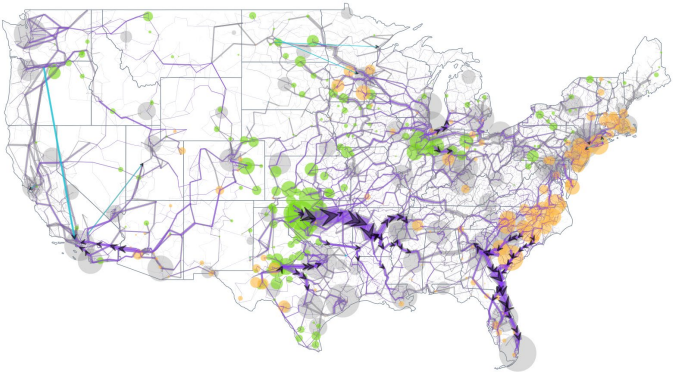

**(B) DESIGN 2A**

With upgrades to the B2B HVDC tie lines in Design 2a, wind energy from the Plains travels along the upgraded AC transmission network to support demand centers all the way from Detroit to the Rocky Mountains.

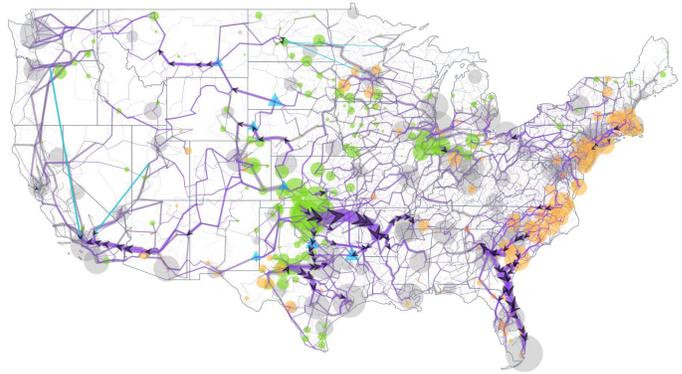

**(C) DESIGN 2B**

With three new HVDC lines crossing the East-West seam in Design 2b, wind energy from the Plains jumps across the East-West seam to support the demand centers on the West Coast along with others around the East.

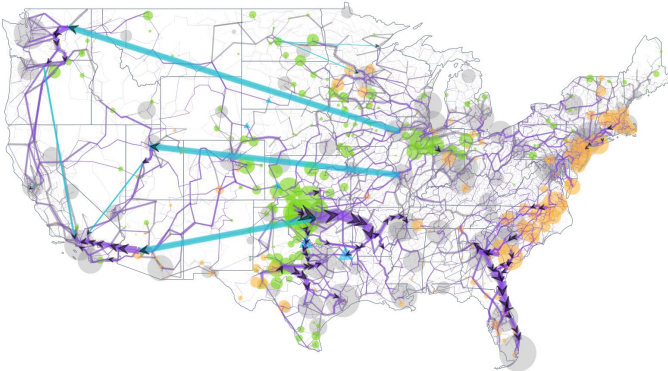

**(D) DESIGN 3**

With a full HVDC overlay in Design 3, wind energy from the Plains supports a wide range of demand centers on the West Coast along with others around the East, passing through ERCOT but not delivering any energy there in this particular hour.

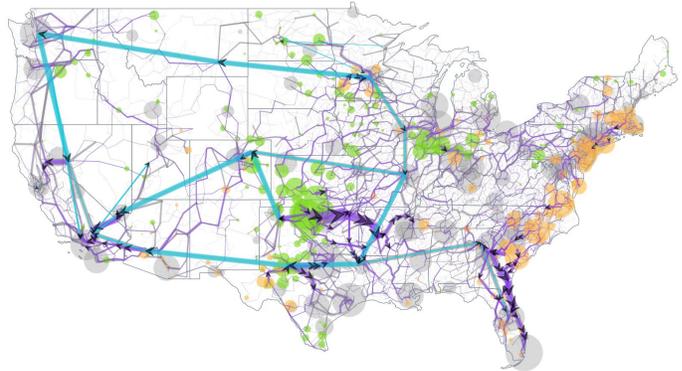

— HVDC power flow   — AC power flow   ● Solar generation   ● Wind generation   ● Demand

\* The line thickness of every transmission line is proportional to the power flow over the line. To help visualize the aggregate flow patterns, all DC lines and any AC transmission line with more than 3,000 MW flowing over it has an arrow added at the receiving end of the line.

FIG. 20

# Making Use of Solar Energy from the West in the Afternoon

At 4:00pm CST on November 2nd, the sun has set in the East but is still up in the West, leading to large amounts of solar energy crossing the East-West seam in Designs 2a, 2b, and 3 to support the late afternoon demand in the Eastern Interconnection.*

**(A) DESIGN 1**

With a limited cross-seam exchange in Design 1, solar energy from the West traverses primarily to the Western demand centers or is curtailed.

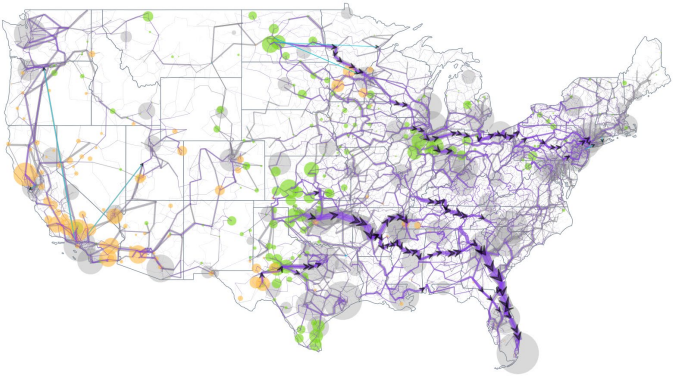

**(B) DESIGN 2A**

With upgrades to the B2B HVDC tie lines in Design 2a, solar energy from the West travels along the upgraded AC transmission network to support demand centers all the way from the West Coast to the Plains.

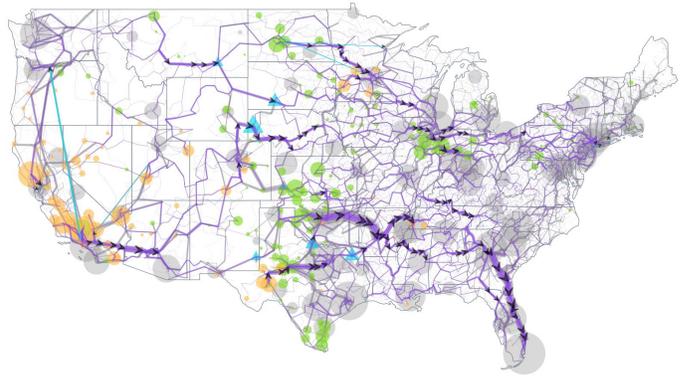

**(C) DESIGN 2B**

With three new HVDC lines crossing the East-West seam in Design 2b, solar energy from the West jumps across the East-West seam to support demand centers as far East as Detroit.

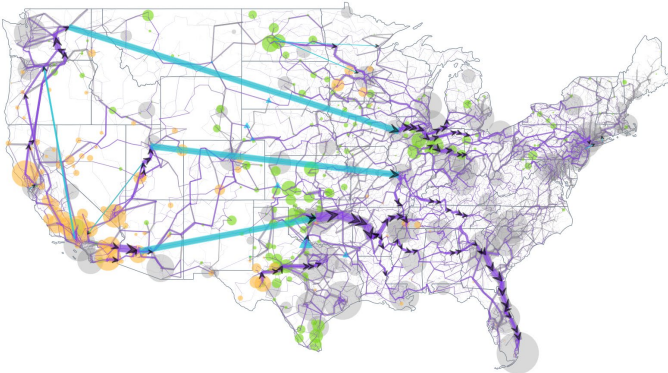

**(D) DESIGN 3**

With a full HVDC overlay in Design 3, solar energy from the West supports a wide range of demand centers across the U.S., this time arguably reaching all the way from the West Coast to Florida, passing through ERCOT but not delivering any energy there in this particular hour.

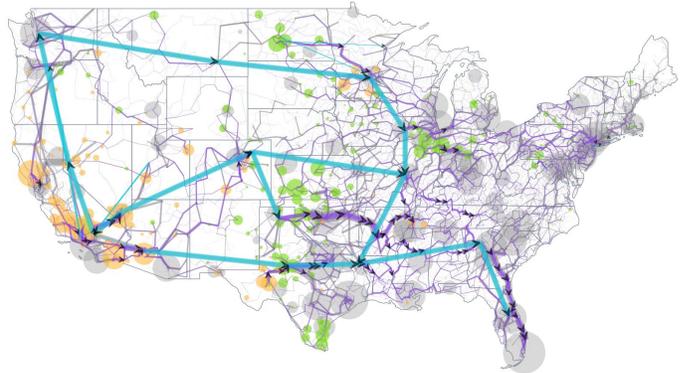

— HVDC power flow     — AC power flow     ● Solar generation     ● Wind generation     ● Demand

* The line thickness of every transmission line is proportional to the power flow over the line. To help visualize the aggregate flow patterns, all DC lines and any AC transmission line with more than 3,000 MW flowing over it has an arrow added at the receiving end of the line.

## 4.5 - Results Unique to Design 2a

The upgraded HVDC back-to-back facilities in Design 2a provide a path for cross-seam sharing of complementary renewable resources, which results in a substantial increase in AC transmission to be built in the Western Interconnection and near the upgraded back-to-back facilities as compared to Design 1. Conversely, the Eastern Interconnection sees substantial reductions in the AC transmission upgrades compared to Design 1, as shown in Figure 21.

Of the HVDC scenarios, Design 2a features the least total energy transferred over the East-West seam, as shown in Table 10. However, it also features the greatest net East-to-West energy transfer (36.1 TWh over the simulated year). Despite this, the Western Interconnection uses approximately twice as much natural gas in Design 2a as compared to Designs 2b and 3; this increased natural gas usage is primarily in Arizona, California, and Nevada, suggesting that the power flowing across the seam from the East is less capable of reaching the load centers in the Southwest.

FIG. 21

### Transmission Changes Between Designs 1 and 2a

Upgraded HVDC B2B facilities in Design 2a cause more AC transmission upgrades in the Western Interconnection and less throughout the Eastern Interconnection as compared to Design 1.

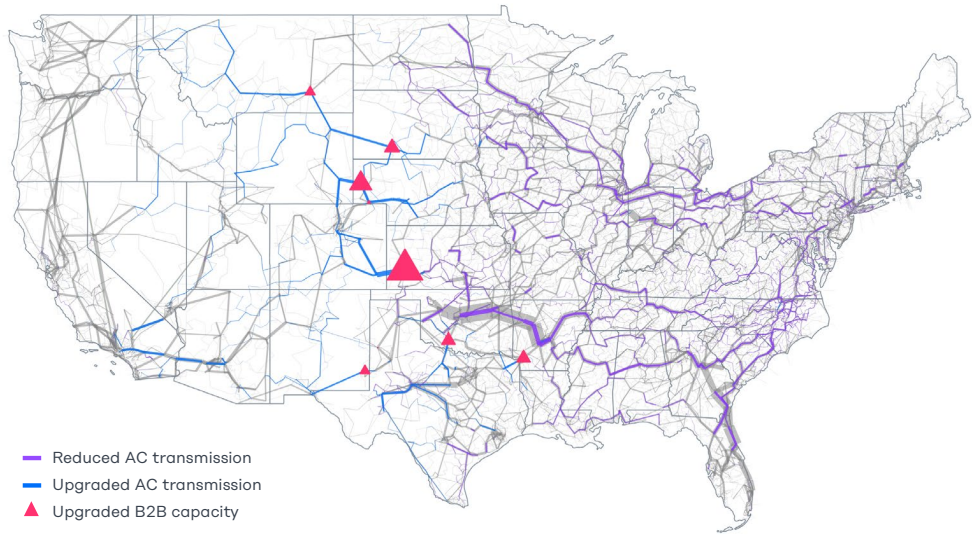

— Reduced AC transmission
— Upgraded AC transmission
▲ Upgraded B2B capacity

FIG. 22

### Patterns of East-ERCOT Power Flow in Design 2a

The pattern of power transfer along the East-ERCOT seam in Design 2a is sensitive to grid conditions, causing varied daily and hourly patterns. Positive power flow indicates exports from East to ERCOT and negative power flows indicate exports from ERCOT to East.

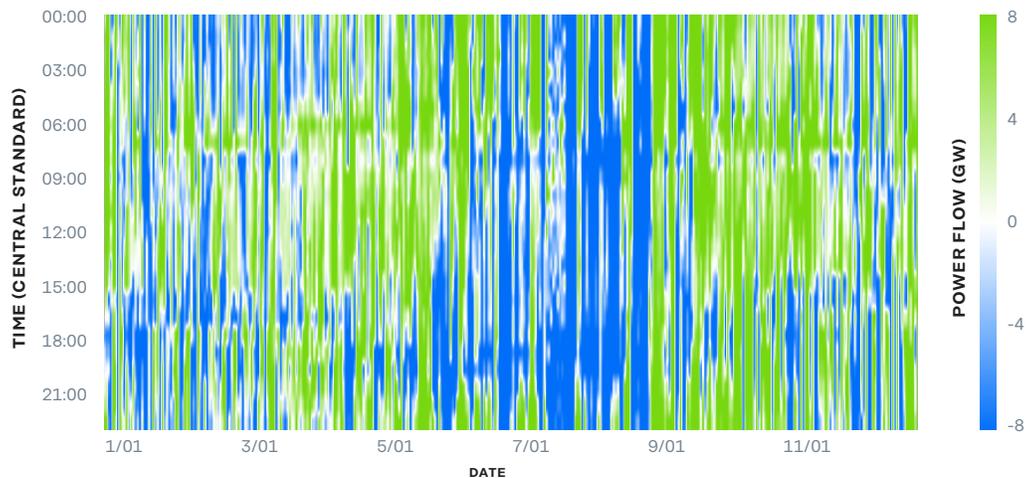

Along the East-ERCOT seam, the upgraded HVDC back-to-back facilities allow for a substantial increase in total energy transferred, jumping from 6.9 TWh in Design 1 to 59.7 TWh in Design 2a. The Eastern Interconnection sends roughly 4 TWh more to ERCOT than in reverse, causing the noticeable drop in total energy generation within ERCOT, shown in Figure 9C. Also, the pattern of energy transfer varies substantially across the hour of the day and the day of the year, as shown in Figure 22.

## 4.6 - Results Unique to Design 2b

The addition of the three HVDC lines and upgraded HVDC back-to-back facilities in Design 2b again causes a substantial increase in AC transmission to be built in the Western Interconnection as compared to Design 1, although this time primarily along pathways leading to the terminals of the new HVDC lines. While the Eastern Interconnection again sees substantial reductions in the AC transmission upgrades relative to Design 1, there are a few segments near the other ends of the HVDC lines that are further upgraded, as can be seen in Figure 23. In total, this design upgrades less AC transmission overall compared to Design 1: 26% of 2020 capacity in terms of MW-miles, as compared to 36% for Design 1.

Of the HVDC scenarios, Design 2b features the highest total energy transferred over the East-West seam at 242 TWh for the simulated year, as shown in Table 10. The structure of the HVDC lines allows for easier access to the demand centers in both the West and the middle of the Eastern Interconnection. The three HVDC lines carry 175.9 TWh, or 73% of the total energy transfer across the East-West seam for Design 2b.

FIG. 23

**Transmission Changes Between Designs 1 and 2b**

Three new HVDC lines and upgraded HVDC B2B facilities in Design 2b require additional AC transmission upgrades along the pathways to the terminals of the new HVDC lines relative to Design 1.

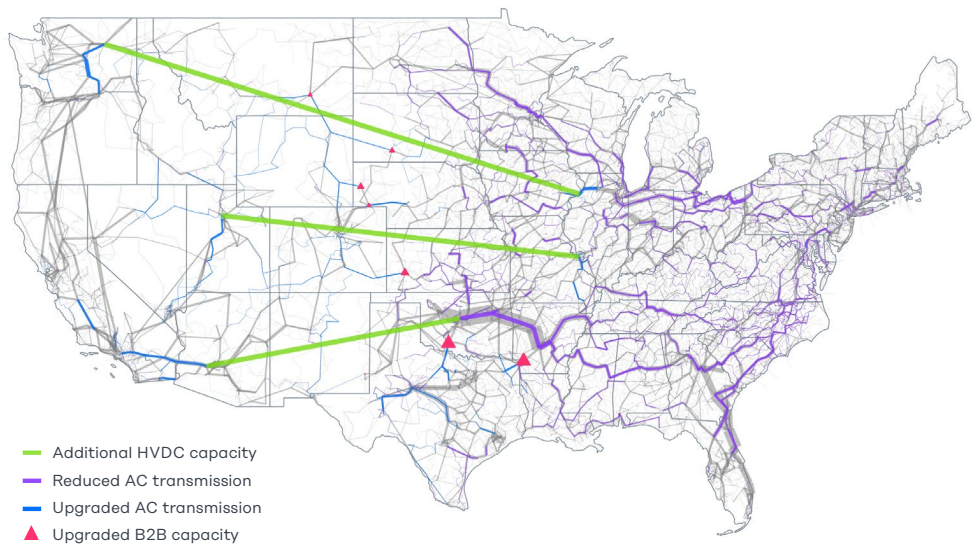

- Additional HVDC capacity
- Reduced AC transmission
- Upgraded AC transmission
- Upgraded B2B capacity

FIG. 24

**Patterns of East-ERCOT Power Flow in Design 2b**

The pattern of power transfer along the East-ERCOT seam in Design 2b is sensitive to grid conditions, causing varied daily and hourly patterns. Positive power flow indicates exports from East to ERCOT and negative power flows indicate exports from ERCOT to East.

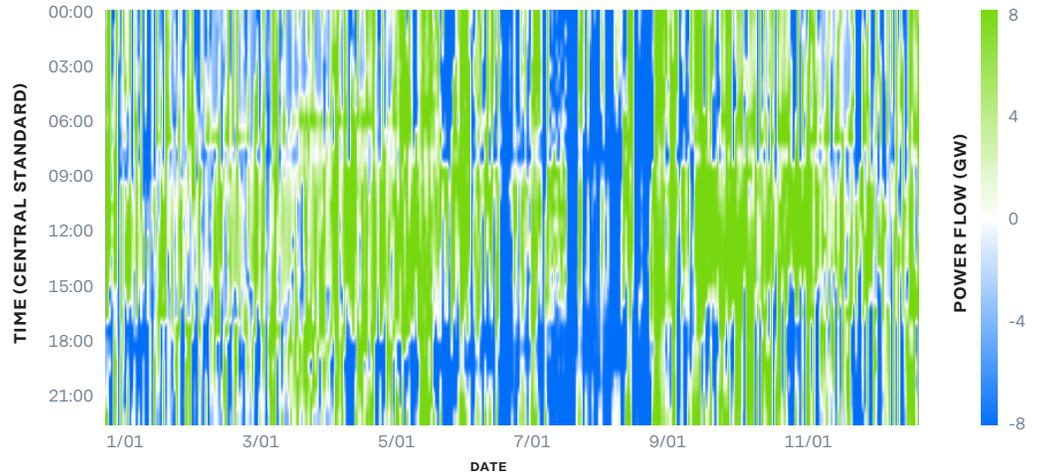

Similar to Design 2a, the upgraded HVDC back-to-back facilities along the East-ERCOT seam allow for a substantial increase in total energy transferred there, jumping from 6.9 TWh in Design 1 to 57.3 TWh in Design 2b. In this case, the Eastern Interconnection sends roughly 7 TWh more to ERCOT, again causing a noticeable drop in the total energy generation within ERCOT, shown in Figure 9C. Once again, the pattern of energy transfer varies substantially across the hour of the day and the day of the year, as shown in Figure 24.

## 4.7 - Results Unique to Design 3

Of all the designs, Design 3 requires the fewest AC transmission upgrades: transmission upgrades increase 23% over 2020 capacity, compared to 36% in Design 1. The differences in the transmission network upgrades between Design 1 and Design 3 can be seen in Figure 25. Examining this map, a few patterns emerge:

— In some regions, HVDC transmission upgrades appear to be a direct substitute for AC transmission upgrades; for example, the transmission corridor from Georgia to Florida.

— In some regions, HVDC transmission upgrades appear to be an indirect substitute for AC transmission upgrades; for example, the HVDC line from the Texas Panhandle to Colorado (and from there, onward to the rest of the HVDC network) seems to reduce the need for AC transmission upgrades from Oklahoma to Arkansas.

— On a continental scale, the HVDC network changes the balance of where renewable energy is sourced in order to meet the country's clean energy goals. The Western Interconnection exports more clean energy (primarily solar), reducing the need for transmission upgrades in the Eastern Interconnection, since the country's goals can be met while tolerating more curtailment in the East.

— Similarly, wind energy from the Plains in the Eastern Interconnection has a more direct path to the Southeast and Florida, as evidenced by the net energy flows depicted in Figure 26.

As in the other HVDC upgrade scenarios, the balance of energy transfer over the simulated year is East-to-West, but the West exports substantial amounts of power to the East during the daytime hours. Net energy transfer across the network is shown in Figure 26. The ERCOT grid is a net exporter to both the Eastern and Western Interconnections, as shown in Table 10, exporting 24.2 TWh to the Eastern Interconnection and 25.3 TWh to the Western Interconnection while importing 7.6 TWh and 15.7 TWh in return, respectively, over the simulated year.

FIG. 25

## Transmission Changes Between Designs 1 and 3

The HVDC overlay in Design 3 requires the least amount of AC transmission upgrades, primarily upgrading AC paths that connect to the HVDC lines.

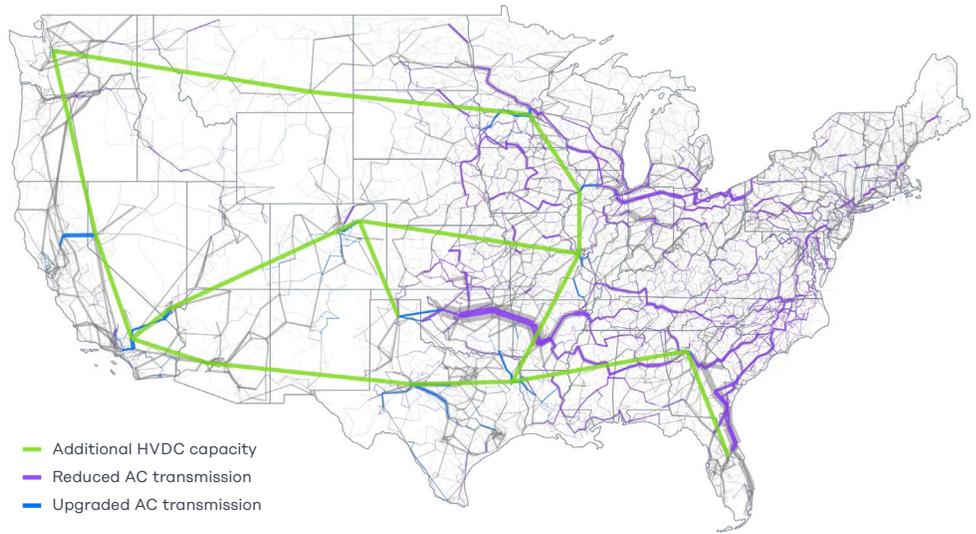

FIG. 26

## Net Energy Flows in Design 3

Net energy flows in Design 3 send wind energy from the Plains to the Southeast and Florida.

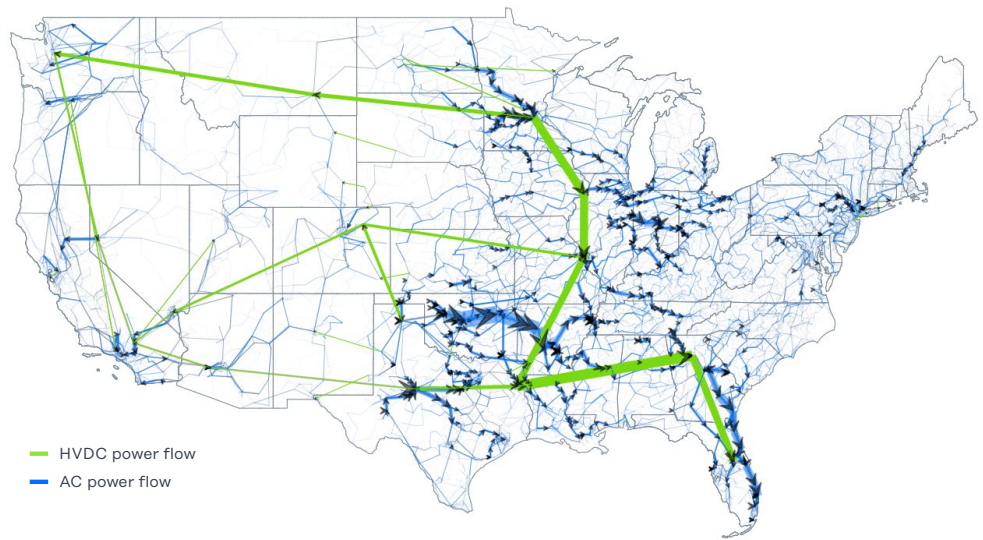

FIG. 27

# Patterns of Power Flow Into, Out of, and Through ERCOT in Design 3

ERCOT is a net exporter of power in Design 3, usually exporting at night and importing during the day. One major difference in Design 3 relative to Designs 2a and 2b is how the energy flowing on the HVDC lines in some hours will 'pass through' ERCOT. This accounts for 27% of the capacity factor utilization of the HVDC lines that connect in Sweetwater, TX.

**(A) EAST-ERCOT**

Power flow over the East-ERCOT seam in Design 3, excluding 'pass through' to and from the West. Positive power flow indicates exports from East to ERCOT and negative power flows indicate exports from ERCOT to East.

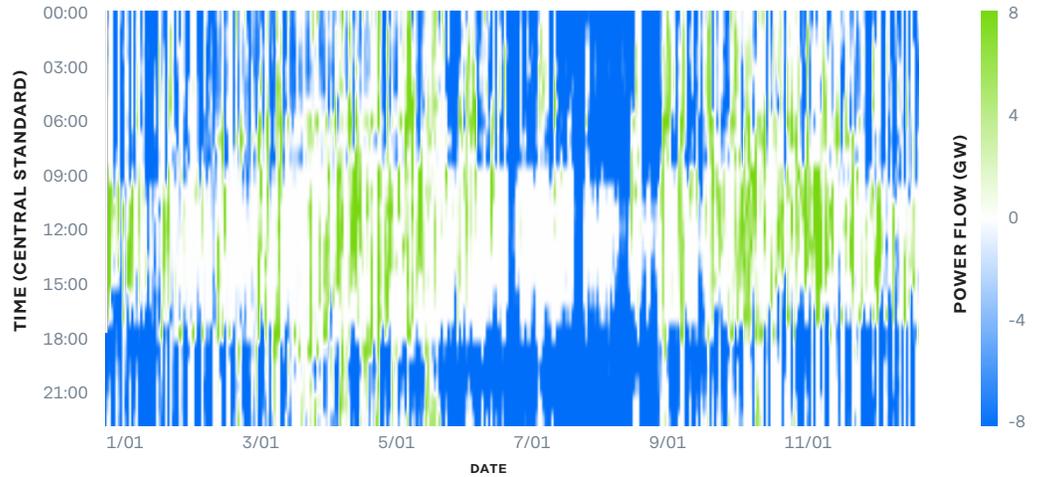

**(B) WEST-ERCOT**

Power flow over the West-ERCOT seam in Design 3, excluding 'pass through' to and from the East. Positive power flow indicates exports from West to ERCOT and negative power flows indicate exports from ERCOT to West.

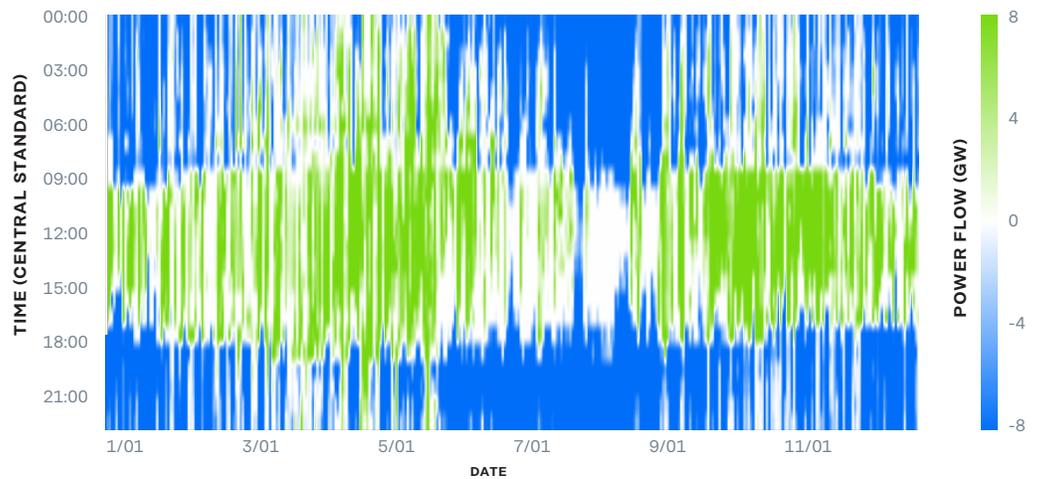

**(C) ERCOT PASS THROUGH**

ERCOT 'pass through', with power flowing East to West. Positive power flow indicates exports from East to West (through ERCOT) and negative power flows indicate exports from West to East (through ERCOT).

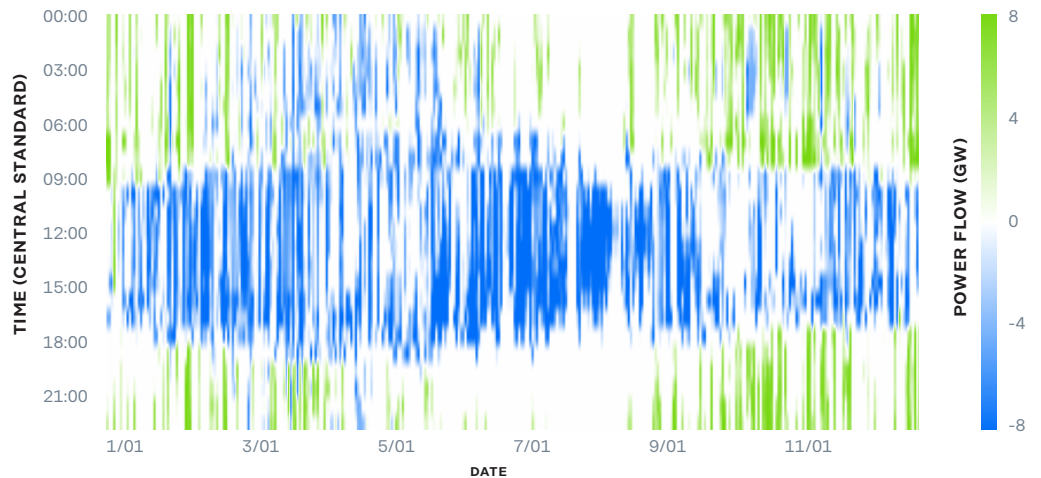

ERCOT mostly exports to the East at night, as shown in Figure 27A, with power relatively balanced during the day; the exception is during the summer months, where ERCOT sometimes exports the full HVDC link capacity (8 GW) for days at a time. ERCOT also mostly exports to the West at night, as shown in Figure 27B, with the greatest exports occurring during the summer. During the spring and fall months, ERCOT receives substantial imports from the West during the daytime. Power flow over the ERCOT HVDC lines is moderately correlated with the imbalance of instantaneous renewable penetration between ERCOT and the other interconnections, as shown in Figure 28.

FIG. 28

## Correlation Between Renewable Generation Differences and Power Flow Across ERCOT Interconnection Seams

Power flow from ERCOT to the Eastern and Western Interconnections is moderately correlated ($r^2$ = 0.63) with the difference in renewable generation shares between the interconnections. The purple circles indicate the hours of November 2nd, a particularly variable day, where the two HVDC lines that connect in Sweetwater, TX are fully utilized in the same and opposite directions at different times throughout the day. See Figure 29 for examples of each.

● = November 2nd

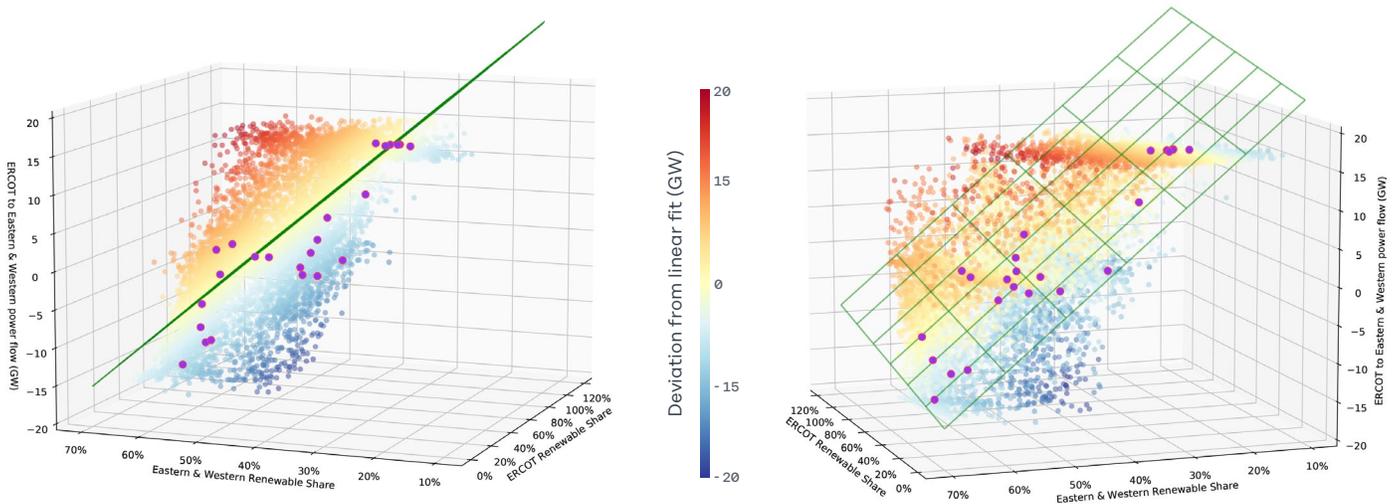

One major difference in Design 3 relative to Designs 2a and 2b is how the energy flowing on the HVDC lines in some hours will 'pass through' ERCOT. This accounts for 27% of the capacity factor utilization of the HVDC lines that connect in Sweetwater, TX. The hourly and daily patterns of this 'pass through' are shown in Figure 27C. A few hourly power flow examples, again from November 2nd as originally discussed in Section 4.4, are shown in Figure 29 with a focus on the ERCOT region that has a full import, export, and 'pass through' in both directions at different hours on this day.

FIG. 29

## Snapshot: Power Flow Across ERCOT Interconnection Seams on a Highly Variable Day

On November 2nd, the ERCOT region has a full import, export, and 'pass through' in both directions at different hours of the day.

**(A) EAST-TO-WEST 'PASS THROUGH' AT 8:00AM CST**

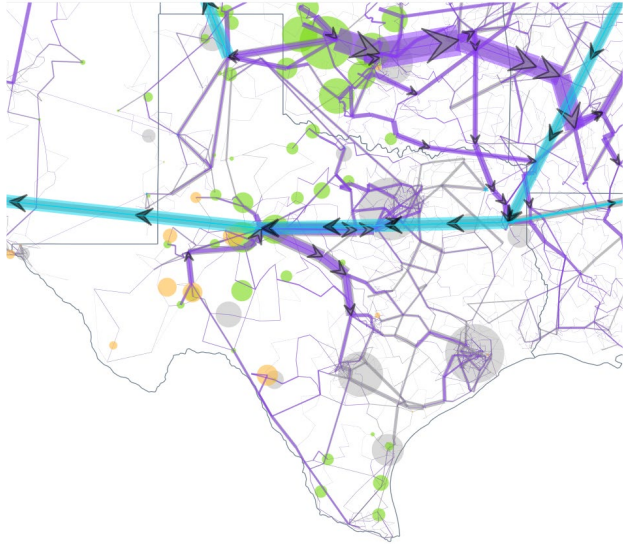

**(B) WEST-TO-EAST 'PASS THROUGH' AT 4:00PM CST**

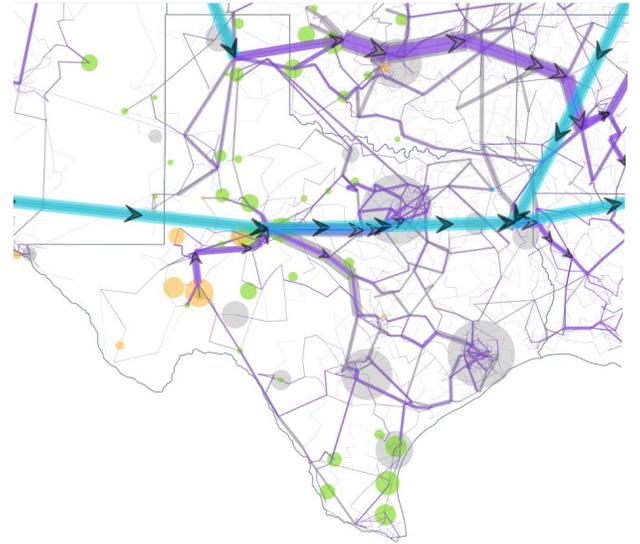

**(C) FULL IMPORT TO ERCOT AT 9:00AM CST**

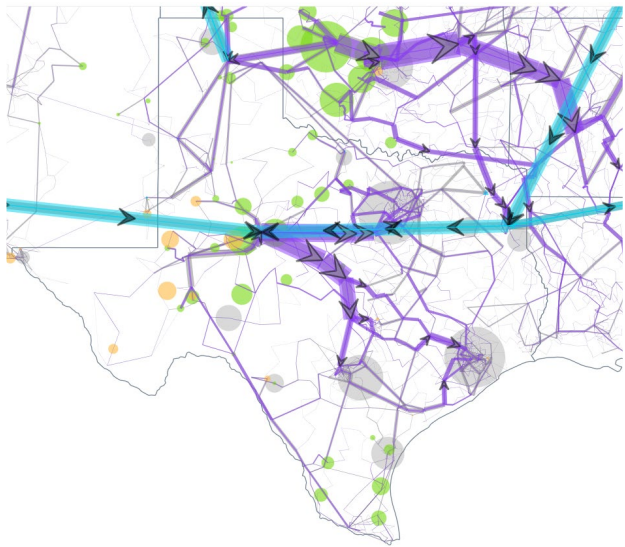

**(D) FULL EXPORT FROM ERCOT AT 6:00PM CST**

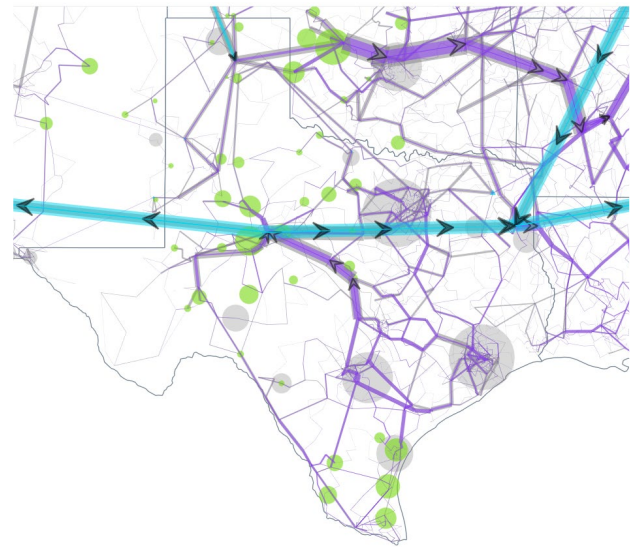

— HVDC power flow   — AC power flow   ● Solar generation   ● Wind generation   ● Demand

## 4.8 - Common Corridors for Transmission Upgrades

To support the renewable capacity buildout proposed in this report, certain U.S. transmission corridors require large capacity upgrades regardless of the Macro Grid design selected. Figure 30 reveals these commonly upgraded corridors, with line thickness representing the minimum amount of transmission capacity added across all four Macro Grid designs. As discussed earlier, the Oklahoma to Memphis corridor is clearly in need of expanded transmission. Corridors between Georgia and Florida, through the upper Midwest, and spanning Texas (even after their successful Competitive Renewable Energy Zones (CREZ) process in the early 2010s) also benefit from substantial transmission upgrades across each Macro Grid design.

FIG. 30

### Transmission Upgrades Selected Across All Macro Grid Designs

These U.S. transmission corridors require large capacity upgrades regardless of the Macro Grid design selected.

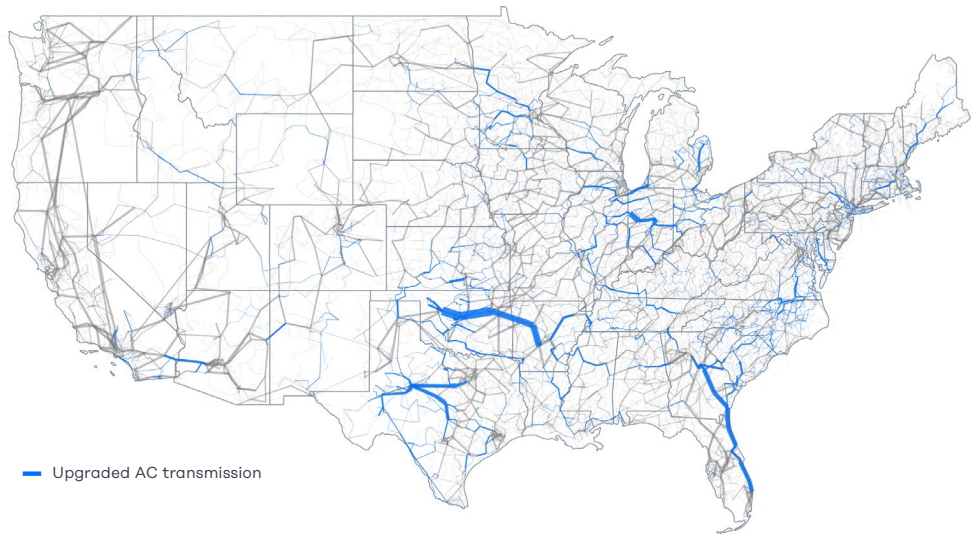

— Upgraded AC transmission

These common upgrades sum to 56 TW-miles, representing at least half of the upgrades for each design, and suggest that many of the AC transmission upgrades required to integrate new clean energy capacity are robust to the design of any HVDC enhancements to the Macro Grid backbone.

## 4.9 - State-Level Details

As discussed in Section 1.2, open-source energy data and models provide transparency and reproducibility, which is particularly important in the energy policy-making process. For this model, high spatial resolution allows for a deeper exploration into the details for each state under different scenario conditions. Two brief examples are highlighted here, and all detailed data has been open-sourced for further investigation.[20]

Figure 31A shows the changes in the generation mix and energy amounts between scenarios in Montana. For the 2020 scenario, Montana is a net energy exporter given its abundance of hydro and coal generation. In the 'ambitious goals' scenarios, Montana takes full advantage of its strong wind resource to continue being a net energy exporter, despite a steep reduction in coal generation. For the 'current goals' scenario, Montana does not realize as much wind generation capacity as in the 'ambitious goals' scenarios, resulting in an annual production that nearly matches the state's annual demand.

On the other hand, states with no clean energy goals that depend on coal and natural gas see their total energy generation drop substantially as their neighbors increase their clean energy capacity. As can be seen in Figure 31B, Louisiana is one such state, going from being a net energy exporter in the 2020 and 'current goals' scenarios to a net energy importer in the 'ambitious goals' scenarios. It should be noted that these scenario designs were finalized prior to Louisiana signing an executive order targeting a 50% reduction in economy-wide emissions by 2030.[48]

FIG. 31

### State-Level Generation Mix Examples

High spatial resolution allows for a deeper exploration into the details for each state under different scenario conditions, with two example states here.

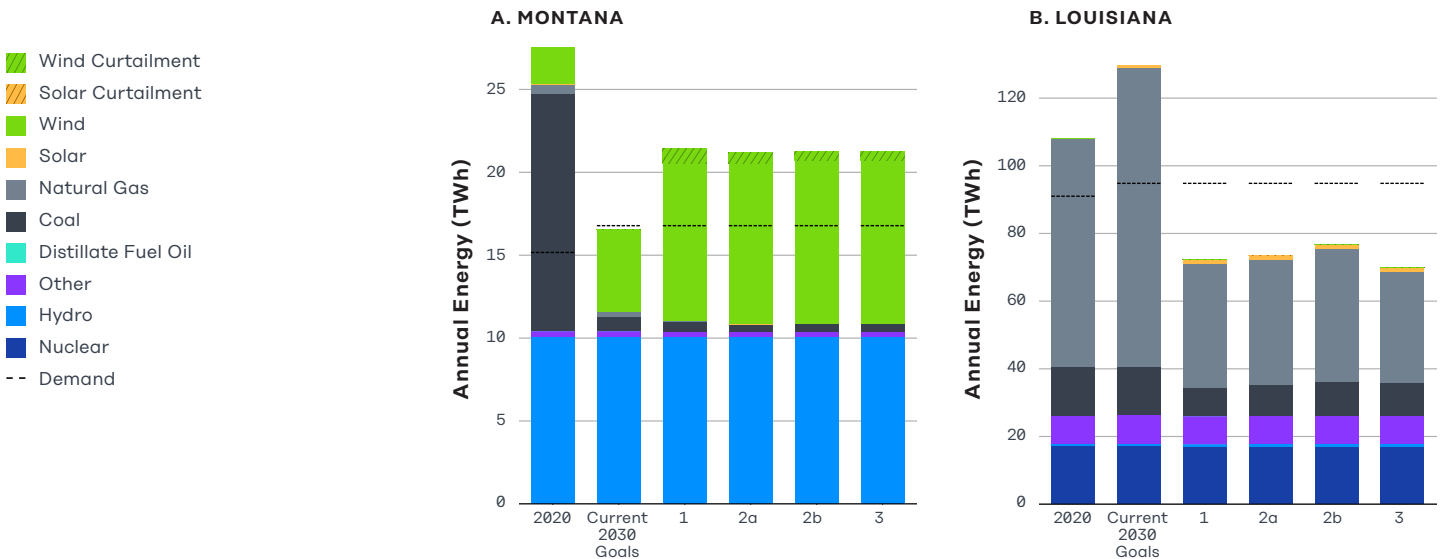

The impact of the shift in generation by source and location is also reflected in the balance of payments from consumers and to generators, as shown in Table 12, which reflects payments based on locational marginal prices (LMPs). Although there are many other aspects besides just LMPs that affect the ultimate balance of payments (e.g., long-term power purchase agreements, capacity market payments, and payments for providing ancillary services), this table illustrates the payments that would be made in a centralized energy market and how these payments could be affected by ambitious buildouts of new renewable generation capacity. In the 'current goals' scenario, payments to generators rise by similar amounts in all states, but payments by consumers grow faster in states without goals.* In the 'ambitious goals' scenarios, payments from consumers fall in all states as the share of zero-marginal-cost generation increases, but fall faster in the states with goals. Similarly, payments to generators fall in all states, but they fall by a greater amount for generators in states without goals.

TABLE 12

## Balance of Payments

The impact of large renewable buildouts on wholesale electricity prices (results shown for Design 3, locational marginal prices only) paid by consumers and received by generators.

|  | STATES WITH 2030 GOALS | STATES WITHOUT 2030 GOALS |
|---|---|---|
| Current grid: Payments from consumers | 91.0 $B | 33.7 $B |
| 2030 Current Goals: Payments from consumers | 100.2 $B | 41.2 $B |
| 2030 Current Goals: Payments from consumers (difference) | +10% | +22% |
| 2030 Ambitious Goals: Payments from consumers | 62.5 $B | 25.6 $B |
| 2030 Ambitious Goals: Payments from consumers (difference) | -31% | -24% |
| Current Grid: Payments to generators | 88.7 $B | 33.1 $B |
| 2030 Current Goals: Payments to generators | 92.5 $B | 34.8 $B |
| 2030 Current Goals: Payments to generators (difference) | +5.2% | +4.4% |
| 2030 Ambitious Goals: Payments to generators | 58.3 $B | 16.9 $B |
| 2030 Ambitious Goals: Payments to generators (difference) | -34% | -49% |

* The difference in payments by consumers and payments to generators is the 'congestion surplus,' a by-product of transmission congestion.

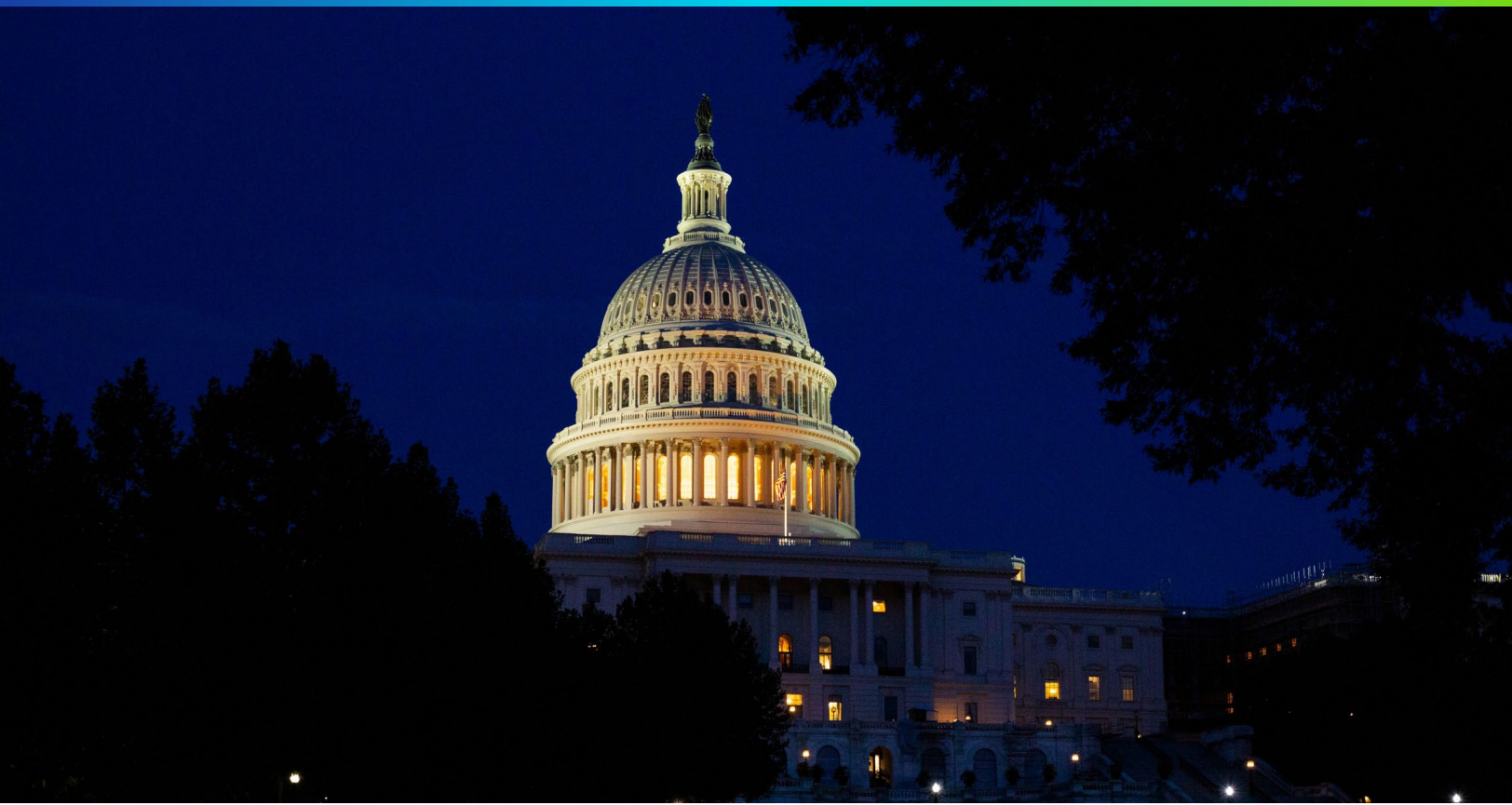

# 5.0 Policy Implications

Strong transmission policies are required to achieve the Macro Grid needed to meet the ambitious goal of an electric grid powered by 70% clean energy by 2030. For whichever style of Macro Grid is chosen, the needs are clear: improve the reliability, operating efficiency, and resilience of the U.S. power system; enable the integration of more variable renewable energy sources such as solar and wind; and help provide the necessary infrastructure for wide-area power exchange across the country. Congress should clarify and strengthen the authority of the Federal Energy Regulatory Commission (FERC) to require regional, interregional, and interconnection-level transmission planning to help make the Macro Grid a reality. FERC should coordinate the Macro Grid planning process, taking into account state energy policies, including clean energy goals, as well as utility resource plans.

One major challenge of building a Macro Grid is the disparate nature of the transmission planning and permitting process. State and local authorities along the path of a transmission line have primary jurisdiction, as they should. However, following up on the Energy Policy Act of 2005, Congress should further enhance FERC's authority to quickly approve projects and permitting in electric-transmission corridors deemed in the national interest. This would allow FERC to resolve disputes and mitigate unreasonable delays while maintaining critical environmental considerations in transmission planning and development. FERC and other federal agencies should assist states in the development of mutually acceptable routes, which will also prevent some disputes from occurring in the first place.

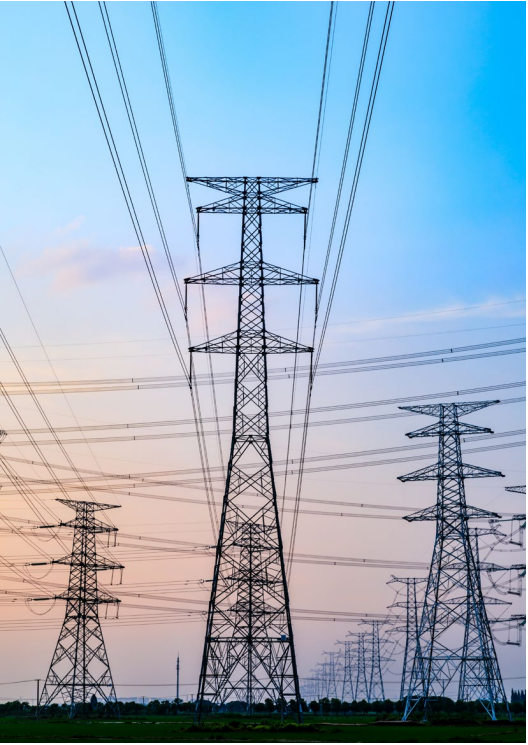

Additional federal agencies that should be engaged fully are the Department of Energy's four Power Marketing Administrations (PMAs): Bonneville Power Administration (BPA), Western Area Power Administration (WAPA), Southwestern Power Administration (SWPA), and Southeastern Power Administration (SEPA). The PMAs already own tens of thousands of miles of high-voltage transmission lines, have already been granted financing and development authority by Congress, and have considerable experience developing, owning, and operating transmission lines in the public interest. Thus, they are in the position to plan, build, own, and operate substantial parts of the Macro Grid. Another option is to combine high-voltage transmission projects with other projects requiring similar rights-of-way. Interstate highways and rail lines could have high-voltage transmission lines co-located, streamlining the environmental reviews across all projects if jointly planned.

Macro Grid policy should work to advance both AC and DC lines without bias. With AC lines allowing local communities and networks to enjoy reliable energy and DC lines enabling more efficient transmission over long distances, each technology should be implemented where appropriate. Particularly for long-distance HVDC lines, which for technical and economic reasons have a limited number of connections to the AC grid, some regions are likely to simply host a line and may not see a direct connection to nor direct benefit from the line because adding a local converter station might cause the project to no longer be economically viable. In those cases, to encourage support from states that do not have a direct connection, a policy that considers the system-wide benefits of transmission lines in their planning processes could provide financial support to the states hosting interstate lines, spreading the collective benefits to all that are impacted by the line.

Ultimately, one, if not the most, critical question falls to who pays for new transmission lines. Most RTOs use a cost allocation method that focuses on beneficiaries, which is understandable given that determining cost causation is incredibly difficult, even within just one region of the electric grid. Expanding to a Macro Grid creates more challenges, not least of which is the modeling inconsistencies between different regions. Interregional collaboration and cost allocation are difficult to move forward without a jointly built and maintained model providing a trusted source to determine the interregional beneficiaries. The model used in this study, developed by Breakthrough Energy Sciences to be open-access and publicly available, could potentially fill such a role. FERC should take actions to further encourage data openness and accessibility across incumbent parties to remove the potential burden on technical studies.

> To support whoever pays in the end, the federal government should provide upfront financing of regionally beneficial lines and provide financial incentives to encourage the development and deployment of a Macro Grid.

Another strategy employed by many RTOs is a system of 'participant funding,' whereby transmission needed to serve large renewable resource areas gets assigned to one or a handful of individual generators seeking to interconnect.[49] This can place an insurmountable burden on generators at the front of the interconnection queue, delaying, if not completely preventing, further development of that renewable resource area. One policy that would remove this obstacle is an interregional cost allocation scheme that allows for proactive transmission construction that is financed in part by load and other beneficiaries around the system, rather than individual generators alone, based on the expected benefits observed from a transmission planning model.

To support whoever pays in the end, the federal government should provide upfront financing of regionally beneficial lines and provide financial incentives to encourage the development and deployment of a Macro Grid. This will help smooth out the initial cost increases until the benefits, which often do not materialize immediately, have time to be realized. This upfront financing will minimize or perhaps eliminate any rate shocks ratepayers might experience, which many states would be likely to resist. These incentives could include investment tax credits for developers of new high-voltage interregional lines, loans akin to those administered by the Transportation Infrastructure Finance and Innovation Act (TIFIA), or bonds similar to Competitive Renewable Energy Bonds (CREBs), that helped finance renewable energy and transmission development.

Last but not least, Macro Grid policy should encourage both regulated and market-based 'merchant' transmission business models. Transmission remains both a natural monopoly and public good, so regulated transmission is needed to build an efficient and reliable Macro Grid at scale. At the same time, encouraging merchant transmission will help offset regulatory costs through voluntary capacity reservations by market participants and enable robust participation by both the public and private sectors.

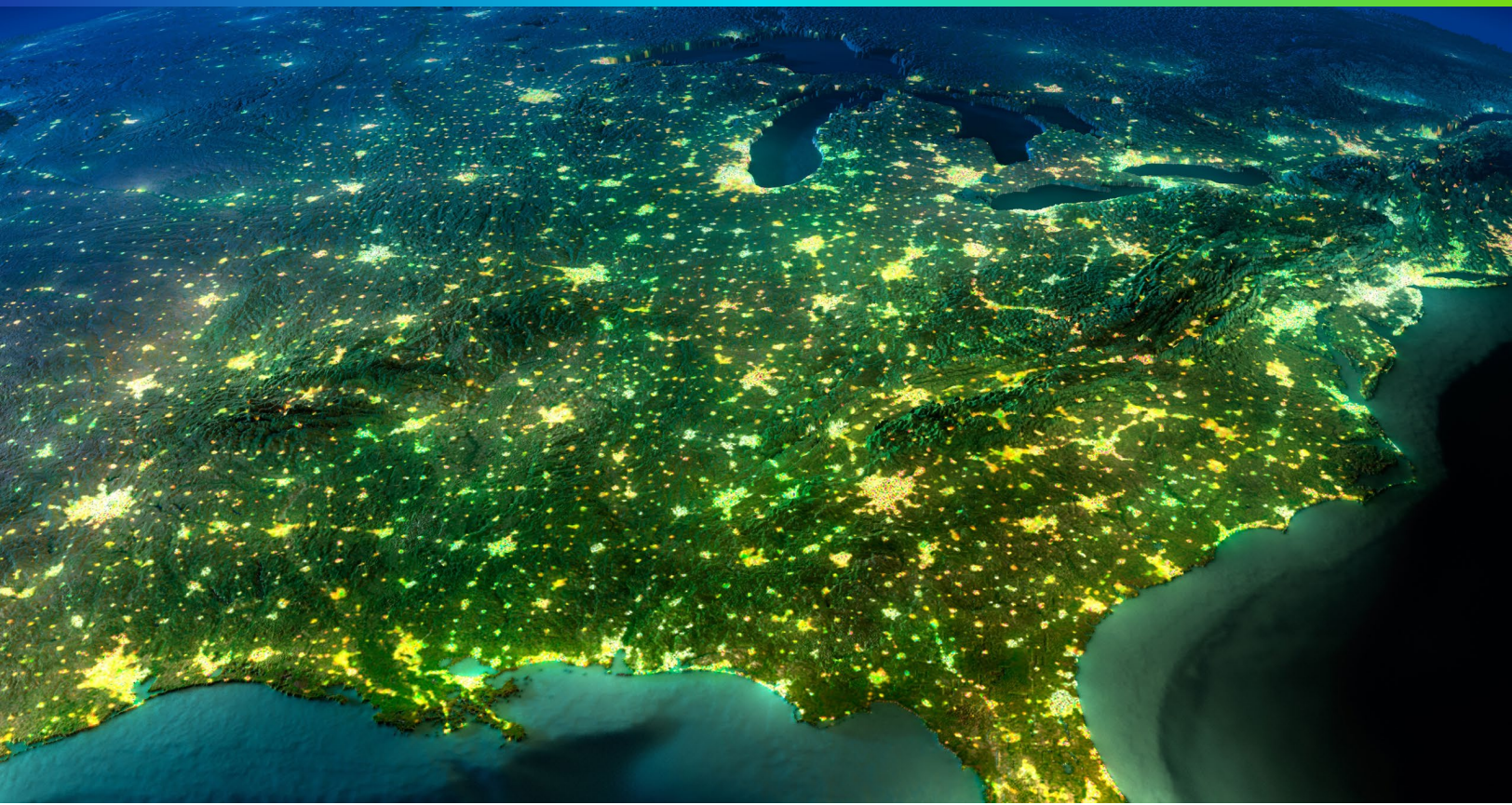

# 6.0 Opportunities for Future Work

While the heuristic method used to upgrade transmission capacity provides a realistic approach for identifying potential upgrades, the selection of transmission upgrades would likely be improved by using an integrated expansion model strictly based on mathematical optimization. The heuristic model presented in this report allows for only a few potential investment trajectories to be considered via upgrades to existing transmission corridors. Implementing a pure mathematical optimization capacity expansion model that explores the option of adding new transmission corridors along with upgrading existing transmission corridors would allow for a much larger solution space of investments to be explored. This would ensure that an optimal (i.e., minimum cost) set of transmission capacity upgrades would be implemented. The Breakthrough Energy Sciences team is currently implementing a generation and transmission capacity expansion model. An expansion optimization model will not only be useful to find economically efficient solutions and identify potential new issues, but will also be capable of suggesting process improvements from a top-down perspective for planning studies and cost allocation methods. This will facilitate better interregional coordination and help unlock potential synergistic benefits among different balancing areas, which is becoming increasingly important under a higher renewable penetration future.

The synthetic grid model presented in this report features some simplifications that are not fully representative of true power system operations. The model is solved as a multi-period DCOPF, with the results rolling over the course of the year. Additionally, this model does not take unit commitment nor security constraints into account. In particular, when integrating the inverter-based solar and wind generation sources at high penetrations, there is likely to be additional system stability concerns for maintaining system security. Finally, ancillary services, such as operating reserves and frequency regulation, are not accounted for as they have relatively limited impact on the main focus of this work, which is on bulk energy deliverability by the transmission backbone. Despite not being co-optimized, operating reserves are determined following the simulation and found to be greater than 15% during all hours. While not considered in this study, these ancillary services would likely reveal further benefits, particularly for designs with upgraded HVDC capacity, due to reductions in operating costs.[4] The inclusion of unit commitment, security constraints, and ancillary services, which would each make the open-source model more representative of power system operations, is left to future work.

For this study, only profiles of the 2016 weather year, including solar, wind, and hydro generation profiles along with the demand profiles, are used. However, the true weather in the year 2030 is unknowable, and a good Macro Grid design should perform well under a variety of weather year realizations. Areas of future work could include simulating multiple weather years to determine the impact on different Macro Grid designs and designing Macro Grids to be robust to various weather years while still meeting clean energy goals.

While this report focuses on the use of solar and wind energy and the development of a Macro Grid for achieving decarbonization goals, there are many other avenues that could have an impact as well. One such avenue that this report does not consider is energy storage. Energy storage has the ability to shift stochastic renewable energy from times of excessive generation to times of high demand. Declines in the costs of battery energy storage have made intraday storage much more economically feasible.[3] Additionally, there is increasing interest in the production and storage of hydrogen for use in the industrial and transportation sectors and as a form of long-duration storage. Other types of long-duration storage, including flow batteries, molten salts, and underground pumped hydro, are actively being explored with the target of reducing multi-day long-duration energy storage costs to competitive levels. Another area that this report does not consider is the impact of wide-scale electrification. As other sectors work towards their own decarbonization goals, it is expected that many processes that currently rely on coal or natural gas will transition to electricity. These changes will not only impact the electricity demand profiles of consumers, but will also introduce new flexibility to grid operators. Flexibility presented by demand-side resources will have the ability to reduce energy consumption and defer transmission capacity upgrades.[50] The use of energy storage and demand-side flexibility are two areas of forthcoming future work.

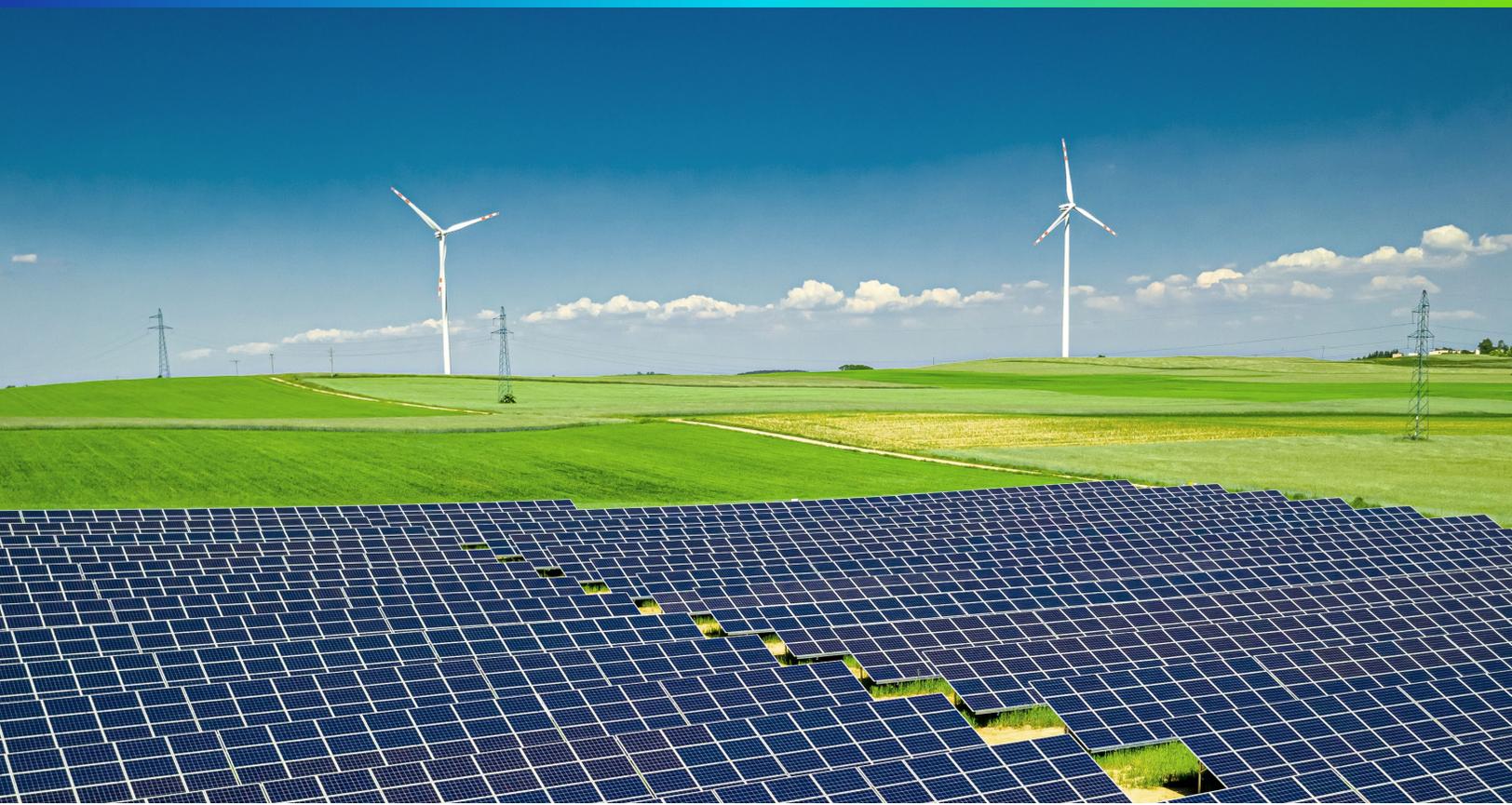

# 7.0 Conclusion

Achieving the deep decarbonization targets recommended by the UN IPCC 2018 report[1] will require major changes to the contemporary U.S. electric grid. One of the pathways to deep decarbonization requires states to adopt clean energy goals that are much more ambitious, with a vision that includes the addition of large quantities of clean generation, including but not limited to solar and wind, and the accompanying transmission capacity to deliver that clean generation to the U.S. demand centers. This report discusses four potential Macro Grid designs that could help the U.S. reach an electric grid powered by 70% clean energy by 2030. The designs that were explored are:

— Design 1, featuring upgrades to AC transmission within each interconnection. Clean energy goals are achieved with the most limited interchange between the three interconnections.

— Design 2a, which provides capacity upgrades to the current back-to-back converter stations between the interconnections. The AC transmission network within each interconnection is upgraded as well.

— Design 2b, which has three new HVDC lines across the seam between the Western and Eastern Interconnections. This design also includes upgrades to the existing back-to-back converter stations and AC transmission network.

- Design 3, which features a nationwide HVDC network with sixteen new HVDC lines. This HVDC network, which spans each of the three interconnections, provides the most robust cross-seam transmission capacity among the designs. Similar to the other three designs, AC transmission is also upgraded within each interconnection.

While all of these designs enable the same quantity of renewable generation to reach consumers, and feature similar reductions in fuel costs (46-47%) and emissions of $CO_2$ (42%), $NO_X$ (38-39%), and $SO_2$ (29-31%) compared to the 'current goals' scenario, there are differences in how these results are achieved. The design with no upgrades in cross-seam HVDC transmission (Design 1) requires the largest increase in AC transmission network capacity (36%), while the design with the country-spanning HVDC network (Design 3) requires the least (23%). However, these variations do not result in substantial differences in the overall value proposition for investing in these Macro Grid designs, and the simple payback periods are very similar across designs, even under different assumptions about the value of avoided $CO_2$ emissions.

Strong transmission polices are required to achieve the Macro Grid needed to meet the ambitious goal of an electric grid powered by 70% clean energy by 2030. FERC should take a greater role in coordinating regional, interregional, and interconnection-level transmission planning, in cooperation with the Department of Energy's existing Power Marketing Administrations. New policies can ensure that the costs of new transmission lines are distributed among all beneficiaries, and that jurisdictions hosting new transmission lines without a direct benefit are still compensated for their contribution. The federal government could also support the financing of these new lines via a combination of tax credits and loan programs. In concert, a suite of new federal policies could improve the efficiency of transmission planning and markets and enable robust participation by both the public and private sectors.

Ultimately, the distinction between an HVDC or an AC Macro Grid design is less important from an energy delivery perspective in order to meet these ambitious goals and assist the U.S. in decarbonizing the electric grid.

> **In all cases, building a Macro Grid that unlocks the geographic diversity of U.S. renewable resources and delivers clean energy to major demand centers is of critical importance to accomplishing ambitious clean energy goals and accelerating the U.S. on the path towards economy-wide deep decarbonization.**

# References

1. United Nations Intergovernmental Panel on Climate Change. *Special Report: Global Warming of 1.5 °C.* 2018. www.ipcc.ch/sr15/ (Accessed: Oct. 28, 2020.)

2. National Conference of State Legislatures. *State Renewable Portfolio Standards and Goals.* www.ncsl.org/research/energy/renewable-portfolio-standards.aspx (Accessed: Oct. 2, 2020.)

3. National Renewable Energy Laboratory. *Annual Technology Baseline.* https://atb.nrel.gov/ (Accessed: September 28, 2020.)

4. Aaron Bloom et al. *The Value of Increased HVDC Capacity Between Eastern and Western U.S. Grids: The Interconnections Seams Study.* Submitted to IEEE Transactions on Power Systems, available at: www.nrel.gov/docs/fy21osti/76850.pdf (2020)

5. Dale L Osborn. "Designing self-contingent HVDC systems with the AC systems." In: *2016 IEEE Power and Energy Society General Meeting (PESGM)*. IEEE. 2016, pp. 1–4.

6. Electric Reliability Council of Texas. *2019 State of the Grid Report.* 2019. www.ercot.com/content/wcm/lists/197391/2019_ERCOT_State_of_the_Grid_Report.pdf (Accessed: October 2, 2020.)

7. Federal Energy Regulatory Commission. *2017 Transmission Metrics.* www.ferc.gov/sites/default/files/2020-04/transmission-investment-metrics_0.pdf (Accessed: Oct. 28, 2020.)

8. Aaron Bloom. "Interconnections Seam Study." Presented at TransGrid-X 2030, July 26, 2018 in Ames, IA.

9. Gregory Brinkman et al. *Interconnections Seam Study.* www.nrel.gov/docs/fy21osti/78161.pdf (Accessed: October 24, 2020.)

10. National Renewable Energy Laboratory. *Eastern Renewable Generation Integration Study.* www.nrel.gov/grid/ergis.html (Accessed: October 2, 2020.)

11. David Corbus et al. "Eastern Wind Integration and Transmission Study." In: NREL www.nrel.gov/docs/fy09osti/46505.pdf - CP-550-4650513 (2010), pp. 1–8.

12. GE Energy. *Western Wind and Solar Integration Study.* Tech. rep. National Renewable Energy Lab (NREL), Golden, CO (United States), 2010.

13. Debra Lew et al. *Western Wind and Solar Integration Study Phase 2.* Tech. rep. National Renewable Energy Lab (NREL), Golden, CO (United States), 2013.

14. National Renewable Energy Laboratory. *North American Renewable Integration Study.* www.nrel.gov/analysis/naris.html (Accessed: October 2, 2020.)

15. American Council on Renewable Energy. *Macro Grid Initiative.* https://acore.org/macro-grid-initiative (Accessed: September 28, 2020.)

16. Stefan Pfenninger et al. "Opening the black box of energy modelling: Strategies and lessons learned." In: *Energy Strategy Reviews 19* (2018), pp. 63–71.

17. Joseph F. DeCarolis et al. "Leveraging Open-Source Tools for Collaborative Macro-energy System Modeling Efforts." In: *Joule* 4 (2020), pp. 1–4.

18. Armando Luis Figueroa-Acevedo et al. "Design and Valuation of High-Capacity HVDC Macrogrid Transmission for the Continental US." In: *IEEE Transactions on Power Systems* (2020.)

19. Armando Luis Figueroa-Acevedo. "Opportunities and benefits for increasing transmission capacity between the US eastern and western interconnections." PhD dissertation. Iowa State University, 2017.

20. Breakthrough Energy Sciences. https://science.breakthroughenergy.org/

21. National Renewable Energy Laboratory. *Regional Energy Deployment System Model.* www.nrel.gov/analysis/reeds/index.html (Accessed: September 28, 2020.)

22. Stuart Cohen et al. *Regional Energy Deployment System (ReEDS) Model Documentation: Version 2018.* Tech. rep. NREL/TP-6A20-72023. Golden, CO: National Renewable Energy Laboratory, 2019. www.nrel.gov/docs/fy19osti/72023.pdf

23. Texas A&M University. *Electric Grid Test Cases.* https://electricgrids.engr.tamu.edu/electric-grid-test-cases

24. *Reliability test system of the Grid Modernization Laboratory Consortium.* https://github.com/GridMod/RTS-GMLC

25. Yixing Xu et al. "U.S. Test System with High Spatial and Temporal Resolution for Renewable Integration Studies." In: 2020 IEEE Power & Energy Society General Meeting (PESGM), Montreal, QC, 2020, pp. 1-5.

26. Yixing Xu, et al. *U.S. Test System with High Spatial and Temporal Resolution for Renewable Integration Studies, 2021.* Zenodo. https://zenodo.org/record/3530898

27. Energy Information Agency. *Form 923, Power Plant Operations Report.*

28. Energy Information Agency. *Form 860, Annual Electric Generator Report, and Form 860M, Monthly Update to the Annual Electric Generator Report.*

29. Chris Kavalec et al. *California Energy Demand 2018-2030 Revised Forecast.* Tech. rep. CEC-200-2018-002-CMF. California Energy Commission, 2018. www.energy.ca.gov/data-reports/reports/integrated-energy-policy-report/2017-integrated-energy-policy-report/2017-iepr

30. *2020 ERCOT System Planning: Long-Term Hourly Peak Demand and Energy Forecast.* Tech. rep. Electric Reliability Council of Texas, 2019. www.ercot.com/content/wcm/lists/196030/2020_LTLF_Report.pdf

31. ISO New England. *Forecast Report of Capacity, Energy, Loads, and Transmission.* Tech. rep. ISO New England, 2019. www.iso-ne.com/system-planning/system-plans-studies/celt

32. Douglas J. Gotham et al. *MISO Energy and Peak Demand Forecasting for System Planning.* Tech. rep. State Utility Forecasting Group at Purdue University, 2018. www.purdue.edu/discoverypark/sufg/miso/reports-presentations.php

33. NERC. *2018 Long-Term Reliability Assessment.* Tech. rep. NERC, 2018. www.nerc.com/pa/RAPA/ra/Reliability%5C%20Assessments%5C%20DL/NERC_LTRA_2018_12202018.pdf

34. NYISO. *2018 Power Trends: New York's Dynamic Power Grid NYISO.* Tech. rep. NYISO, 2018. www.nyiso.com/documents/20142/2223020/2018-Power-Trends.pdf

35. PJM Resource Adequacy Planning Department. *PJM Load Forecast Report.* Tech. rep. PJM, 2020. www.pjm.com/-/media/library/reports-notices/load-forecast/2020-load-report.ashx?la=en

36. State of California. *SB-100 California Renewables Portfolio Standard Program: Emissions of Greenhouse Gases.* 2018. https://leginfo.legislature.ca.gov/faces/billTextClient.xhtml?bill_id=201720180SB100

37. National Renewable Energy Laboratory. *Standard Scenarios: Mid-Case Scenario.* https://openei.org/apps/reeds (Accessed: July 2, 2020.)

38. Midcontinent Independent System Operator. *MISO Transmission Expansion Plan.* https://cdn.misoenergy.org//Final%20Draft%20MTEP20%20%20Chapter%203%20-%20Policy%20&%20Economic%20Studies485665.pdf (Accessed: October 28, 2020.)

39. California Independent System Operator. *ISO Board Approved 2019-2020 Transmission Plan.* www.caiso.com/Documents/ISOBoardApproved-2019-2020TransmissionPlan.pdf (Accessed: October 28, 2020.)

40. CAISO. *Western Energy Imbalance Market.* www.westerneim.com/Pages/About/default.aspx (Accessed: October 12, 2020.)

41. *2018 Long-term System Assessment for the ERCOT Region December 2018.* Tech. rep. Electric Reliability Council of Texas, 2018. http://www.ercot.com/content/wcm/lists/144927/2018_ LTSA_Report.pdf

42. National Renewable Energy Laboratory. *Interconnections Seam Study.* www.nrel.gov/analysis/seams.html (Accessed: October 2, 2020.)

43. Russell Gold. *Superpower: One Man's Quest to Transform American Energy.* Simon & Schuster, 2019.

44. *Transmission Cost Estimation Guide: MTEP2020.* Tech. rep. MISO, July 3, 2020.

45. Amol Phadke et al. *2035 The Report.* 2020. www.2035report.com (Accessed: Oct. 28, 2020.)

46. Yuqiang Zhang et al. "Co-benefits of global, domestic, and sectoral greenhouse gas mitigation for US air quality and human health in 2050." In: *Environmental Research Letters* 12.11 (2017).

47. Toon Vandyck et al. "Air quality co-benefits for human health and agriculture counterbalance costs to meet Paris Agreement pledges." In: *Nature communications* 9.1 (2018).

48. Emma Penrod. *Louisiana governor puts state on path to net-zero emissions by 2050.* www.utilitydive.com/news/louisiana-governor-puts-state-on-path-to-net-zero-emissions-by-2050/583924 (Accessed: October 10, 2020.)

49. Midcontinent Independent System Operator. *Generator Interconnection.* www.misoenergy.org/planning/generator-interconnection (Accessed: October 28, 2020.)

50. Trieu Mai et al. *Electrification Futures Study: Scenarios of Electric Technology Adoption and Power Consumption for the United States.* Tech. rep. NREL/TP-6A20-71500. Golden, CO: National Renewable Energy Laboratory, 2018. www.nrel.gov/docs/fy18osti/71500.pdf